\documentclass[a4paper,fleqn,usenatbib]{mnras}
\usepackage{newtxtext,newtxmath}
\pdfoutput=1 
\usepackage[T1]{fontenc}
\usepackage{ae,aecompl}
\usepackage{graphicx}	
\usepackage{amsmath}	
\usepackage{amssymb}	
\usepackage[flushleft]{threeparttable}
\usepackage{booktabs,caption}
\usepackage{rotating}
\usepackage{enumitem}
\usepackage{pdflscape}
\usepackage{gensymb}
\usepackage{xspace}
\usepackage{geometry}
\usepackage{xcolor}

\usepackage[section]{placeins}

\newcommand{\ie}{{\it i.e.}\xspace}
\newcommand{\eg}{{\it e.g.}\xspace}

\newcommand{\e}{\ensuremath{\,{\!=\!}\,}\xspace}
\newcommand{\sm}{\ensuremath{\,{\rm M_{\odot}}}\xspace}
\newcommand{\slu}{\ensuremath{\,{\rm L_{\odot}}}\xspace}
\newcommand{\ml}{\ensuremath{\,{\rm M_{\odot}\,L_{\odot}^{-1}}}\xspace}

\newcommand{\si}{\ensuremath{\,{\!\sim\!}\,}\xspace}
\newcommand{\Mpc}{\ensuremath{\,{\rm Mpc}}\xspace}
\newcommand{\kpc}{\ensuremath{\,{\rm kpc}}\xspace}

\newcommand{\ikpc}{\ensuremath{\,{\rm kpc^{-1}}}\xspace}
\newcommand{\pc}{\ensuremath{\,{\rm pc}}\xspace}

\newcommand{\iyr}{\ensuremath{\,{\rm yr^{-1}}}\xspace}
\newcommand{\Myr}{\ensuremath{\,{\rm Myr}}\xspace}
\newcommand{\Gyr}{\ensuremath{\,{\rm Gyr}}\xspace}
\newcommand{\kms}{\ensuremath{\,{\rm km}\,{\rm s}^{\rm -1}}\xspace}
\newcommand{\s}{\ensuremath{\,{\rm s}}\xspace}
\newcommand{\km}{\ensuremath{\,{\rm km}}\xspace}

\newcommand{\as}{\ensuremath{\,{\rm arcsec}}\xspace}

\newcommand{\cm}{\ensuremath{\,{\rm cm}}\xspace}
\newcommand{\icmsq}{\ensuremath{\,{\rm cm^{-2}}}\xspace}

\newcommand{\mags}{\ensuremath{\,{\rm mag}}\xspace}
\renewcommand{\arcmin}{\ensuremath{\,{\rm arcmin}}\xspace}

\newcommand{\HI}{{\textnormal{H}}{\small \textnormal{I}}\xspace}

\newcommand{\mud}{\ensuremath{\mu_{\delta}}\xspace}

\newcommand{\muas}{\ensuremath{\mu_{\alpha^{*}}}\xspace}
\newcommand{\muasz}{\ensuremath{\mu^{\circ}_{\alpha^{*}}}\xspace}
\newcommand{\mudz}{\ensuremath{\mu^{\circ}_{\delta}}\xspace}

\newcommand{\masyr}{\ensuremath{\,{\rm mas\,yr^{-1}}}\xspace}
\newcommand{\rvir}{\ensuremath{r_{\rm vir}}\xspace}
\newcommand{\Rvir}{\ensuremath{R_{\rm vir}}\xspace}
\newcommand{\z}{\ensuremath{{\rm z}}\xspace}
\newcommand{\vlosgsr}{\ensuremath{\vec{v}_{\rm los}^{\rm GSR}}\xspace}

\newcommand{\utgsr}{\ensuremath{\vec{u}_{\rm t}^{\rm GSR}}\xspace}
\newcommand{\Nutgsr}{\ensuremath{| \vec{u}_{\rm t}^{\rm GSR}|}\xspace}

\newcommand{\Nutgsrbas}{\ensuremath{| \vec{u}_{\rm t\,BAS}^{\rm GSR}|}\xspace}
\newcommand{\Nutgsrrps}{\ensuremath{| \vec{u}_{\rm t\,RPS}^{\rm GSR}|}\xspace}

\newcommand{\utgc}{\ensuremath{\vec{u}_{\rm t}^{\rm GC}}\xspace}

\newcommand{\vlosgc}{\ensuremath{\vec{v}_{\rm los}^{\rm GC}}\xspace}
\newcommand{\unvlosgc}{\ensuremath{\hat{v}_{\rm los}^{\rm GC}}\xspace}

\newcommand{\PA}{\ensuremath{{\rm PA}}\xspace}
\newcommand{\PAO}{\ensuremath{{\rm PA_{\rm offset}}}\xspace}

\newcommand{\PAF}{\ensuremath{{\rm PA_{\rm flat}}}\xspace}

\newcommand{\PAOV}{\ensuremath{{\rm PA_{\rm offset}\!=\!3\degree}}\xspace}
\newcommand{\PATV}{\ensuremath{{\rm PA_{\rm tail}\!=\!-10\degree}}\xspace}
\newcommand{\PAFV}{\ensuremath{{\rm PA_{\rm flat}\!=\!-21\degree}}\xspace}

\usepackage{scalerel}
\usepackage{tikz}
\usetikzlibrary{svg.path}

\definecolor{orcidlogocol}{HTML}{A6CE39}
\tikzset{
  orcidlogo/.pic={
    \fill[orcidlogocol] svg{M256,128c0,70.7-57.3,128-128,128C57.3,256,0,198.7,0,128C0,57.3,57.3,0,128,0C198.7,0,256,57.3,256,128z};
    \fill[white] svg{M86.3,186.2H70.9V79.1h15.4v48.4V186.2z}
                 svg{M108.9,79.1h41.6c39.6,0,57,28.3,57,53.6c0,27.5-21.5,53.6-56.8,53.6h-41.8V79.1z M124.3,172.4h24.5c34.9,0,42.9-26.5,42.9-39.7c0-21.5-13.7-39.7-43.7-39.7h-23.7V172.4z}
                 svg{M88.7,56.8c0,5.5-4.5,10.1-10.1,10.1c-5.6,0-10.1-4.6-10.1-10.1c0-5.6,4.5-10.1,10.1-10.1C84.2,46.7,88.7,51.3,88.7,56.8z};
  }
}

\newcommand\orcidicon[1]{\href{https://orcid.org/#1}{\mbox{\scalerel*{
\begin{tikzpicture}[yscale=-1,transform shape]
\pic{orcidlogo};
\end{tikzpicture}
}{|}}}}

\usepackage{hyperref}

\newcommand{\ch}{}

\newcommand{\chn}{}

\renewcommand{\arraystretch}{1.3}
\defcitealias{Bland-Hawthorn2016}{BG16}
\defcitealias{Adams2018}{AO18}

\title[First in-fall or backsplash MW satellites?]{Dwarfs in the Milky Way halo outer rim: first in-fall or backsplash satellites?}

\author[M. Bla\~{n}a et al.]
{Mat\'ias Bla\~{n}a $^{1,2}$\thanks{Web:\href{http://matiasblana.github.io}{matiasblana.github.io} E-mail:mblana@mpe.mpg.de} \orcidicon{0000-0003-2139-0944},
Andreas Burkert $^{1,2,4}$  \orcidicon{0000-0001-6879-9822},
Michael Fellhauer $^{3}$   \orcidicon{0000-0002-3989-4115},
\newauthor
Marc Schartmann $^{1,2,4}$   \orcidicon{0000-0003-1318-8631},
Christian Alig$^{2}$
\\
$^{1}$ Max-Planck-Institut f\"ur extraterrestrische Physik, Gie\ss enbachstra\ss e 1, D-85748 Garching bei M\"unchen, Germany,\\
$^{2}$ Universit\"ats-Sternwarte, Fakult\"at f\"ur Physik, Ludwig-Maximilians-Universit\"at M\"unchen, Scheinerstraße 1, D-81679 M\"unchen, Germany\\
$^{3}$ Departamento de Astronom\'ia, Universidad de Concepci\'on, Avenida Esteban Iturra s/n Casilla 160-C, Concepci\'on, Chile\\
$^{4}$ Excellence Cluster ORIGINS, Boltzmannstr. 2, D-85748 Garching bei M\"unchen, Germany}

\date{Accepted 2020 July 16. Received 2020 June 20; in original form 2020 March 9}

\pubyear{2020}

\begin{document}
\label{firstpage}
\pagerange{\pageref{firstpage}--\pageref{lastpage}}
\maketitle

\begin{abstract}
Leo T is a gas-rich dwarf located at $414\kpc$ $(1.4\Rvir)$ distance from the Milky Way (MW) and it is currently assumed to be on its first approach. 
Here, we present an analysis of orbits calculated backward in time for the dwarf 
with our new code 
{\sc delorean},
exploring a range of systematic uncertainties, \eg\,MW virial mass and accretion, 
M31 potential, and cosmic expansion.
We discover that orbits with tangential velocities in the Galactic Standard-of-Rest frame lower than $\Nutgsr\!\leq\! 63^{+47}_{-39}\kms$ result in backsplash solutions, 
i.e. orbits that entered and left the MW dark matter halo in the past,
and that velocities above $\Nutgsr\!\geq\!21^{+33}_{-21}\kms$ result in wide orbit backsplash solutions with a minimum pericenter range of $D_{\rm min}\!\geq\!38^{+26}_{-16}\kpc$, 
which would allow this satellite to survive gas stripping and tidal disruption. 
\chn{Moreover, new proper motion estimates match with our region of backsplash solutions.}
We applied our method to other distant MW satellites, finding a range of gas stripped backsplash solutions for the gas-less Cetus and Eridanus II, 
providing a possible explanation for their lack of cold gas, 
while only first in-fall solutions are found for the \HI rich Phoenix I.
We also find that the cosmic expansion can delay their first pericenter passage when compared to the non-expanding scenario.
This study explores the provenance of these distant dwarfs and provides constraints on the environmental and internal processes that shaped their evolution and current properties.
\end{abstract}

\begin{keywords}
Local Group -- methods: numerical -- galaxies: dwarfs -- galaxies: individual:LeoT
\end{keywords}


\section{Introduction}
\label{sec:intro}
The transient \HI gas-rich dwarf galaxy Leo T, discovered by \citet{Irwin2007}, is currently in the outskirts of the Milky Way, at $D_{\odot}\e409^{+29}_{-27}\kpc$
from the Sun  \citep{Clementini2012}
with a Galactic-Standard-of-Rest (GSR) line-of-sight (LOS) stellar velocity of
$v^{\rm GSR}_{\rm los, \star}\e-65.9 \pm 2.0\kms$ \citep[][recalculated for the MW Galactocentric coordinates adopted in this work]{Simon2007}.
\citet{Adams2018}{ (hereafter \citetalias{Adams2018})} performed deep \HI observations of Leo T that show with exquisite detail its gas density and kinematic properties (see Fig.\ref{fig:LeoT:HImap}), which present features that suggest an ongoing interaction with the Milky Way gaseous halo through ram pressure stripping. This makes Leo T not only an ideal laboratory to study the formation and evolution of dwarfs \citep{Read2016}, but also a probe to study the properties of the hot halo of the Milky Way, which is the environment where these dwarfs live \citep{Grcevich2009,Gatto2013,Belokurov2017}.

An observed property of dwarfs in the Milky Way and in M31 is that, excluding the Magellanic Cloud satellites, most satellites located within the MW's virial radius ($\Rvir\e282\pm30 \kpc$ for a virial mass of $M_{\rm vir}\e1.3\pm0.3\times10^{12}\sm$ \citet{Bland-Hawthorn2016}{ hereafter \citetalias{Bland-Hawthorn2016}}) show very little or no neutral gas; while most of the dwarfs located beyond the host's virial radius show gas-to-stellar-light ratios larger than one \citep{Blitz2000,McConnachie2012,Spekkens2014}, with some interesting exceptions such as Cetus I, and Tucana.
The gas-loss in dwarfs would be a natural result of environmental interactions with the host galaxy through ram pressure stripping, tidal disruption and UV background gas evaporation \citep{Mori2000,Sawala2012,Simpson2018,Buck2019,Hausammann2019}, as well as internal effects, such as stellar feedback \citep{Read2016}. \\\\

\ch{An important prediction from cosmological galaxy simulations 
 is  the existence of two types of populations of satellites located near the virial radius of a host galaxy at redshift zero: 
 field satellites that are currently falling for the first time into the halo of the host, and another population of satellites that are currently on their second in-fall, called "backsplash" satellites, 
which are found in a Local Group context \citep{TeyssierM2012,Garrison-Kimmel2017}, and in galaxy clusters \citep{Gill2005,Lotz2019,Haggar2020}.}
The backsplash population can represent an important fraction between 30 and 50 per cent of the satellites located  between one and twice the virial radius at redshift zero \citep{Simpson2018, Rodriguez2018,Buck2019}. 
Among the backsplash population they find gas-rich satellites, as well as gas-poor satellites, which would depend on internal processes and also on the amount of gas stripping that the satellite endured during its orbital path. 
And, more importantly, \citet{Buck2019} show that the backsplash population and the first in-fall population can have similar gas fractions however, their baryon to dark matter fractions can be quite different, 
as the former could have lost up to 50 per cent of their initial dark matter masses due to tidal stripping \citep[see also][]{VandenBosch2018}. 
This makes determining the orbital history of the satellites extremely important to understand the main drivers of their evolution \citep{Tonnesen2019,Hausammann2019}, and to use them as probes of the gaseous halo of the Milky Way.
Furthermore, from cosmological Milky Way type simulations presented in \citet{TeyssierM2012,Simpson2018,Buck2019} are predicted the likelihoods of MW satellites being a backsplash system depending on their distances and LOS velocities with respect to the host galaxy, finding for example for Leo T a likelihood of 50 to 70 per cent.
\ch{Interesting as well is that Leo T is located near the first caustic or splashback radius of MW-type cosmological simulated galaxies \citep{Deason2020}.}
However, cosmological galaxy simulations provide only statistical comparisons of properties and evolution histories of dwarfs and their hosts, 
while here we use observations of dwarfs to calculate different orbits, exploring the parameter space of several uncertainties, such as the MW virial mass and more.\\
\section{Leo T observations and properties}
\label{sec:obs}
\begin{table}
\begin{threeparttable}
\caption{Main properties of Leo T}
\begin{center}
\begin{tabular}{lr}\hline \hline
${\rm RA} $ & $09^{\rm h}34^{\rm m} 53^{\rm s}\!.4$ $^{(1)}$\\  
${\rm Dec} $ & $+17^{\rm o}03'05''$  $^{(1)}$\\ 
$D_{\rm \odot} $ & $409^{+29}_{-27}\kpc$ $^{(2)}$\\ 
$D^{\rm GC} $ & $414^{+29}_{-27}\kpc$  $^{(3)}$ \\ 
$v^{\odot}_{\rm los, \star} $ & $38.1\!\pm\!2.0\kms$ $^{(4)}$\\ 
$v^{\rm GSR}_{\rm los, \star} $ & $-65.9\!\pm\!2.0\kms$ $^{(*,3,4)}$\\ 
$\muas$ & $-0.01\pm0.05\masyr$ $^{(5)}$\\ 
$\mud$ & $-0.11\pm0.05\masyr$ $^{(5)}$\\ 
$\muasz\left(\utgsr\e0\kms\right)$ & $-0.0150\masyr$ $^{(*,3)}$\\ 
$\mudz\left(\utgsr\e0\kms\right)$ & $-0.1153\masyr$ $^{(*,3)}$\\ 
$X^{\rm GC},Y^{\rm GC},Z^{\rm GC}$ & $(-250,-169,283)\kpc$ $^{(3)}$\\ 
$V_{X}^{\rm GC},V_{Y}^{\rm GC},V_{Z}^{\rm GC}$ & $(39.1,27.2,-45.5)\kms$ $^{(*,3)}$ \\\hline 
$M_{\rm V}$ & $-8.0\mags$ $^{(5)}$\\ 
$L_{\rm V}$ & $1.41\!\pm\!\times10^5\slu$ $^{(5)}$\\ 
$R_{\rm h}^{\rm V} $ & $73\!\pm\!8\as$ $(145\!\pm\!15\pc)$  $^{(6)}$\\  
$M^{\star}_{\rm half}$ & $1.05^{+0.27}_{-0.23}\times10^5\sm$ $^{(7)}$\\  
$\sigma_{\rm los, \star} $ & $7.5\!\pm\!1.6\kms$ $^{(4)}$\\ 
$M^{\rm dyn}_{\rm half}$ & $7.6\pm3.3\times10^6\sm$ $^{(3,4)}$ \\ \hline 
$M^{\rm HI}$ & $3.8\!\pm\!0.4\times10^5\sm$ $^{(8,3)}$\\  
$R_{\rm HI}$ & $106\!\pm\!10\as$ $(210\!\pm\!20\pc)$ $^{(3)}$\\   
$M^{\rm HI}_{\rm Pl}$ & $4.1\!\pm\!0.4\times10^5\sm$ $^{(3)}$\\  
$\Sigma^{\rm HI}_{\rm Pl}$ & $3.48\!\pm\!0.33\times10^6\sm\kpc^{-2}$ $^{(3)}$\\ 
$N^{\rm HI}_{\rm Pl}$ & $4.34\!\pm\!0.45\times10^{20}\cm^{-2}$ $^{(3)}$\\ 
$\rho^{\rm HI}_{\rm Pl}$ & $1.35\!\pm\!0.13\times10^7\sm\kpc^{-3}$ $^{(3)}$\\ 
$n^{\rm HI}_{\rm Pl}$ & $0.54\!\pm\!0.05\cm^{-3}$ $^{(3)}$\\\hline 
$M^{\rm gas}$ & $5.2\!\pm\!0.5\times10^5\sm$ $^{(8,3)}$ \\ 
$M^{\rm gas}_{\rm Pl}$ & $5.5\!\pm\!0.5\times10^5\sm$ $^{(3)}$\\ 
$\Sigma^{\rm gas}_{\rm Pl}$ & $4.64\!\pm\!0.45\times10^{6} \sm\kpc^{-2}$ $^{(3)}$\\ 
$\rho^{\rm gas}_{\rm Pl}$ & $1.79\!\pm\!0.17\times10^{7} \sm\kpc^{-3}$ $^{(3)}$\\ 
$R_{\rm Pl}^{\rm gas}$ & $97.81\!\pm\!0.03\as$ $(193.94\!\pm\!0.06\pc)$ $^{(3)}$\\   
$r_{\rm 3D-half,\, Pl}^{\rm gas}$ & $130.41\!\pm\!0.04\as$ $(258.59\!\pm\!0.08\pc)$ $^{(3)}$\\   
$v^{\rm \odot}_{\rm los, gas} $ & $39.6\!\pm\!0.1\kms$ $^{(8)}$\\ 
$v^{\rm GSR}_{\rm los, gas} $ & $-64.4\!\pm\!0.1\kms$ $^{(8,3)}$ \\\hline\hline 
\end{tabular}
\begin{tablenotes}
\small
\item References: 
(1) \citet{Irwin2007}, 
(2) \citet{Clementini2012}, 
(3) calculated in this publication or re-calculated from estimations in the literature that are re-scaled to an Heliocentric distance of Leo T of 409\kpc,
using a conversion from Heliocentric to Galactocentric (GC) or Galactic Standard of Rest (GSR) coordinates with the solar values presented in Section \ref{sec:met:orbit}.
(*) values when the GSR tangential velocity is assumed to be zero ($\Nutgsr=0\kms$).
(4) \citet{Simon2007}, 
(5) \citet{McConnachie2020}, 
(6) \citet{DeJong2008}, 
(7) \citet{Weisz2012},
(8) \citet{Adams2018}, 
Variables and symbols are explained in the main text.
\end{tablenotes}
\end{center}
\label{tab:LeoT:prop}
\end{threeparttable}
\end{table}

In this paper we investigate the possible origin of Leo T by studying backsplash orbital solutions as well as the first in-fall solutions.
For this we develop a new method and a software called \textsc{delorean} to  calculate orbits backwards in time considering a range of scenarios.
We also apply our new code to study the orbits of the dwarfs Cetus, Phoenix I and Eridanus II. 
The paper is ordered as follows:
In Sec. \ref{sec:obs} we detail the main properties and observations of LeoT.
In Sec.\ref{sec:met} we explain our new software to setup and calculate the orbits, and the method to analyse them. 
The results are presented and discussed in Section \ref{sec:res}, concluding then in Section \ref{sec:con}.

We present the main properties of Leo T in Table \ref{tab:LeoT:prop}.
Leo T is a gas rich dwarf located at $1.45\Rvir$ distance from the Galactic center, located at $D_{\odot}\e409\kpc$ \citep{Clementini2012}. 
\citet{DeJong2008} estimate that Leo T possesses an old stellar population with a V-band luminosity of $L_{\rm V}\e8.9\times10^4\slu$ within 5 half-light radii and also a younger population of stars (between $\si200\Myr$ and 1\Gyr in age) with $L_{\rm V}\e5.1\times10^4\slu$. 
They estimate a combined luminosity of $L_{\rm V}\e1.4\times10^5\slu$, with the younger stars contributing approximately with 10 per cent of the total stellar mass. 
\citet{Weisz2012} find a stellar mass of $10^5\sm$ within one half-light radius.
The younger population is evidence that Leo T still can form stars, although there is no strong evidence of molecular gas so far \citepalias{Adams2018}. 
This is consistent with star formation history studies that show a small and fluctuating star formation rate
with an average of $\si10^{-5}\sm\iyr$, with two peaks of high rates at 1-2 and 7-9\Gyr ago \citep{Weisz2012,Clementini2012}.

The systemic GSR stellar velocity of  Leo T is $v^{\odot}_{\rm los, \star}\e-65.9\pm2.0\kms$. 
The stellar proper motion $\muas$, $\mud$ of this dwarf is unknown.
Taking a zero tangential velocity in the GSR frame ($\utgsr\e0\kms$) transforms into the proper motion values of
$\muasz\e-0.0150\masyr$ and $\mudz\e-0.1153\masyr$, which is just the proper motion in the direction of Leo T due to the motion of the Sun relative to the Galactic center.
The stellar and the gas kinematics indicate that this dwarf is dark matter dominated, as determined by a line-of-sight velocity dispersions of between 7 and 8\kms (Table \ref{tab:LeoT:prop}), implying a dynamical mass within 400\pc of $10^6-10^7\sm$ \citep{Simon2007,RyanWeber2008,Faerman2013,Adams2018,Patra2018},
with a circular velocity $V_{\rm c}$ between $7$ and 14\kms, similar to other dwarfs \citep{McConnachie2012}.\\

Leo T has a relatively massive \HI reservoir \citep[][ \citetalias{Adams2018}]{RyanWeber2008,Grcevich2009,Faerman2013} with an estimated \HI mass of $3.8\times10^5\sm$, which is 5 per cent lower than the original value reported in \citetalias{Adams2018} ($4.1\times10^{5}\sm$), because we have rescaled the observations to the latest distance estimate of 409\kpc, 
instead of the 420\kpc used in the original publication. 
Taking a helium to hydrogen gas mass ratio of 0.33 we obtain a gas mass of $M^{\rm gas}\e5.2\pm0.5\times10^5\sm$, which does not include the ionised gas that could be surrounding the dwarf.
This results in a \HI-mass-to-stellar-light ratio of $M_{\rm \textsc{H}I}/L_{\rm V}\e2.7\ml$ and a gas-mass-to-stellar-light ratio of $M_{\rm gas}/L_{\rm V}\e3.6\ml$.
Assuming a stellar mass-to-light ratio of $2\ml$  would give us a gas-mass-to-stellar-mass ratio of $M_{\rm \textsc{H}I}/M_{\star}\si1.3$.

To estimate the central density and surface density we calculate the surface density profile fitting the \HI map of \citetalias{Adams2018} with \textsc{ellipse} \citep{Jedrzejewski1987}, shown in Fig.\ref{fig:LeoT:HIprof}, and fitted a Plummer function to obtain the central gas density and the Plummer radius of the gas (Table \ref{tab:LeoT:prop}), which we use in Section \ref{sec:met:rps} to calculate the thermal pressure. 
We also interpolate the \HI profile of the observations to calculate $R_{\rm HI}$, the radius where $\Sigma^{\rm HI}\left(R_{\rm HI}\right)\e1\sm\pc^{-2}\e1.248\times10^{20}\icmsq$ \citep{Broeils1997,WangJ2016},
with errors estimated from the 10 per cent uncertainties in the \HI mass.

\begin{figure}
\begin{center}
\includegraphics[width=8.6cm]{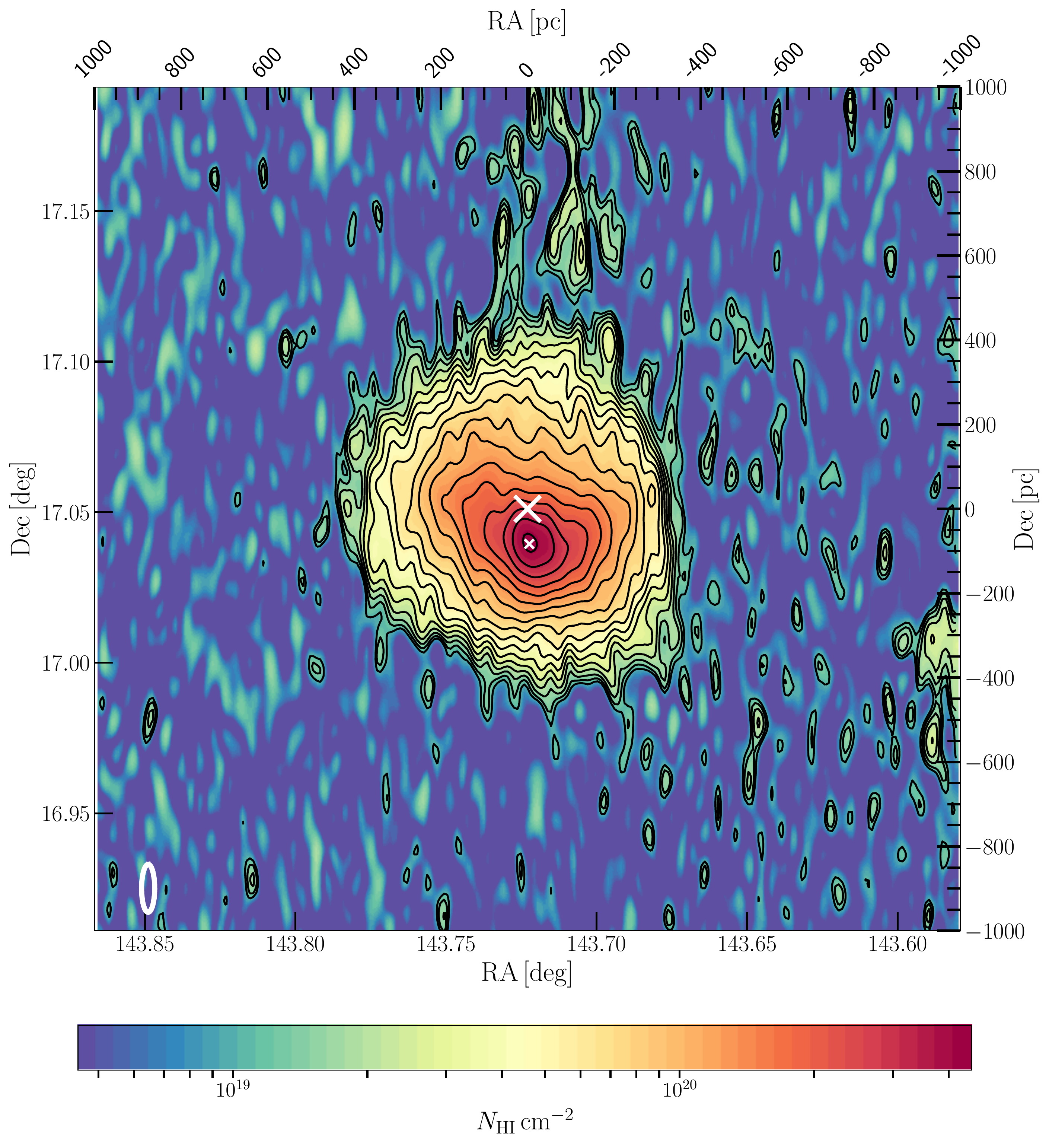}\\
\includegraphics[width=8.6cm]{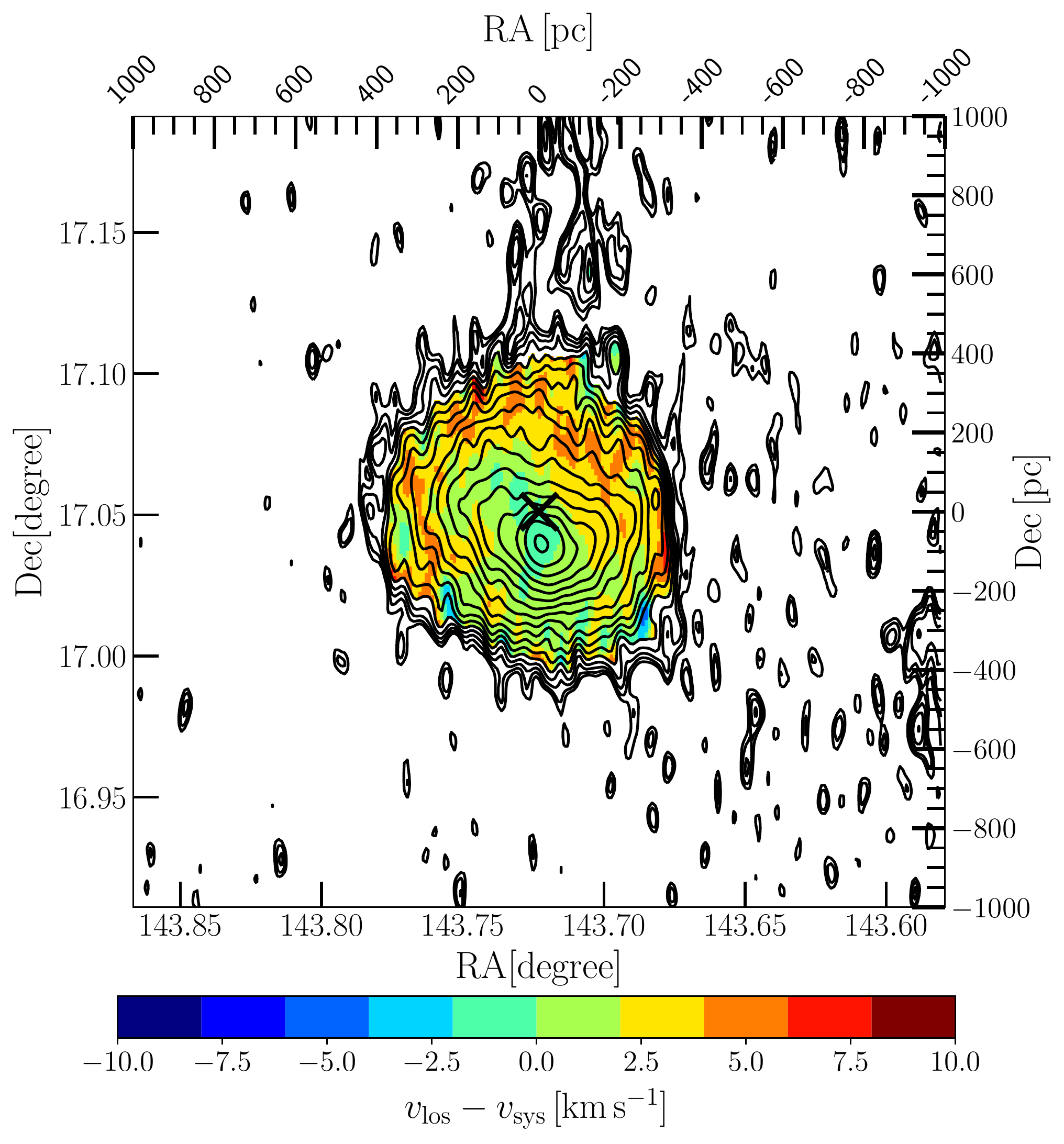}
\vspace{-0.4cm}
\caption{\HI observations from \citetalias{Adams2018} performed with the Westerbork Synthesis Radio Telescope. 
The optical or stellar center is at the origin of the axes in pc units and is marked in both panels by a large cross.
Top panel: column density map and iso-contours between $10^{19}$ and $4.2\times10^{20}\cm^{-3}$ logarithmically spaced every $0.1 {\rm dex}$.
The small cross marks the \HI density peak location.
The beaming size is marked with the white ellipse (bottom left).
Bottom panel: column density iso-contours and \HI line-of-sight velocity map where we subtracted the stellar systemic velocity $v^{\odot}_{\rm los, \star}$.}
\label{fig:LeoT:HImap}
\end{center}
\end{figure}

The \HI distribution in Leo T also reveals a very interesting morphology.
The observations of \citetalias{Adams2018} in Fig. \ref{fig:LeoT:HImap}, performed with the WSRT telescope (Westerbork Synthesis Radio Telescope),
reveal exquisite details in the \HI density and kinematics. We list some features reported by \citetalias{Adams2018} below:
\begin{enumerate}[leftmargin=0.3cm]
\item \ch{the central gas iso-contours are systematically more compressed towards the southern side, with a compressed edge to the west as well, showing a trapezoidal flattened shape. 
This could be produced by a bow shock as the dwarf moves through the MW hot halo, producing the observed flattening which would be then perpendicular to the direction of motion of the dwarf.
Under this assumption we measure the orientation of this flattening by fitting ellipses and finding the position angle of the minor axis of the ellipse at $\PAFV$, which we can use later to constrain the orientation of the orbits on the sky.}

\item \ch{the \HI velocity map shows a gradient from south to north of about $\si16\kms\ikpc$, reaching up to +5\kms from the systemic velocity in the northern part, which would be then lagging behind. 
This is comparable to the circular velocity value $V_{\rm c}\left(0.25\kpc\right)\approx13\kms$.}

\item \ch{the high \HI surface density peak is shifted in projection $83\pc\,(42\as)$ south from the optical center,
at ${\rm RA}\e143.722\degree$ and ${\rm Dec}\e17.0394\degree$. 
\citet{RyanWeber2008} report a similar offset of $40\as$ with the Giant Meterwave Radio Telescope (GMRT) and also previous WSRT measurements.
Furthermore, the deeper WSRT observations of \citetalias{Adams2018} report that the global \HI distribution is indeed centered near the stellar center and that only the inner region is shifted south. 
A similar offset has been observed in the dwarf Phoenix I as well.
If this offset is related to the interaction between this dwarf and the MW corona, we could use its orientation to constrain its motion on the sky.
We find that the line connecting the optical center with the \HI density peak is at a position angle at $\PAOV$.}

\item \ch{finally, there is an unreported faint tail of \HI material collected in clumps of 
different sizes extending to the north from the optical center, at approximately $\PATV\pm10\degree$.
While the tail substructure is at the limit of detection, considering the additional \HI features which have a similar alignment in $\PA$ ,
builds a convincing scenario of a hydrodynamical interaction between Leo T and the MW corona,
where the origin of the tail could be the dwarf's gas being stripped and that is trailing the dwarf.
We will further explore this with full hydrodynamical simulations in a future publication.}

\end{enumerate}

\begin{figure}
\begin{center}
\includegraphics[width=8.6cm]{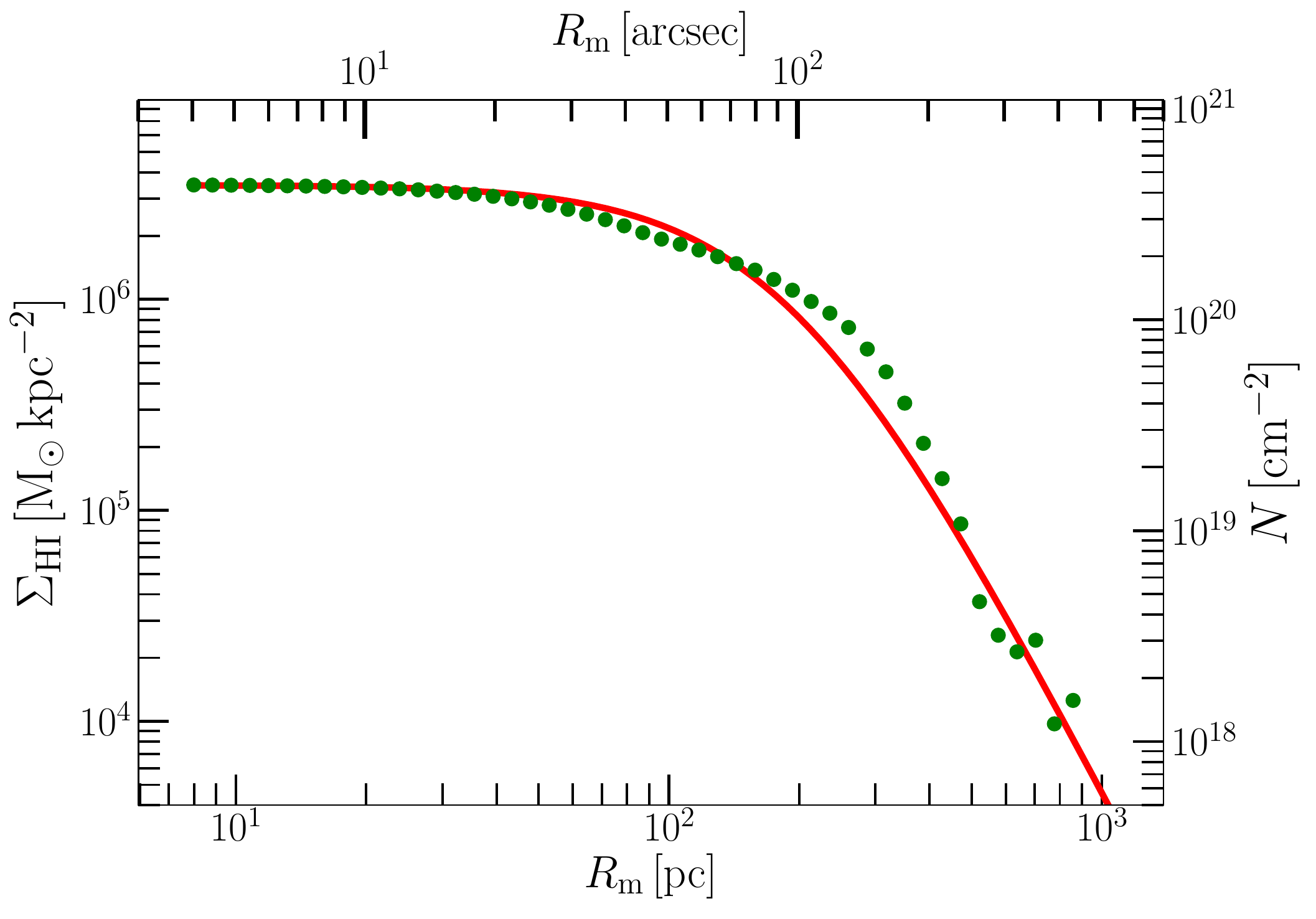}
\vspace{-0.5cm}
\caption{Azimuthally averaged \HI surface mass profile as function of the ellipse major axis of ellipses fitted to the \HI observations \citepalias{Adams2018} (green dots) and the Plummer fit (red curve), with the fit parameters shown in Table \ref{tab:LeoT:prop}.}
\label{fig:LeoT:HIprof}
\end{center}
\end{figure}

We decide to use the stellar (optical) center and the stellar systemic LOS velocity of Leo T ($v^{\odot}_{\rm los, \star}$) as our orbital initial conditions,
given that the kinematics of the gas shows perturbations, such as a \HI velocity gradient and the offset between the stars and the gas.
Furthermore, given that Leo T is dark matter dominated, the stellar kinematics should remain almost unaffected by the gravitational force of the perturbed gas density arising from the ram pressure. Moreover, we note that the \HI systemic velocity is only $1.5\kms$ slower than the stellar one.

\section{Method}
\label{sec:met}
In this paper we want to find constraints for the current tangential velocity of Leo T by 
calculating different orbits backwards in time and estimating for what values of the tangential velocity 
the orbit would have experienced a gas stripping from the Milky Way weak enough to allow the satellite to keep its gas,  
and which velocities the gas would have been completely removed.
We also estimate for what values of the tangential velocity we can find orbits that would have entered the dark halo of the Milky Way in the past.
In the following section we explain our new orbital integrator code, and the setup to explore the space of free parameters such as tangential velocities with different directions and magnitudes, and the scenarios that explore different gravitational potentials, dynamical friction effects, and effects due to the cosmic expansion. 

\subsection{Orbit integrator code}
\label{sec:met:orbit}

To perform the orbital exploration forward or backwards in time for the dwarfs or objects, we developed an orbit integrator code in python called 
Dwarfs and clustErs orbitaL integratOR codE and ANalysis or more simply \textsc{delorean}\footnote{soon available in \href{http://matiasblana.github.io}{http://matiasblana.github.io}}, which has three modules.

The first module converts the Heliocentric coordinates of the dwarfs (or object) to Galactic-Standard-of-Rest ($\rm GSR$) coordinates and to cartesian Galactocentric coordinates (GC) using \textsc{astropy} routines \citep{TheAstropyCollaboration2013,TheAstropyCollaboration2018}, where we use
a solar distance to the Galactic Center of $R_{\rm 0}\e8.2\kpc$, 
a solar height above the galactic disk of
$z_{0}\e25\pc$, and a GC solar motion of $\vec{v}^{\rm GC}_{\odot}=(11\!\pm\!1,248\!\pm\!3,7.3\!\pm\!0.5)\kms$ \citepalias{Bland-Hawthorn2016}.
The code has the option to explore heliocentric proper motion values pre-defined by the user,
or it can construct the vector $\utgsr$, which is the velocity 
vector in the GSR frame at the current position of the object that is tangential to the LOS velocity, \ch{following a prescription explained in the appendix \ref{sec:ap:int}.}
We explore different directions of $\utgsr$ and magnitudes $\Nutgsr$.
For Leo T we explore 36 different directions of the tangential velocity \utgsr (every $\Delta \PA\e10\degree$) 
and we add two more directions in the grid (see Section \ref{sec:res:main:sky}). We sample the magnitude ($\Nutgsr$) from 0  to 10\kms every 0.5\kms, 
then 12 to 180\kms every 2\kms, from 180 to 300\kms every 10\kms  and finally from 300 to 350\kms every 50\kms, exploring then in total a grid with 4218  values of \utgsr.\\

In the second module the code uses a leap-frog scheme to integrate the orbits backward or forward in time, with an option for variable time step $(\Delta t)$, 
where we use this to calculate the orbits $12\Gyr$ backwards in time.
For efficiency we use as default in our setup a time step of $\Delta t\e1\Myr$, testing the accuracy of the orbit calculation with $\Delta t\e0.1\Myr$ and with a variable time step scheme.
It has options to use a drift-kick-drift or a kick-drift-kick scheme. 
The orbits are calculated as test particles moving within a setup of a combination of gravitational potentials that are available in the code.
Each setup or scenario of potentials is defined here as a case, which is specified in Section \ref{sec:met:pot}. 
For the cases that simultaneously include the potentials of the MW and M31, the code pre-computes M31 and MW orbiting each other in a two-body scheme, 
to then compute the orbits of the dwarfs as test particles moving in the also moving and joint potentials of MW and M31 (see case 3 in section \ref{sec:met:pot}).
\ch{Furthermore, our code \textsc{delorean} can also integrate orbits backward or forward in time in an expanding universe (or contracting if it's time reversed) which is further explained in the appendix \ref{sec:ap:int}}.\\

In the third module the code has routines to calculate dynamical and hydrodynamical quantities analytically after the orbit calculations are finished.
We determine the tidal radius as function of time as:
\begin{align}
R^{\rm tidal}(D,t)&=D(t)\left(\frac{m_{\rm Sat}}{3M_{\rm host}(D,t)}\right)^{1/3}
\label{eq:rtidal}
\end{align}
where 
$m_{\rm Sat}$ is the satellite's virial mass,
$D(t)$ is the distance between the host galaxy (MW) and the satellite as function of time during the orbit (comoving coordinates in case a cosmological scheme is chosen), 
where $M_{\rm host}(D,t)$ is the mass of the host (MW) enclosed within $D(t)$.
This module also calculates the analytical ram-pressure-stripping force and other hydrodynamical quantities measured for each orbit, which are explained in Section \ref{sec:met:rps}.


\begingroup
\setlength{\tabcolsep}{6pt} 
\renewcommand{\arraystretch}{1.} 
\begin{table}
\begin{threeparttable}
\caption{Setup of main parameters for all cases.}
\begin{center}
\begin{tabular}{cccccc}\hline \hline
 MW  & $M_{\rm vir}[\times10^{12}\sm]$ & $R_{\rm vir}[\kpc]$ &  $c$   \\\hline
 cases                &     $1.3$       &     $288$        &  8.6   \\
 subcases  a      &    $1.0 $       &     $264$          &   8.4       \\
 subcases  b      &    $1.6 $       &     $308$            &  8.8        \\  \hline\hline
 cases & MW potential    &   M31? & Cosmo? \\\hline
     1     &     static        &     no       &     no      \\
     2     &   evolving      &     no      &  no    \\   
     3     &    static          &    yes     &  no   \\     
     4     &   evolving      &    yes     &  no  \\   
     5     &    static          &    no       & no   \\ 
   cos3 &       static        &   yes      &  yes     \\ 
   cos4 &       evolving    &   yes      &  yes   \\ \hline
\end{tabular}
\begin{tablenotes}
\small
\item \textbf{Notes:} by static or evolving MW potential, we mean that the MW parameters
$M_{\rm vir}$, $R_{\rm vir}$ and $c$ are constant in time (static), or evolve with redshift.
M31?: informs if the orbits of M31 and the MW were pre-computed and their orbits and potentials included in the orbital calculation of the satellite.
Cosmo?: informs if the orbits were calculated using the cosmological equations of motion (Section \ref{sec:met:orbit}).
\end{tablenotes}
\end{center}
\label{tab:cases}
\end{threeparttable}
\end{table}
\endgroup

\subsection{Setups for the gravitational potentials}
\label{sec:met:pot}
In order to quantify the orbital variations due to uncertainties in the gravitational potential,
we consider in total 13 cases or setups for our modelling. 
\ch{These cases consider extreme scenarios, including for example a constant virial mass which implies an instantaneous mass accretion at high redshift, 
and other scenarios with a redshift dependent MW mass accretion rate according to extended Press-Schechter models. 
We included more variations by changing the final MW virial mass and concentration, the influence of the Andromeda galaxy, the dynamical friction, and the cosmic expansion.}
We summarise the main properties of these cases in Table \ref{tab:cases}.
The details of each case are the following:
\begin{itemize}[leftmargin=0.2cm]
\item Cases 1, 1a, 1b: here we consider a static Milky Way potential with a NFW dark matter halo with a virial mass of $M_{\rm vir}=1.3\pm0.3\times10^{12}\sm$ \citepalias{Bland-Hawthorn2016}. We note that this virial mass estimation includes the mass contribution from all the satellites living within the halo. We use a constant virial radius of $\Rvir=288\kpc$ estimated from $M_{\rm vir}$ for redshift zero ($\z\e0$), calculated as in \citet{MoBoWh2010} with the virial radius in physical units defined as:
\begin{align}
\rvir&=\left(\frac{M_{\rm vir}(\z)}{{\small 4/3}\,\pi\rho_{\rm crit}(\z)\,\Omega_{\rm m}(\z)\,\Delta_{\rm vir}(\z)}\right)^{1/3},
\label{eq:rvir}
\end{align}
where the cosmological parameters $\rho_{\rm crit}$, $\Omega_{\rm m}$, $\Delta_{\rm vir}$ are the critical density, the matter to critical density ratio, and the spherical collapse over-density criterion $\Delta_{\rm vir}\approx\left(8\pi^2+82\epsilon-39\epsilon^2\right)\left(\epsilon+1\right)^{-1}$ with $\epsilon=\Omega_{\rm m}-1$. The comoving virial radius is then $\Rvir=\rvir/a(\z)$. 
The concentration parameter is $c=8.6$, which is obtained from the halo concentration-mass relation from \citet{Correa2015a}. 
We include a Plummer potential for the inner spheroid component with a mass of $M_{\rm s}=5\times10^{9}\sm$ \citep{Bovy2015} and a scale-length of $0.3\kpc$, and a Miyamoto-Nagai disk potential with mass $M_{\rm disk}=5.5\times10^{10}\sm$ that encompasses the stellar and the gaseous disks masses \citepalias{Bland-Hawthorn2016}, with a scale-length of $2.5\kpc$ and scale-height of $0.3\kpc$. In cases 1a, and 1b (and in all the following cases with subcases a and b) we explore the uncertainties in $M_{\rm vir}$, $c$ and \Rvir, using $1.0\times10^{12}\sm$, $8.8$, $264\kpc$ for subcases (a), and $1.6\times10^{12}\sm$, $8.4$, $308\kpc$ for subcases (b).

\item Cases 2, 2a, 2b: we explore a time varying MW potential, where the halo is accreting mass as function of redshift. For this we use the software \textsc{commah} that provides solutions to the semi-analytical extended Press-Schechter formalism \citep{Press1974,Bond1991} and the halo mass accretion history models fitted to cosmological simulations to estimate evolution of the halo properties as function of redshift (\z) \citep{Correa2015a,Correa2015b,Correa2015c}. We use the values of the setup of case 1 for the current ($\z=0$) virial mass, radius and concentration parameter to calculate the variation of these parameters as function of redshift ($\z$): $c(\z)$, $M_{\rm vir}(\z)$, $R_{\rm vir}(\z)$, as shown in Fig.\ref{fig:cosmopar}.
For the disk and spheroid components we assume that their masses ($M_{\rm comp}$) change with redshift (or lookback time) in the same proportion as the halo \ie $M_{\rm comp}(\z)=M_{\rm comp}(0)\,M_{\rm halo}({\z})/M_{\rm halo}(0)$.

\item Cases 3, 3a, 3b: we explore the influence of the gravitational potential of Andromeda (M31) on the orbits of the satellite, as it orbits the MW. 
For this we setup the MW as in case 1 and we set an NFW potential for M31 and pre-compute the orbits 12\Gyr backward in time for MW and M31 orbiting each other.
The virial mass of M31 is usually assumed to be larger than in the MW
however, the low abundance of tracers at large radii up to 500\kpc,
result in substantial uncertainties in the measurement of $M_{\rm vir}$,
with values in the literature of $M_{\rm vir}=1.6\pm0.6\times10^{12}\sm$ \citep{Watkins2010} or 
$M_{\rm vir}= (0.8 -1.1)\times10^{12}\sm$ \citep{Tamm2012}.
Here we use the upper value of \citet{Tamm2012}, choosing $M_{\rm vir}= 1.1\times10^{12}\sm$ and $\Rvir=266.3\kpc$, which 
still remains within other estimates in the literature. 
We show in Section \ref{sec:res:main} that, while the potential of M31 has an impact on the MW satellite orbits, is the virial mass of the MW what more strongly dominates the orbital properties at first order.
For the pre-computation of the orbits of the MW and M31, we use constant virial masses and radii, as M31 and the MW are far enough that both systems are attracted by the whole regions that later assemble each in the MW and M31.
For the orbit of M31 we use the heliocentric line-of-sight velocity of $v^{M31, \odot}_{\rm los}=-301\!\pm\!1\kms$ \citep{Courteau1999} and the proper motions 
$\mu^{\rm M31}_{\alpha*}=64\!\pm\!18\times10^{-3}\masyr$ and $\mu^{\rm M31}_{\delta}=-57\!\pm\!15\times10^{-3}\masyr$ \citep{VanderMarel2019}. 
Then, we calculate the orbits of the satellite in the moving potentials of the MW and M31. 
We also test MW-M31 orbits with different tangential velocities for M31 (including zero tangential velocity), finding small differences in the orbits of Leo T.

\item Case 4, 4a, 4b: we compute the orbits of the satellite in the MW-M31 moving potentials, 
but where the MW potential is also changing with redshift due to the mass accretion as case 2. 
As with the previous case, for the pre-computation of the orbits of the MW and M31, 
we use a constant MW virial mass, as M31 is far enough that it is attracted by the whole region that later assembles the MW.
Case 4 corresponds to our fiducial scenario, as this includes the effects of the most relevant quantities.

\item Case 5: we setup a scenario as case 1, but where we include a deceleration term due to the dynamical friction, using the Chandresekhar approximation \citep[eq. 8.6 in ][]{Binney2008}. 
We note that this term accelerates an object if the orbit is calculated backwards in time. 
Case 5c: as in case 5, but we setup a scenario with an accreting MW as in case 2.
We also test the friction effect using full N-body simulations (see Section \ref{sec:res:dynfr}).

\item Cases cos3 and cos4: We explore the effects of the cosmic expansion in the calculation of the orbits of the dwarfs for cases 3 and 4 (see Section \ref{sec:res:cosmo}). 
\end{itemize}

\ch{While currently we do not include major merger events in the modelling, 
we expect that our wide range of MW mass models and their time dependence can reflect the impact of these merger events.
The satellite orbits that we explore here go to large distances, spending most of their time near their apocenters, 
at around 1\Rvir and 2.6\Rvir, and therefore their orbits are mostly affected by the MW total mass. 
Accretion events such as Sagittarius \citep{Fellhauer2006,Niederste-Ostholt2010,Ruiz-Lara2020} and Sausage/Enceladus \citep{Belokurov2018, Deason2018, Helmi2018, Haywood2018} have progenitors with estimated masses between 0.1 to 10 per cent of the MW virial mass, 
and we explored variations of radial mass profiles larger than this.
For example, for the extreme cases: case 1b has a static MW mas model with a constant dynamical mass within 
20\kpc of $M(R<20\kpc)\e15.3\times10^{10}\sm$, while the mass accreting MW case 2a has a lower 
mass of $7.0\times10^{10}\sm$ at $T=-8\Gyr$, at the time of the closest pericenter passages of radial satellite orbits (see \ref{sec:res:main}), 
and even lower at $-10\Gyr$ with $4.7\times10^{10}\sm$.}

\ch{Despite the large mass variations in the center, 
we find that the main orbital constraints do not have extreme changes depending on the case (\ref{sec:res:main}).
Therefore, we expect that merger events in the MW center will vary the main orbital constraints within our range of result. 
There is however a probability that these satellites might have had a close interaction with a merger 
event at early times, but it would be unlikely given that these long distance orbits have fast pericenter 
passages, spending there only \si400\Myr.
Moreover, despite all of this, even if such close interaction actually occurred, it would have likely transpired in the 
central region of the Milky Way, and given the advantage in our method where the orbits of the satellites 
are calculated backwards in time, they would still be correct until this event in the past.}

\begin{figure}
\begin{center}
\includegraphics[height=7.5cm]{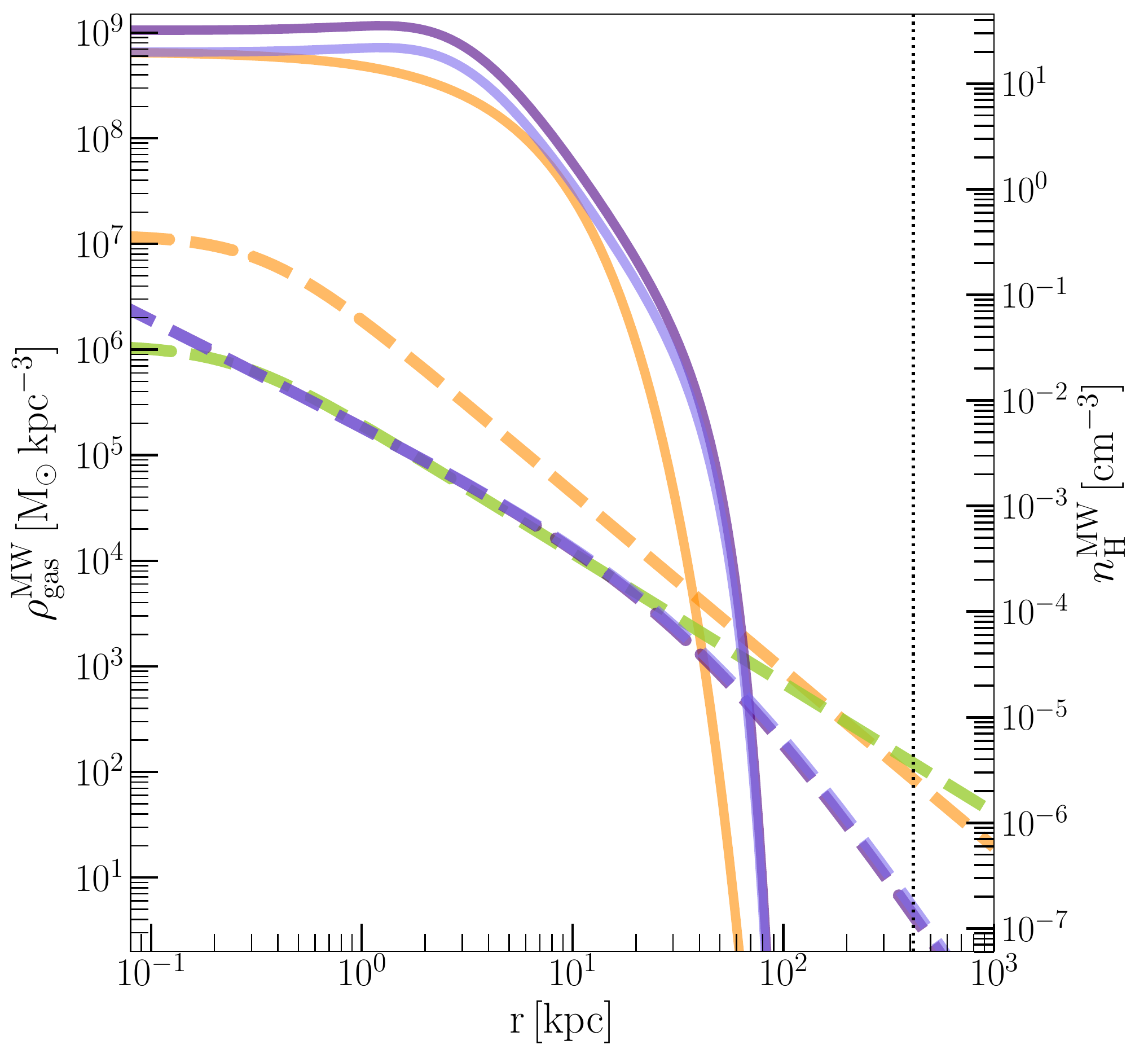}
\vspace{-0.4cm}
\caption{Analytical Milky Way gas mass density and hydrogen number density profiles.
The hot gas halo density models are marked with dashed curves, and 
the gaseous disk density models in the plane ($z\e0\kpc$) with solid curves. 
We show the setup i with yellow curves, \ie the MW hot halo beta model \citep{Salem2015} and the gaseous disk \citep{Kalberla2009}, and 
the setup ii (time varying hot halo density and cold gas disk) at two times: -7.5\Gyr (dark violet) and -10\Gyr (light violet). 
For comparison, we show a hot halo model from \citet{Miller2015} in green.
The vertical dotted line marks the current distance of Leo T.}
\label{fig:gasMW}
\end{center}
\end{figure}

\subsection{Gas stripping estimation}
\label{sec:met:rps}

We estimate the ram pressure experienced by the satellite for each orbit using the analytical estimators from \citet{Mori2000}. 
They find that a gas-rich satellite galaxy that moves through the medium of its host galaxy will lose its gas if the ram pressure ($P_{\rm R}$) that the satellite experiences by the environment of the host becomes larger than the thermal pressure that allows the dwarf to retain its gas ($P_{\rm T}$), \ie if the satellite is ram pressure stripped by the host we will have  $P_{\rm R}/P_{\rm T}>1$.
The ram pressure is then given by
\begin{align}
P_{\rm R}&=\rho^{\rm medium}_{\rm gas} V^2
\label{eq:frps}
\end{align} 
where $\rho^{\rm medium}_{\rm gas}$ is the gas density of the interstellar or intergalactic medium of the host galaxy (MW), through which the satellite moves with a velocity $V$. 
We consider two scenarios for the gas density of the Milky Way (see Figure \ref{fig:gasMW}): 
\begin{itemize}[leftmargin=0.2cm]

\item \ch{Setup i): 
in this setup we use the Milky Way  \HI disk model from \citet[][see their Fig. 4]{Kalberla2009}.
This is an exponential disk with a shift of 9.8\kpc, scale of 3.15\kpc and central density of $n^{\HI}_{\rm o}\e0.9\cm^{-3}$. It has a flaring vertical exponential profile with scale $h_{\rm o}\e0.15\kpc$ 
which fits well the radial and vertical \HI distribution between 5 and 35\kpc, beyond which the \HI disk becomes faint.
We assign a rotational velocity to the gaseous disk assuming that it rotates with the same speed as the circular velocity derived from the total gravitational MW potential. 
The direction of rotation is the same as in the MW, a clockwise rotation in the Galactocentric frame.
For the corona or hot halo gas density profile we use the fiducial beta model from \citet{Salem2015}.
We note that the hot halo density is carved out where the gaseous disk density is larger, and vice versa. 
We use then this medium to calculate the ram pressure variable $P_{\rm R,1}$ along each orbit.}

\item Setup ii) we consider a time varying gas density medium with parameters motivated by the NIHAO simulations \citep{WangJ2016}.
Here we set a hot gas halo density profile that follows the dark matter distribution in the outer parts with a hot gas to dark matter mass ratio of $0.16$ \footnote{private communication}.
For the cold gaseous disk we use a Miyamoto-Nagai gas disk with a scale length and height of $2.5\kpc$ and $0.1\kpc$, but we set up an extreme scenario 
where the gaseous disk also contains the mass of the stellar disk ($5.5\times10^{10}\sm$). 
The assumption here is that at a high redshift, most of the baryons were in the form of gas. 
\ch{Given that the MW \HI disk beyond 35\kpc becomes very shallow and that the Miyamoto-Nagai density profile in the plane
 is still large at large radii (100\kpc,) we reduce the disk density multiplying by a factor ($e^{(-R/R_{\rm o})^3}$) with scale of $R_{\rm o}\e40\kpc$, resulting in a disk density that 
equals the hot halo at 70\kpc (see Fig. \ref{fig:gasMW}).}
Additionally, for the cases with MW mass accretion, we vary the gas mass of the gaseous disk and hot halo as function of redshift, as explained in cases 2 and 4. 
We use this medium to calculate the ram pressure variable $P_{\rm R,2}$.
The main differences between both gas model setups are that the setup i is constant in time and that it
 reaches slightly lower densities in the disk, while setup ii evolves with time and in the outer part the hot halo density profile drops faster than the beta model of setup i.
We note that the density profiles extend beyond the virial radius, and that we adopt a helium to hydrogen gas mass ratio of 0.25.
\end{itemize}

\ch{We also tested different MW gas medium models finding results similar for our velocity constraints, considering for example for setup i a Miyamoto-Nagai for the \HI disk with a mass of $5\times10^9\sm$ \citepalias{Bland-Hawthorn2016} 
\citep[see also][]{Bovy2013} and a scale length of $2.5\kpc$ and height of $0.1\kpc$, which has a \HI density in the center of the disk 5 times smaller than the adopted exponential disk.
We also tested the MW hot halo model from \citet{Miller2015} where we take the parameters of the most massive hot halo (see Fig. \ref{fig:gasMW}).
The models explored here have a range of coronal densities similar to other estimations:
$n_{\rm H}\approx3\times10^{-4}\cm^{-3}$ at 60\kpc \citep{Belokurov2017}, 
$1 \times 10^{-4}\cm^{-3}$ at 70-120\kpc \citep{Grcevich2009}, 
 and $(1.3-3.6)\times10^{-4}\cm^{-3}$ at 50-90\kpc \citep{Gatto2013}.}\\

Now we need to calculate the thermal pressure that allows the satellite to retain its gas.
For this we use the estimate of \citet{Mori2000}, where the thermal pressure is: 
\begin{align}
P_{\rm T}&= \frac{{\rm G}\, M_{\rm o} \rho^{\rm sat}_{\rm core}}{3 r_{\rm core}}
\label{eq:fres}
\end{align}
with ${\rm G}$ being the gravitational constant and
$\rho^{\rm sat}_{\rm core}$ is the central gas density of the satellite, 
where we use the value from our Plummer fit $\rho^{\rm gas}_{\rm Pl}$.
The scale of the dwarf's dark matter core is $r_{\rm core}$, 
which we approximate by using the de-projected stellar half light radius $r_{\rm h}\e4/3 R_{\rm h}^{\rm V}$.
This is motivated by dwarf galaxy formation simulations that show that a bursty star formation can generate a dark matter cored with a similar size to the stellar distribution \citep{Ogiya2014a,Read2016}.
$M_{\rm o}$ is the dynamical mass within $r_{\rm core}$, where we use the virial relation \citep{Wolf2010}
\begin{align}
M^{\rm dyn}_{\rm half}&=3G^{-1}\,(4/3)R^{\rm V}_{\rm h}\sigma_{\rm los,\star}^2
\label{eq:mdyn}
\end{align} 
and the stellar dispersion $\sigma_{\star}$ \citep{Simon2007}, 
finding with the latest distance estimate a dynamical mass within the half-light radius of $M^{\rm dyn}_{\rm half}\e7.6\pm 3.3\times10^6\sm$. 
This gives us a thermal pressure of $P_{\rm T}\e1.0\times 10^9\sm\kpc^{-3}\km^{2}\s^{-2}$.\\

\begin{figure}
\begin{center}
\includegraphics[width=8.4cm]{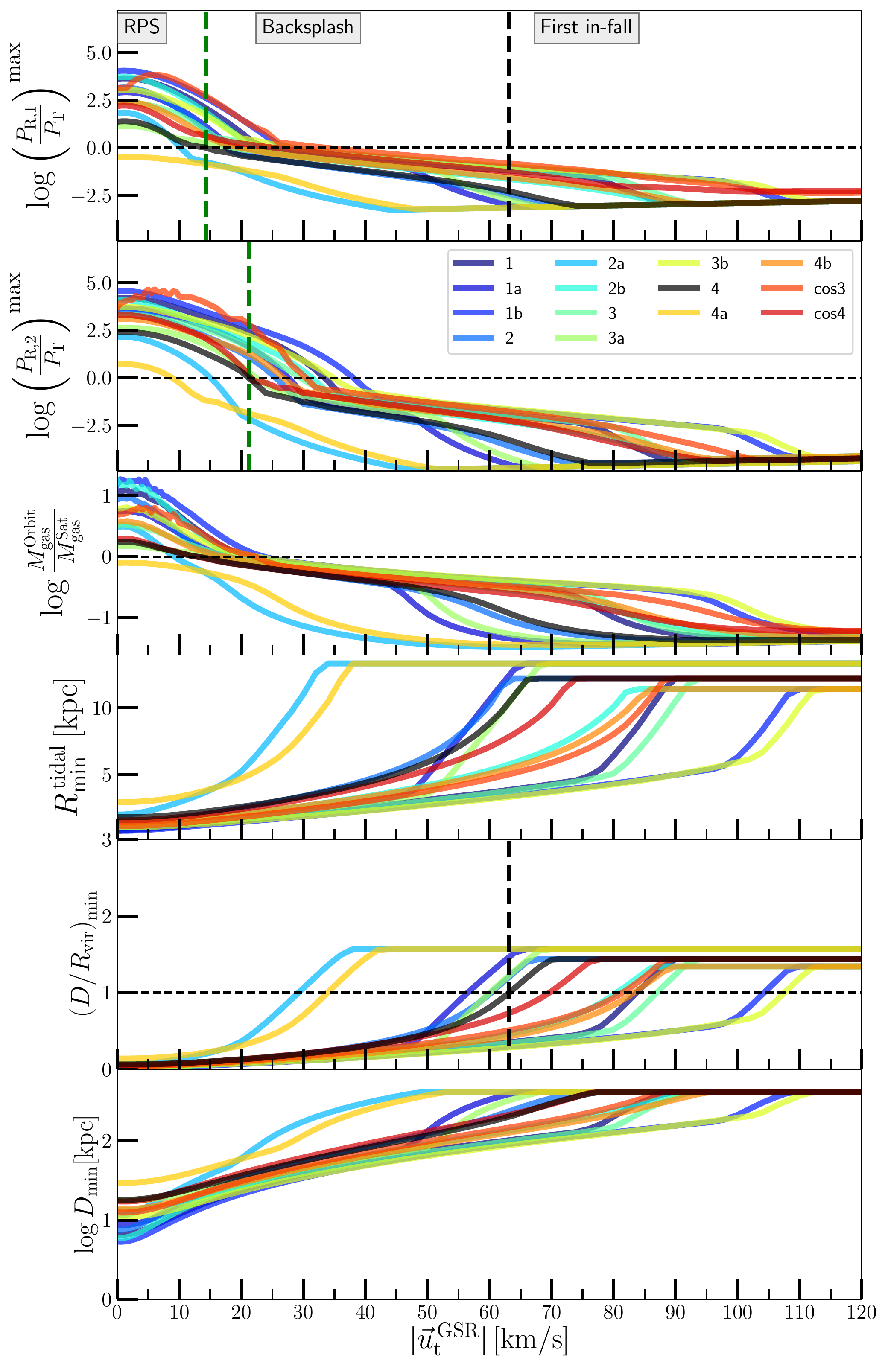}
\vspace{-0.5cm}
\caption{The main parameters taking the median of the parameter for different directions of $\utgsr$ and for different cases (see Table \ref{tab:cases}) as function of \Nutgsr. 
See the main text in Section \ref{sec:res:main} for the definitions of the parameters.
Each coloured line corresponds to a case labeled in the second top panel, where we note
that our fiducial case 4 (accreting MW and M31 potential) is shown in black. 
From this figure we identify three main regions in \utgsr;
The region RPS where the MW ram pressure is larger than the thermal pressure of the satellite.
Each direction and case has a slightly different value of \Nutgsrrps.
We mark the median of $P_{\rm R}$ of our fiducial case 4 taking all directions of the tangential velocities finding $\Nutgsr=14^{+26}_{-14}\kms$ for $P_{\rm R,1}$ (top panel)
and $21^{+33}_{-21}\kms$ for $P_{\rm R,2}$ (second panel) (dashed green vertical lines).
The error range considers the maximum and minimum threshold velocity values found among different directions of \utgsr and different cases.
The backsplash region with orbits that passed through the MW dark matter halo,
which we mark for case 4 at the median value and range of $\Nutgsr=63^{+47}_{-39}\kms$ (dashed black vertical line).
And the first in-fall region where orbits never entered the halo. 
We present the complete results for different directions of \utgsr in Fig. \ref{fig:orbs:par:leot:ang}.}
\label{fig:orbs:par:leot:av}
\end{center}
\end{figure}

Additionally, \citet{Mori2000} analyse the Kelvin-Helmholtz (KH) instability gas stripping process of a satellite that moves in the host's medium.
This mechanism is not instantaneous, it operates cumulatively with time, stripping the gas of the satellite within a given timescale.
\citet{Nulsen1982} find that the gas mass-loss rate of a satellite through the KH instability is 
\begin{align}
\frac{dM_{\rm gas}}{dt}&= \pi r^2_{\rm core}\,\rho^{\rm medium}_{\rm gas}\,V
\label{eq:KH}
\end{align}
Therefore, as an additional parameter to estimate the gas stripping for each orbit we re-formulate the previous equation as 
\begin{align}
M_{\rm gas}^{\rm Orbit} &= \int dM_{\rm gas}= \int\pi r^2_{\rm core}\,\rho^{\rm medium}_{\rm gas}\,V dt
\label{eq:KH2}
\end{align}
where in practice, we just compute the amount of MW gas collected along the orbit within a tube with a radius of $r_{\rm core}\e300\pc$, which represents roughly the radius of the Leo T \HI distribution: $M_{\rm gas}^{\rm Orbit}$.
Surviving gas rich satellites have orbits where $M_{\rm gas}^{\rm Orbit}/M^{\rm Sat}_{\rm gas}<1$.
Of course, this would be an upper limit, under the assumption that the satellite's gas has not changed much.
If we include the gas that formed the stars in the past we would have $M_{\rm gas}^{\rm Orbit}/(M^{\rm Sat}_{\rm gas}+M^{\star}_{\rm sat})<M_{\rm gas}^{\rm Orbit}/M^{\rm Sat}_{\rm gas}<1$.
Note that the comparison between the gas mass of the satellite and the mass collected along the orbits is approximately equivalent to a comparison between the gas column density of the satellite and the ambient gas column density along the orbit within the area $\pi\,r^2_{\rm core}$.



\section{Results}
\label{sec:res}
\ch{We present our main results in Section \ref{sec:res:main},
with a further analysis of the backsplash orbital solutions in Section \ref{sec:res:main:bas} and the gas stripped solutions in Section \ref{sec:res:main:rps},
including examples of orbits.
This is followed by Section \ref{sec:res:cosmo}, where we explore in more detail the effects of the cosmic expansion in the orbital calculation.
In Section \ref{sec:res:dynfr} we explore the effects of dynamical friction and tidal disruption in the orbital calculation with N-body simulations.
Finally, we apply our method to other distant dwarfs, presenting these results in Section \ref{sec:disc:sat}.}

\subsection{Three main orbital solutions}
\label{sec:res:main}

We analyse a total of 4218 orbits for Leo T, which includes all setup cases for the gravitational potential 
and different directions and magnitudes of the tangential velocity \utgsr.
For each orbit we measure 7 variables as function of \Nutgsr: 

\begin{enumerate}[label=\arabic*),leftmargin=0.3cm]
\item $(D/\Rvir)_{\rm min}$: the minimum of the ratio of the distance between the satellite and the MW center and the MW virial radius.  
\item $P^{\rm max}_{\rm R,1}/P_{\rm T}$: the maximum of the ratio of the ram pressure for each orbit considering the setup i for MW gas model and the thermal pressure of the satellite.
\item $P^{\rm max}_{\rm R,2}/P_{\rm T}$: same ram pressure parameter but for the setup ii MW gas model.
\item $M^{\rm Orbit}_{\rm gas}/M^{\rm Sat}_{\rm gas}$: the ratio of the total gas mass collected along each orbit and the satellite's gas mass.
\item $R^{\rm tidal}_{\rm min}$: the minimum tidal radius of the satellite.
\item $D_{\rm min}$ the minimum distance of the satellite's orbit to the MW center.
\item ${\rm PA}\left(\Delta R\e0.1\degree\right)$ the orientation that each orbit has on the sky at $\Delta R\e0.1\degree$ from the currrent's position of Leo T projected on the sky.
\end{enumerate}

The main trend of each variable as function of \Nutgsr is shown in Fig. \ref{fig:orbs:par:leot:av} with the median of the distributions taken for different directions of \utgsr. 
The values for each direction of \utgsr are shown in Fig. \ref{fig:orbs:par:leot:ang}.
Using now these parameter we can determine if each orbit is a first in-fall orbital solution, a backsplash orbital solution or a gas stripped backsplash orbital solution.

\begin{figure}[ht!]
\begin{center}
\includegraphics[width=8cm]{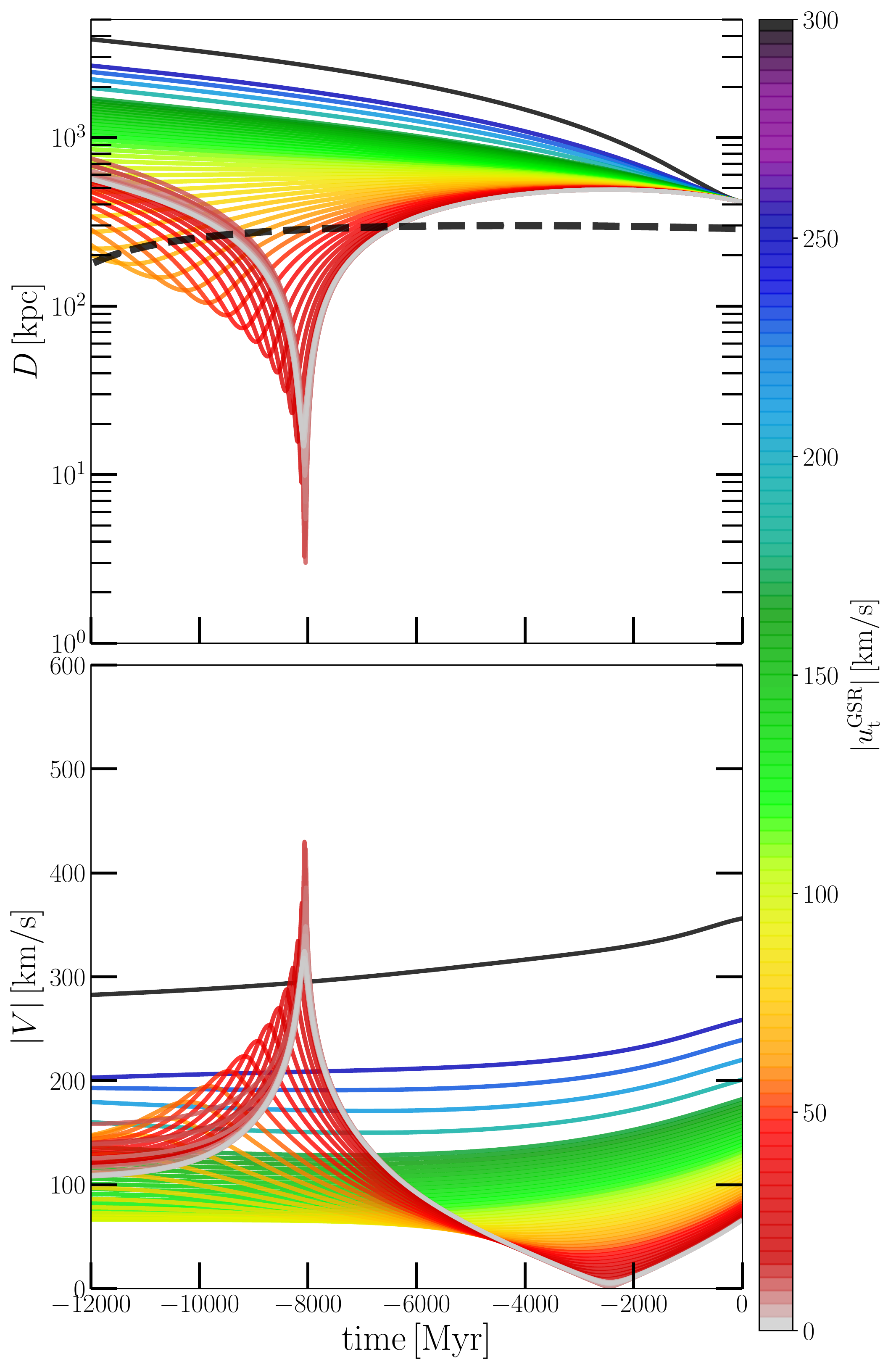}
\vspace{-0.42cm}
\caption{
Galactocentric distance (top panel) and velocity (bottom panel) as function of lookback time for case 4. 
Colours indicate the value of \Nutgsr, showing only orbits pointing in the direction of $\PA=3\degree$ to avoid an overcrowding of lines.
In the top panel we show $R_{\rm vir}$ as function of time (curved dashed line).
All orbits that go within $R_{\rm vir}$ are by definition backsplash orbits, while the ones that remain outside are first in-fall orbits.}
\label{fig:orb:case4:distvel}
\end{center}
\end{figure}

\begin{figure*}{~}
\begin{center}
\includegraphics[width=8.8cm]{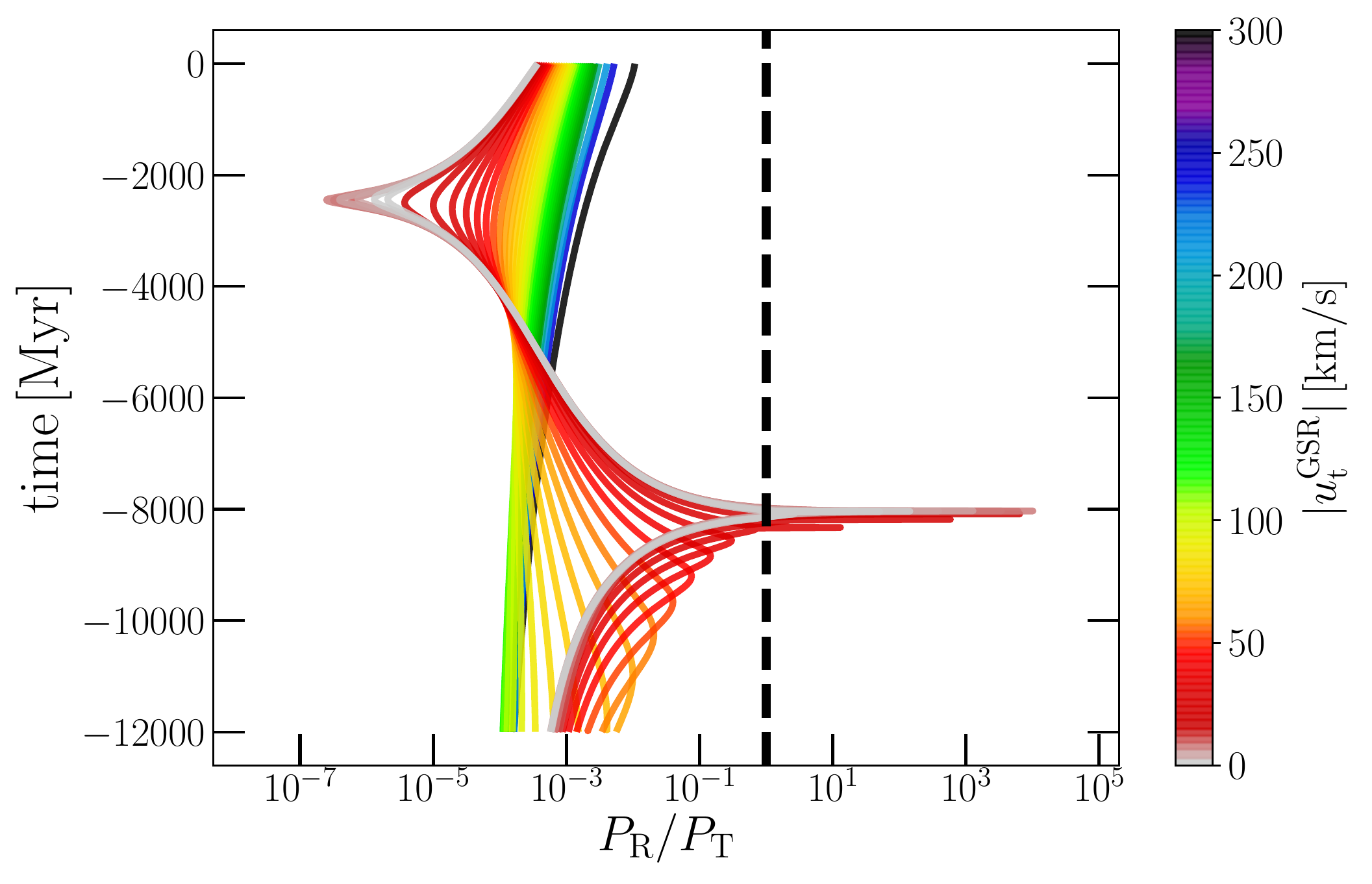}
\includegraphics[width=8.8cm]{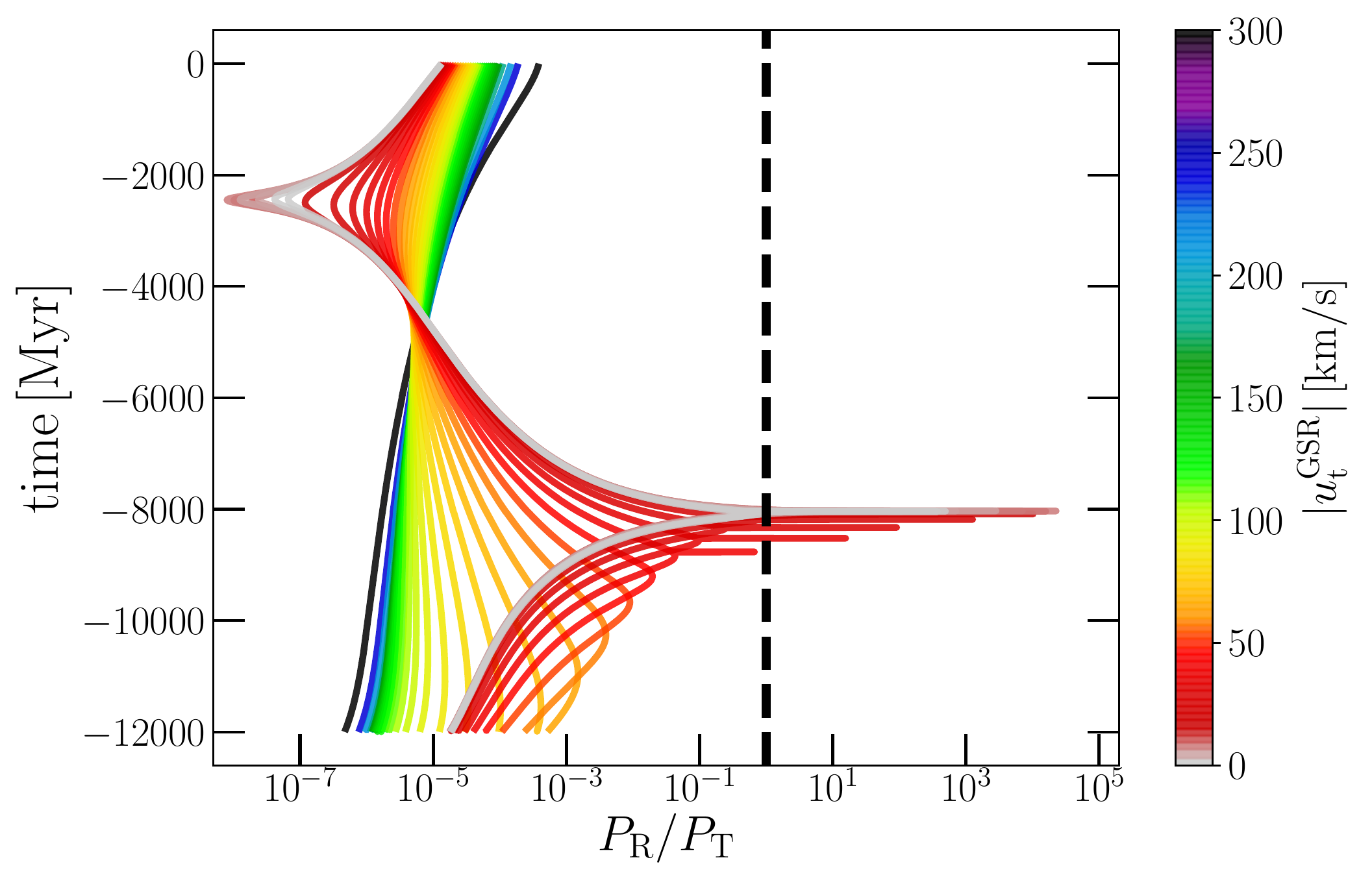}
\vspace{-0.5cm}
\caption{Ram pressure to thermal pressure ratio as function of time for case 4, 
calculated for the setup i of the MW gas medium $P_{\rm R,1}$ (left panel), and the setup ii $P_{\rm R,2}$ (right panel). 
Colours indicate the value of \Nutgsr, showing here only orbits pointing in the direction of $\PA=3\degree$ to avoid an overcrowding of lines.
A ratio of $P_{\rm R}/P_{\rm T}\e1$ is marked with a vertical dashed line. 
Ratios larger than one denote the region where the gas of the satellite would be stripped by the MW.}
\label{fig:orb:case4:rps}
\end{center}
\end{figure*}

\subsubsection{Backsplash and first in-fall orbital solutions}
\label{sec:res:main:bas}

We can classify each orbit as backsplash if the satellite entered the MW dark matter halo in the past and $(D/\Rvir)_{\rm min}\leq1$, or as a first in-fall orbit if $(D/\Rvir)_{\rm min}>1$. 
In Fig. \ref{fig:orb:case4:distvel} we present as example a set of orbits in a direction of \utgsr for our fiducial case 4, 
\ie the setup that includes the potentials of an accreting Milky Way and the M31 potential.
It shows that there is a range of orbits with $\Nutgsr<65\kms$ where the satellite entered the (time varying) virial radius of the Milky Way between $-8$ and $-12\Gyr$ ago, 
while for larger velocities there are found only first in-fall solutions.
Therefore, for each direction of \utgsr, there is a particular value of \Nutgsr, a threshold $\Nutgsrbas$, below which all orbits are backsplash solutions and $(D/\Rvir)_{\rm min}\leq1$,
finding a similar threshold value when we examine different directions and different cases (see case 1 in Fig. \ref{fig:orb:cases}).

We can more easily see the main trend of $(D/\Rvir)_{\rm min}$ as function of \Nutgsr for different cases 
with the the median of $(D/\Rvir)_{\rm min}$ for different directions of \utgsr, which is shown in Fig. \ref{fig:orbs:par:leot:av}.
In the figure we can identify the threshold value for the median $\langle\Nutgsrbas\rangle$ below which we find the backsplash solutions.
The values for each direction are shown in Fig. \ref{fig:orbs:par:leot:ang}, where we find values of $\Nutgsrbas$ that can be larger and smaller than
the medians.

In Table \ref{tab:res:par:leot} we summarise our main results, showing the median threshold values for different cases, as well as the maximum and minimum values found among different directions of \utgsr.
From our fiducial case 4 we obtain that orbits with tangential velocities lower than $\langle\Nutgsrbas\rangle\leq62^{+49}_{-38}\kms$ result in backsplash solutions, 
where the error range corresponds to the maximum and minimum values found for different cases and directions. 
The minimum distance is at the backsplash orbital threshold occurred at $t\left(D_{\rm min}^{\rm BAS}\right)\e-11.9^{+0.3}_{-0.1}\Gyr$.
We discuss how this changes when we include the cosmic expansion in Section \ref{sec:res:cosmo}.
We also provide the values of \Nutgsrbas when we choose the \PA directions that better align with the \HI morphology, as explain in Section \ref{sec:res:main:sky}.
In Table \ref{tab:res:par:leot} we find that the largest variations in the exact value of \Nutgsrbas are due to the variations of the virial mass of the Milky Way. When we compare subcases a) and  b) we can see that the threshold increase and decrease by a factor of 2.
This is simply because the potential of a more (or less) massive halo will require larger (or smaller) velocities to obtain orbits that do not enter the virial radius, and
also because the virial radius of a more (or less) massive halo is larger (or smaller).

The second important effect comes from the MW accretion history, followed by the M31 potential. 
For example, in Table \ref{tab:res:par:leot}, the largest value of \Nutgsrbas is for case 3b with a constant and large value of $M_{\rm vir}\e1.6\times10^{12}\sm$, 
needing then a larger velocity to obtain first in-fall solutions.
The smallest \Nutgsrbas is for case 2a, where $M_{\rm vir}$ is the lowest, and furthermore, it decreases with redshift.
The effects of M31 are noticeable when comparing the tangential velocity threshold between cases 1 and cases 3, where the velocity thresholds are larger for case 3 due to the contribution of the M31's potential. 
But still the MW virial mass plays the major role.
This provides our first constraint for the tangential velocity. 
We see in Fig. \ref{fig:orbs:par:leot:av} and \ref{fig:orbs:par:leot:ang} that the curves of $(D/\Rvir)_{\rm min}$ become almost flat for $\Nutgsr>\Nutgsrbas$.
This is simply because we enter the region of first in-fall solutions, and the minimum values correspond to the current position of Leo T.

It is also possible to see  in Fig. \ref{fig:orb:case4:distvel} that the maximum distance of the backsplash orbits are between $D\e1R_{\rm vir}$ and $D\si4\times R_{\rm vir}$ at a lookback time of $-12\Gyr$ (or $\z\e3.51$). 
This value could be even larger in units of the virial radius if we integrate to higher redshifts, given that $R_{\rm vir}$ decreases as well, because the halo mass decreases with redshift, as described by Eq. \ref{eq:rvir}. 
This range of distances of backsplash orbits is similar to the satellites in the NIHAO cosmological Milky Way-type simulations \citep[see their Fig. 7]{Buck2019}. 
In Section \ref{sec:res:cosmo} we will also show backsplash orbits when we include the cosmic expansion.

\subsubsection{Gas stripped orbital solutions}
\label{sec:res:main:rps}
In Fig. \ref{fig:orb:case4:rps} we show the ram pressure for different orbits of case 4 as function of time and $\Nutgsr$ for a particular direction.
In general, we find that for different cases and different directions of \utgsr, the ram pressure of the orbits is quantitatively and qualitatively similar.
As expected, the largest ram pressure values happen for the most radial orbits of the backsplash solutions, 
with a maximum of $\left(P_{\rm R}/P_{\rm T}\right)^{\rm max}\si10^4$ larger than the thermal pressure of the satellite. 
We also compare both MW gas model setups, where we see that the ram pressure 
of the setup ii MW gas model ($P_{\rm R,2}/P_{\rm T}$ ) reaches slightly larger values in the center than with the setup i ($P_{\rm R,1}/P_{\rm T}$).
The first in-fall solutions experience the largest ram pressure at the current position of Leo T, 
or in some solutions at the apocenters of orbits that passed near but outside of the virial radius.

Similarly to our backsplash analysis, we can classify an orbit as gas stripped if the instantaneous ram pressure overcomes the thermal pressure of the satellite
\ie $\left(P_{\rm R}/P_{\rm T}\right)^{\rm max}\geq1$, or as unstripped if $\left(P_{\rm R}/P_{\rm T}\right)^{\rm max}<1$.
We note that in our setups, because the largest MW gas density is in the center, all the ram pressure stripped orbital solutions are also backsplash solutions.
The only non backsplash ram pressure stripped solutions would first in-fall orbits with extremely large tangential velocities ($\Nutgsr>3000\kms$). 

In Fig. \ref{fig:orbs:par:leot:av} we show the median of $\left(P_{\rm R}/P_{\rm T}\right)^{\rm max}$ for different cases as function of \Nutgsr,
showing the values for different directions of \utgsr in Fig. \ref{fig:orbs:par:leot:ang}.
We can also find a tangential velocity threshold, below which we find only ram pressure stripped orbital solutions (RPS), 
providing a second criterion to constrain the tangential velocity, namely $\Nutgsrrps$.
Interestingly, we find that this threshold does not change much depending on the MW gas model.
In fact, our tests with other MW gas distributions (Section \ref{sec:met:rps}) result in similar threshold values.
This results from the
ram pressure depending linearly with the density medium and quadratically with the velocity.
When the satellite reaches the central regions $R<40\kpc$ and hits the MW gaseous disk, 
it has already a large velocity, between 200 and $400\kms$ (see Fig.\ref{fig:orb:case4:distvel}, or up to 500\kms in case 1 Fig.\ref{fig:orb:cases}),
which increases the ram pressure quadratically.

In Table \ref{tab:res:par:leot} we present the median threshold values for different cases, 
finding for example for our fiducial case 4 that backsplash solutions with $\langle\Nutgsrrps\rangle>21^{+33}_{-21}\kms$ 
would allow Leo T to survive the ram pressure stripping of the Milky Way,
with wide orbits that have a minimum distance of $D_{\rm min}\!\geq\!38^{+26}_{-16}\kpc$.
The error range considers different directions of \utgsr and different cases.
We note that in Table \ref{tab:res:par:leot} we show the threshold values obtained from the MW gas setup ii, 
because the threshold value is slightly larger than with the setup i.
The minimum distance is at the ram pressure threshold occurred at $t\left(D_{\rm min}^{\rm RPS,2}\right)\e-8.4^{+0.2}_{-0.4}\Gyr$. 
We discuss how this changes when we include the cosmic expansion in Section \ref{sec:res:cosmo}.
We also show in the table the threshold $\Nutgsrrps$ for directions that better align with different features in the \HI morphology, as explain in Section \ref{sec:res:main:sky}.

Similarly to the backsplash analysis, in the table we also find that the largest variations of \Nutgsrrps arise from differences in the total mass of the Milky Way,
and from the MW accretion history. 
Particularly, this latter determines the potential in the MW center,
which can bring the orbits of the satellites closer to the MW center where the gas is denser, which then pushes 
region of the the ram pressure solution to larger values of \Nutgsrrps. 
However, even when comparing extreme cases, such as case 1 with the static MW potential and case 2 with the accreting MW, we 
see that  \Nutgsrrps values agree within their ranges.\\

And finally, we also find that the cumulative gas stripping through the KH instability could strip very radial orbits, 
but it is subdominant when compared to the ram pressure (see Section \ref{sec:met:rps}). 
In Fig. \ref{fig:orbs:par:leot:av} and Fig. \ref{fig:orbs:par:leot:ang} we show that the amount of gas collected along 
each orbit is larger than the gas in Leo T only for very radial orbits, which are already within the ram-pressure-stripped region.
This implies that, in this particular scenarios, the KH instability, parametrised as in Eq. \ref{eq:KH2}, 
would be less efficient than the instantaneous ram pressure stripping to remove the gas of the dwarf.

\begingroup
\setlength{\tabcolsep}{3pt} 
\renewcommand{\arraystretch}{1.6} 
\begin{table*}
\scriptsize
\begin{threeparttable}
\caption{Orbital constraints for Leo T}
\begin{center}
\begin{tabular}{|@{\hspace{0.01cm}}c@{}|cccccccccccc}\hline \hline
1 & 2 & 3 & 4 & 5 & 6   & 7 & 8 & 9 & 10 & 11 & 12 & 13 \\\hline
&  $\langle\Nutgsrbas\rangle$ & $\langle D_{\rm min}^{\rm BAS}\rangle$ & ${\Nutgsrbas}$ & $D_{\rm min}^{\rm BAS}$  & $\muas^{\rm BAS}$ & $\mud^{\rm BAS}$ & $\langle\Nutgsrrps\rangle$ & $\langle D_{\rm min}^{\rm RPS}\rangle$  & $\Nutgsrrps$ & $D_{\rm min}^{\rm RPS}$ & $\muas^{\rm RPS}$ & $\mud^{\rm RPS}$ \\
Case 	&  $\kms$ & $\kpc$ & $\kms$ & $\kpc$ &  $\masyr$ & $\masyr$ & $\kms$ & $\kpc$ & $\kms$ & $\kpc$ & $\masyr$ & $\masyr$ \\\hline
1 & $83^{+2}_{-2}$ & $288$ & $84\,(83)$ & $287\,(287)$ & $-0.0164\,(-0.0017)$ & $-0.1587\,(-0.1561)$ & $33^{+13}_{-19}$ & $43^{+16}_{-20}$ & $41\,(33)$ & $52\,(43)$ & $-0.0157\,(-0.0096)$ & $-0.1365\,(-0.1319)$\\
1a & $56^{+2}_{-3}$ & $264$ & $56\,(55)$ & $263\,(263)$ & $-0.016\,(-0.006)$ & $-0.1446\,(-0.1428)$ & $27^{+14}_{-16}$ & $39^{+18}_{-17}$ & $35\,(27)$ & $49\,(39)$ & $-0.0156\,(-0.0106)$ & $-0.1336\,(-0.129)$\\
1b & $104^{+2}_{-2}$ & $308$ & $104\,(103)$ & $308\,(308)$ & $-0.0168\,(0.0016)$ & $-0.1693\,(-0.1662)$ & $38^{+12}_{-20}$ & $46^{+15}_{-22}$ & $45\,(38)$ & $55\,(46)$ & $-0.0158\,(-0.0088)$ & $-0.1389\,(-0.1342)$\\\hline
2 & $60^{+2}_{-2}$ & $197^{+2}_{-5}$ & $60\,(59)$ & $194\,(195)$ & $-0.016\,(-0.0054)$ & $-0.1467\,(-0.1447)$ & $25^{+8}_{-14}$ & $44^{+11}_{-20}$ & $31\,(26)$ & $52\,(44)$ & $-0.0155\,(-0.0108)$ & $-0.1314\,(-0.1281)$\\
2a & $29^{+2}_{-3}$ & $164^{+2}_{-2}$ & $29\,(28)$ & $164\,(164)$ & $-0.0155\,(-0.0104)$ & $-0.1306\,(-0.1295)$ & $15^{+4}_{-9}$ & $38^{+6}_{-13}$ & $18\,(15)$ & $44\,(38)$ & $-0.0153\,(-0.0125)$ & $-0.1251\,(-0.1229)$\\
2b & $80^{+2}_{-2}$ & $244^{+2}_{-7}$ & $81\,(80)$ & $242\,(242)$ & $-0.0164\,(-0.0022)$ & $-0.1571\,(-0.1547)$ & $31^{+8}_{-17}$ & $46^{+11}_{-22}$ & $37\,(31)$ & $55\,(47)$ & $-0.0157\,(-0.0099)$ & $-0.1346\,(-0.131)$\\\hline
3 & $87^{+4}_{-12}$ & $288$ & $91\,(91)$ & $287\,(287)$ & $-0.0166\,(-0.0004)$ & $-0.1624\,(-0.16)$ & $29^{+21}_{-16}$ & $38^{+21}_{-15}$ & $48\,(41)$ & $54\,(43)$ & $-0.0158\,(-0.0083)$ & $-0.1403\,(-0.1358)$\\
3a & $60^{+7}_{-11}$ & $264$ & $67\,(67)$ & $263\,(263)$ & $-0.0162\,(-0.0042)$ & $-0.15\,(-0.1483)$ & $22^{+24}_{-14}$ & $34^{+24}_{-11}$ & $43\,(36)$ & $52\,(40)$ & $-0.0158\,(-0.0091)$ & $-0.138\,(-0.1334)$\\
3b & $107^{+3}_{-12}$ & $308$ & $109\,(109)$ & $308\,(308)$ & $-0.0169\,(0.0025)$ & $-0.1716\,(-0.169)$ & $34^{+19}_{-19}$ & $41^{+19}_{-17}$ & $52\,(45)$ & $56\,(46)$ & $-0.0159\,(-0.0077)$ & $-0.1423\,(-0.1378)$\\\hline
4 & $63^{+6}_{-11}$ & $201^{+15}_{-2}$ & $69\,(69)$ & $209\,(209)$ & $-0.0162\,(-0.0039)$ & $-0.151\,(-0.1492)$ & $21^{+18}_{-12}$ & $38^{+17}_{-13}$ & $39\,(34)$ & $54\,(45)$ & $-0.0157\,(-0.0094)$ & $-0.1357\,(-0.1324)$\\
4a & $33^{+9}_{-10}$ & $164^{+2}_{-2}$ & $43\,(43)$ & $164\,(164)$ & $-0.0158\,(-0.0081)$ & $-0.1376\,(-0.1364)$ & $8^{+22}_{-8}$ & $34^{+16}_{-9}$ & $30\,(26)$ & $49\,(39)$ & $-0.0155\,(-0.0107)$ & $-0.1313\,(-0.1286)$\\
4b & $83^{+4}_{-12}$ & $245^{+13}_{-3}$ & $87\,(87)$ & $257\,(256)$ & $-0.0165\,(-0.001)$ & $-0.1603\,(-0.158)$ & $26^{+18}_{-14}$ & $39^{+18}_{-14}$ & $44\,(39)$ & $56\,(47)$ & $-0.0158\,(-0.0086)$ & $-0.1383\,(-0.1348)$\\\hline
cos3 & $81^{+4}_{-11}$ & $288$ & $81\,(81)$ & $287\,(287)$ & $-0.0164\,(-0.0019)$ & $-0.1573\,(-0.1555)$ & $31^{+11}_{-21}$ & $51^{+13}_{-22}$ & $41\,(38)$ & $59\,(54)$ & $-0.0157\,(-0.0089)$ & $-0.1366\,(-0.1341)$\\
cos4 & $69^{+4}_{-12}$ & $284^{+2}_{-2}$ & $73\,(73)$ & $287\,(286)$ & $-0.0163\,(-0.0032)$ & $-0.1532\,(-0.1513)$ & $21^{+17}_{-12}$ & $42^{+21}_{-14}$ & $38\,(33)$ & $59\,(49)$ & $-0.0157\,(-0.0097)$ & $-0.135\,(-0.1317)$\\\hline\hline
B.V & $63$ & $201$ & $69\,(69)$ & $209\,(209)$ & $-0.0162\,(-0.0039)$ & $-0.151\,(-0.1492)$ & $21$ & $38$ & $39\,(34)$ & $54\,(45)$ & $-0.0157\,(-0.0094)$ & $-0.1357\,(-0.1324)$\\
$\pm\Delta$ & $^{+47}_{-39}$ & $^{+107}_{-36}$ & $^{+39}_{-39}\,\left(^{+40}_{-40}\right)$ & $^{+98}_{-44}\,\left(^{+99}_{-44}\right)$ & $^{+0.0007}_{-0.0007}\,\left(^{+0.0065}_{-0.0065}\right)$ & $^{+0.0204}_{-0.0206}\,\left(^{+0.0197}_{-0.0198}\right)$ & $^{+33}_{-21}$ & $^{+26}_{-16}$ & $^{+12}_{-20}\,\left(^{+10}_{-19}\right)$ & $^{+5}_{-9}\,\left(^{+9}_{-6}\right)$ & $^{+0.0003}_{-0.0002}\,\left(^{+0.0017}_{-0.0031}\right)$ & $^{+0.0106}_{-0.0066}\,\left(^{+0.0095}_{-0.0053}\right)$\\\hline\hline
\end{tabular}
\begin{tablenotes}
\small
\item \textbf{Notes:} 
The zero tangential GSR velocity for Leo T ($\Nutgsr=0\kms$) corresponds to $\muasz\e-0.0150\masyr$ and $\mudz\e-0.1153\masyr$.
Column 1 corresponds to the case scenario.
Col. 2 and 3 lists the value of $\Nutgsr\,^{\rm BAS}$ for the median of $(D/\Rvir)_{\rm min}$ for different directions of \utgsr, and the minimum distance, below which the orbits are backsplash solutions,
with the error range taken from the threshold values for different directions of $\utgsr$, or from the grid resolution.
Col. 4 to 7 correspond to the backsplash solutions of our grid that are close to the direction of $\PAOV$, where we show the velocity, the minimum distance and the proper motion.
In brackets we show the solutions close to the direction of $\PAFV$.
Col. 8 and 9 are the value of $\Nutgsr\,^{\rm RPS}$ and the minimum distance when we take the median of $P_{\rm R,2}/P_{\rm T}$ for different directions of \utgsr. 
Below this value the orbits are ram pressure stripped solutions, with the error range taken from the different directions of $\utgsr$, or from the grid resolution.
Col. 10 to 13 correspond to the RPS solutions of $\utgsr$ closest to the direction $\PAO$, showing the values $\Nutgsr$, minimum distance, and the proper motions.
In brackets we include the values for the direction $\PAF$.
The last two rows: B.V. are the best values selected from our fiducial scenario case 4 with errors taken from the range of maximum and minimum values from different cases, including the range for different  directions of \utgsr.
\end{tablenotes}
\end{center}
\label{tab:res:par:leot}
\end{threeparttable}
\end{table*}
\endgroup


\subsubsection{Tidally disrupted orbits.}
 We calculate this quantity according to Eq. \ref{eq:rtidal}, and assuming a total virial mass for the satellite of $10^{8}\sm$, based on dwarf galaxy formation models in isolation \citep{Read2016}. 
Changing this to $10^{7}\sm$ or $10^{9}\sm$ would imply a change of the tidal radius only by a factor of $\si2$ smaller or larger, respectively. 
As shown in Fig. \ref{fig:orbs:par:leot:av} and Fig. \ref{fig:orbs:par:leot:ang} the value is large enough ($R^{\rm tidal}_{\rm min}>1\kpc$) to avoid the complete disruption of the satellite in the region of $\Nutgsr>20\kms$.
This is corroborated with our N-body simulations in Section \ref{sec:res:dynfr}, where only the outer layers of the dark matter distribution of the dwarf could be stripped, leaving a core that can keep the stellar component of the dwarf bound.

\begin{figure}
\begin{center}
\includegraphics[width=8.5cm]{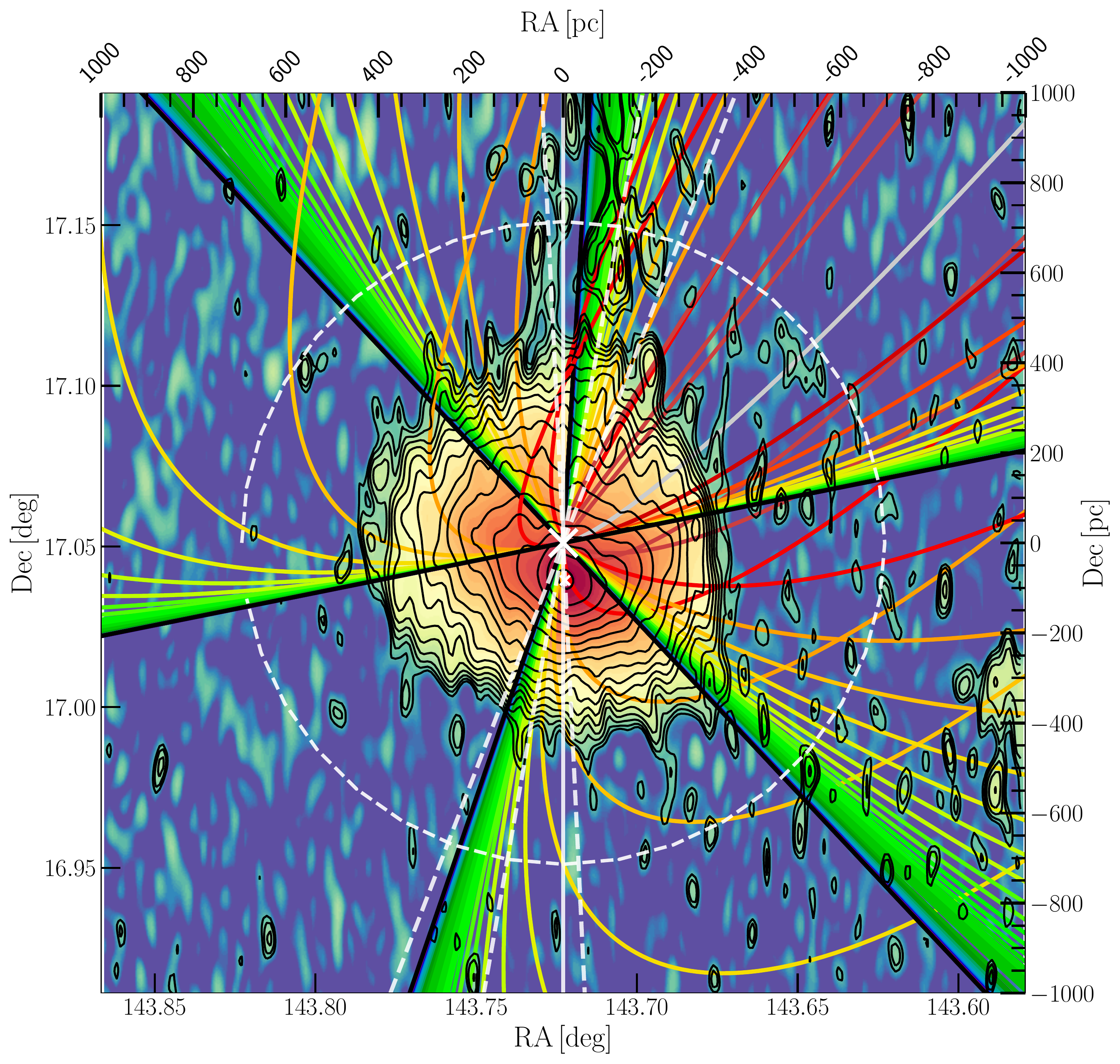}
\includegraphics[width=8.5cm]{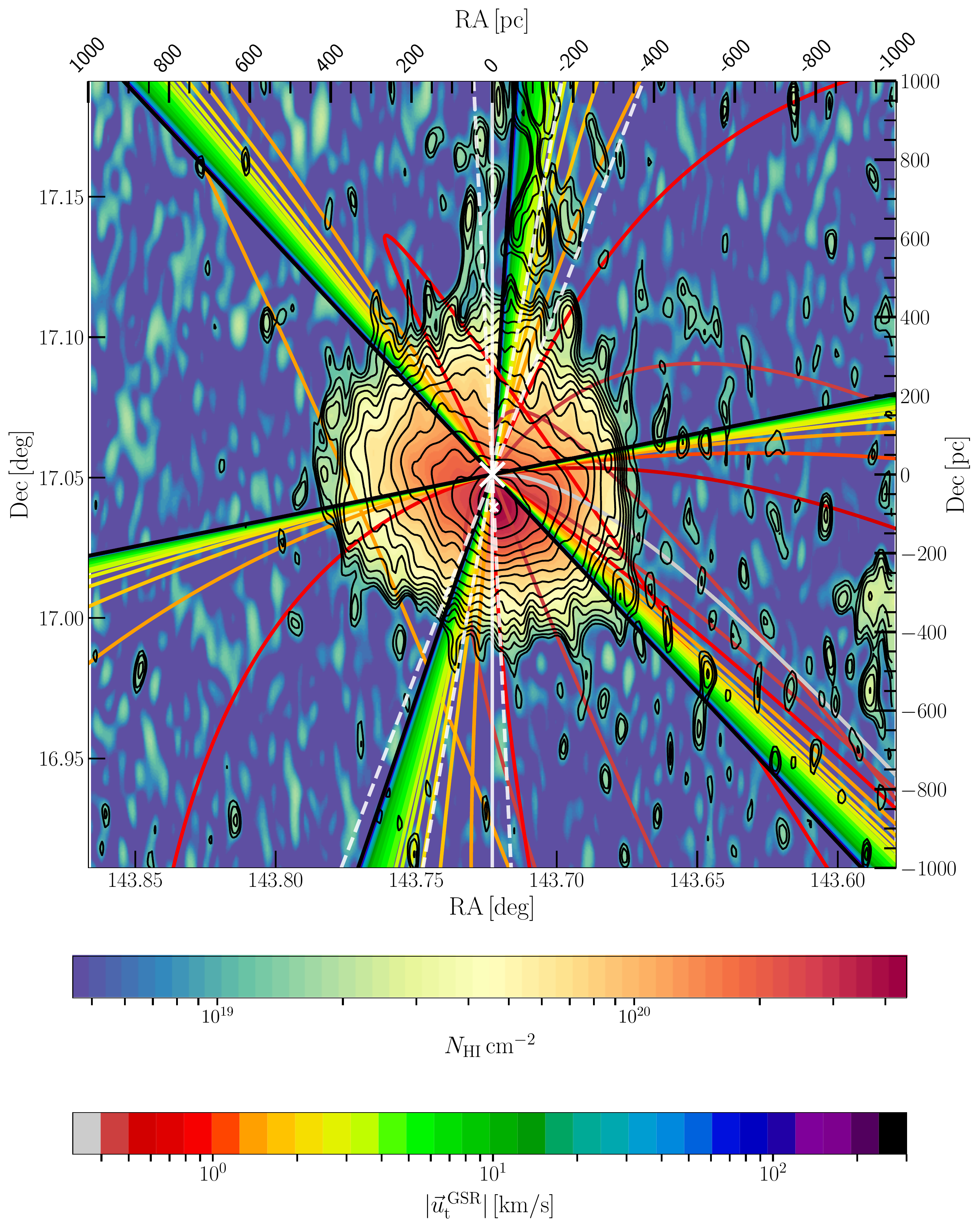}
\vspace{-0.2cm}
\caption{
We show orbits of case 4 (top panel) and of case 2 (bottom panel) projected on the sky, 
showing different directions \utgsr and magnitudes (coloured curves).
We also plot the \HI column density map and contours of \citetalias{Adams2018}.
The optical center is marked with the large cross, the \HI density peak with small cross, and the North-South axis is shown with a white solid vertical line.
We plot a circle with $\Delta R\e0.1\degree\,(713\pc)$ where we measure the \PA of the orbits.
Given that we use a logarithmic velocity color bar, we tag the zero velocity curve ($\Nutgsr\e0\kms$) as the minimum value in the bar ($0.3\kms$)
(grey curve).
Note the way the orbits bend towards the most radial orbits when \Nutgsr decrease to zero.
For case 4 the most radial orbits approach from North West due to the acceleration towards M31, while in case 2 they approach from South West.}
\label{fig:orb:sky}
\end{center}
\end{figure}

\subsubsection{The trajectory of Leo T on the sky and proper motion constraints}
\label{sec:res:main:sky}
In the previous sections we determined constraints for the magnitude of the tangential velocity considering several directions of \utgsr.
Here, we constraint the direction of \utgsr by comparing the orientation of the orbits of Leo T projected on the sky 
with several gas features in Leo T (Fig. \ref{fig:LeoT:HImap}) that we discuss in Section \ref{sec:obs}. 

Particularly, we search for orbits that are aligned with the \HI-stellar offset \PAOV, the \HI tail at $\PATV$
and the \HI flattening at $\PAFV$.
We argue in Section \ref{sec:obs} that these features could be produced by the ram pressure from the MW gas halo, 
which would generate the tail and the offset which would be aligned with the projected orbit of Leo T, and the flattening of the \HI isophotes to the South of Leo T, which would be perpendicular to the projected orbit.

Under this assumption we vary the direction of the tangential velocity $\utgsr$ to find the best alignment of the projected orbit with this offset axis. 
In Fig. \ref{fig:orb:sky} we show some orbits of our fiducial case, case 4, which are projected on the sky together with the \HI map of Leo T, 
where we show a range of directions and magnitudes for $\utgsr$.
The figure shows how the orbits start bending in the direction of the most radial orbits when the values of \Nutgsr decrease below 10\kms. 
Above this value the orbits align well with the given direction of $\utgsr$.
We also include in Fig. \ref{fig:orb:sky} some orbits for a case that does not include the potential of M31 in the calculation (case 2),
to illustrate an interesting difference with the cases that do account for M31:
given that the Milky Way and Leo T have different accelerations towards M31, the resulting orbits are slightly different to that of the cases without M31.
This effect is mostly noticeable for the most radial orbits with $\Nutgsr<1.5\kms$,
where for case 4 the radial orbits approach from the MW from $\PA\e-45\degree$ to Leo T's current position, 
while in the isolated case 2 the orbits come from $\PA\e-100\degree$. 
We note that in Fig. \ref{fig:orb:sky} we have plotted the orbits on the sky from an inertial frame.

\begin{figure}
\begin{center}
\includegraphics[width=8.2cm]{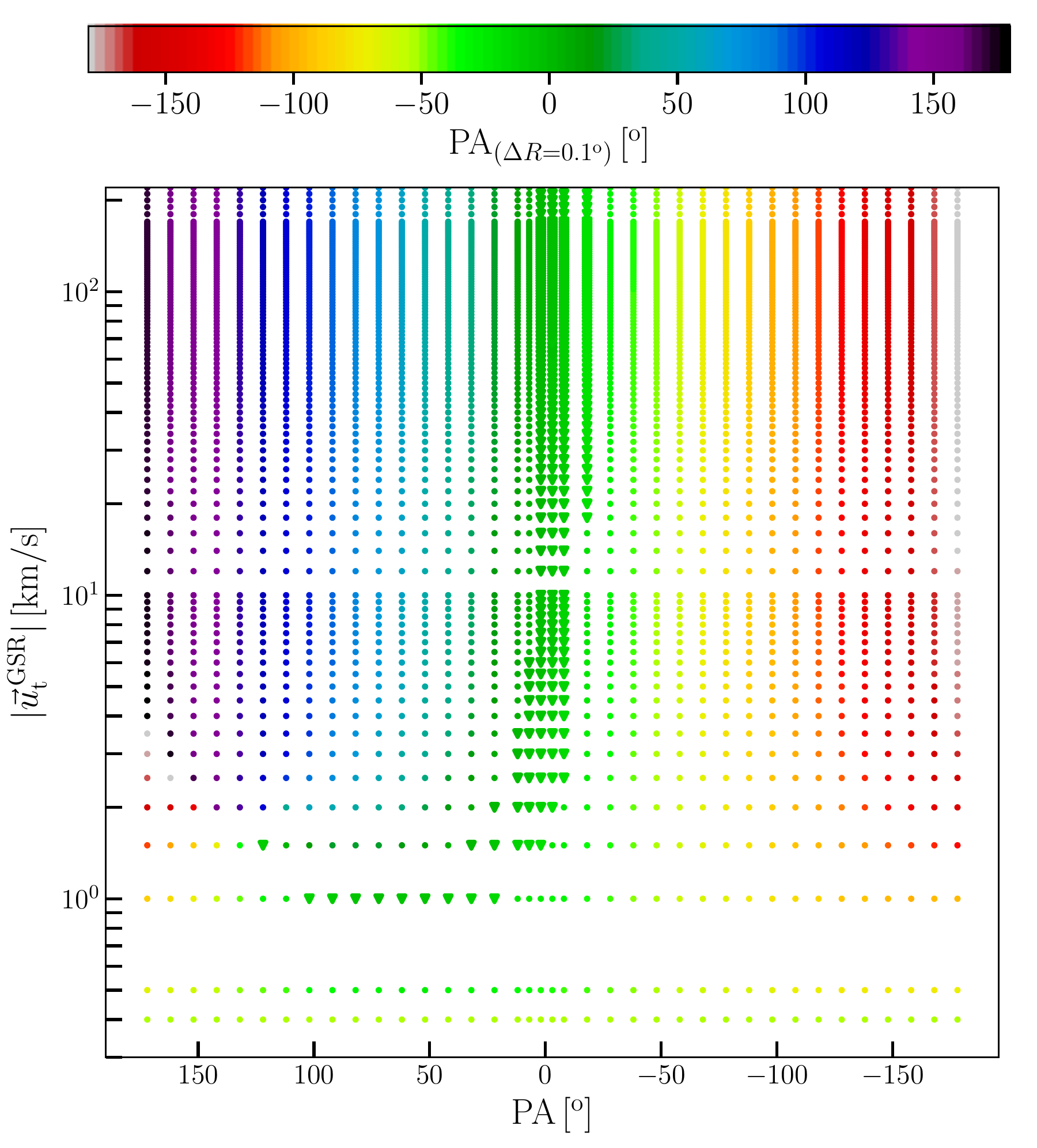}
\includegraphics[width=8.6cm]{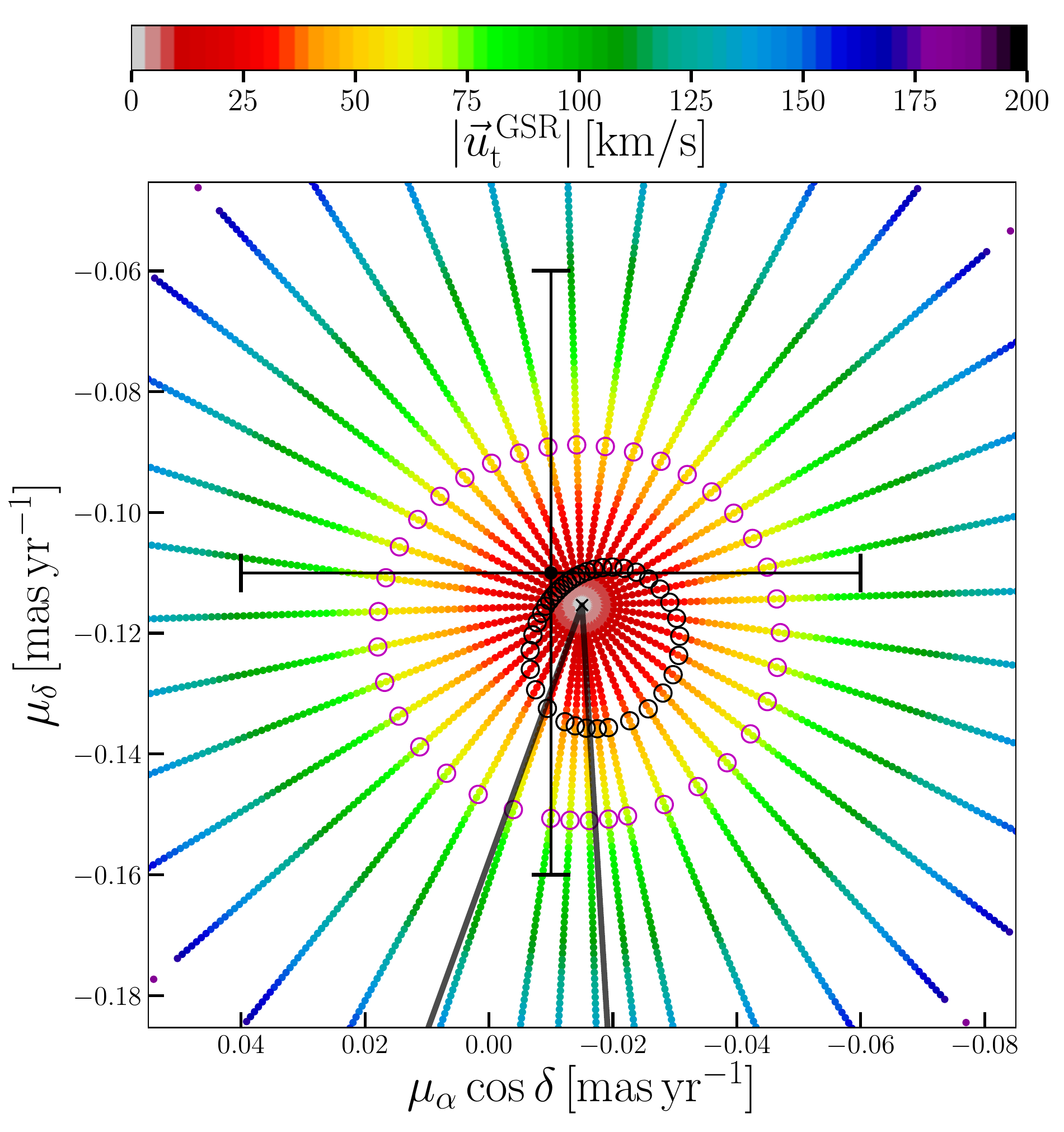}
\vspace{-0.6cm}
\caption{
Top panel: direction of the tangential velocity \utgsr as function of is magnitude \Nutgsr for case 4. 
Given that the velocity color bar scale is logarithmic, we have tagged the zero velocity ($\Nutgsr\e0\kms$) as 0.4\kms in the plot. 
The colors show the position angle of each orbit where it intersects a circle on the sky with radius $\Delta R\e 0.1\degree$ centered on the optical center of Leo T. 
The triangle symbols mark the \PA values that fall within the position angle of the \HI-stellar offset and the \HI flattening, with a few degrees of error range, 
\ie between $\PAOV$ and $\PAFV$.
Bottom panel: proper motions explored for Leo T's current position for case 4 as function of \Nutgsr (colour bar). 
The proper motion when $\Nutgsr=0\kms$ corresponds to $\muas=-0.0150\masyr$, $\mud=-0.1153\masyr$ 
due to the solar motion around the MW (grey region).
Each radial set of points corresponds to different directions of \utgsr. 
The values that better align with the \HI - stellar offset at $\PAO$ and the \HI flattening located between at $\PAF$
are located within the black lines shaping a triangle. 
The black small circles surrounding the center mark the ram pressure stripped region due to $P_{\rm R,2}$,
and the magenta circles mark the backsplash solution region.
The point with error bars correspond to the estimate of \citet{McConnachie2020}.}
\label{fig:OA:propmot}
\end{center}
\end{figure}

To constrain the direction of $\utgsr$ we project all the orbits on the sky as shown in Fig.\ref{fig:orb:sky}, and then measure the position angle of each orbit 
where the orbit intersects a ring on the sky with a radius $\Delta R\e0.1\degree\, (713\pc)$ centered on the position of Leo T determining ${\rm PA}\left(\Delta R\e0.1\degree\right)$.
The result of this measurement is shown in Fig.\ref{fig:OA:propmot} for case 4.
In the plot we mark all the orbits with values of ${\rm PA}\left(\Delta R\e0.1\degree\right)$ that fall within the \PAO and \PAF.
Some solutions are found for larger values of \PA ($\si50 - 120\degree$) and very small tangential velocities 
$\Nutgsr\leq1.5\kms$, but larger than 0.5\kms, because those are very radial orbits, which in projection rapidly 
bend and turn back in direction to the Galaxy. 
In Fig.\ref{fig:OA:propmot} we present the explored proper motions for Leo T as function of \Nutgsr for case 4. 
We mark the regions where we obtain backsplash orbital solutions and gas stripped solutions,
determined as in sections \ref{sec:res:main:bas} and \ref{sec:res:main:rps}.
The proper motions of the grid that generate orbits that best align with the direction of the \HI-stellar offset and the \HI flattening are enclosed by the two lines shaping a wedge.
We note in the figure that the ram pressure region is not centered exactly around $\Nutgsr\e0\kms$. 
This is because there is some angular momentum, given that $\vlosgsr$ is not exactly equal to the radial velocity in Galactocentric coordinates, 
also due to the potential of M31, which perturbs the orbit, and also to the fact that the satellite "hits" the gaseous disk on different regions with different densities and at slightly different times. 

In Table \ref{tab:res:par:leot} we provide for different cases the tangential velocity threshold values \Nutgsrrps and \Nutgsrbas for both the directions, the \HI-stellar offset and the \HI flattening,
providing this values as proper motions as well, with the proper motion errors in the table estimated from the different cases.
In addition, we take the direction along the \HI tail $\PATV$ that lays between the \PA of the \HI offset and the flattening, and obtain the proper motion ranges for different orbital solution.
Given that these selection of orbits is almost align with the north axis, the proper motion in {\rm $DEC$} is what mostly determines the region where the solution is:
\begin{enumerate}[leftmargin=0.5cm]
\item first in-fall solutions for $\mud<-0.1507[\masyr]$.

\item unstripped gas backsplash orbital solutions (BAS) are within the ranges of 
and $-0.1507\leq\mud/[\masyr]\leq-0.1347$.

\item backsplash ram pressure stripped solutions (RPS) are within the range of $-0.1347\leq\mud/[\masyr]\leq-0.1153$.
\end{enumerate}

Furthermore, if the restriction on the direction of the velocity is not imposed, Fig \ref{fig:OA:propmot} can provide constraint on the magnitude of the proper motion for case 4.
\chn{We note that new proper motion estimates for Leo T were published by \citet{McConnachie2020}, who find 
$\muas\e-0.01\pm0.05\masyr$ and $\mud\e-0.11\pm0.05\masyr$, with large errors due to the large distance to Leo T and systematic errors.
These estimates locate Leo T in the region of gas surviving backsplash orbital solution, but due to the large error some first in-fall solutions are also considered (see Fig \ref{fig:OA:propmot}).
Moreover, the proper motion values estimated from the \HI morphology still lay within the errors of the observed proper motion.}

\subsection{Effects of the cosmic expansion on the orbits}
\label{sec:res:cosmo}

\begin{figure*}
\begin{center}
\includegraphics[width=8.8cm]{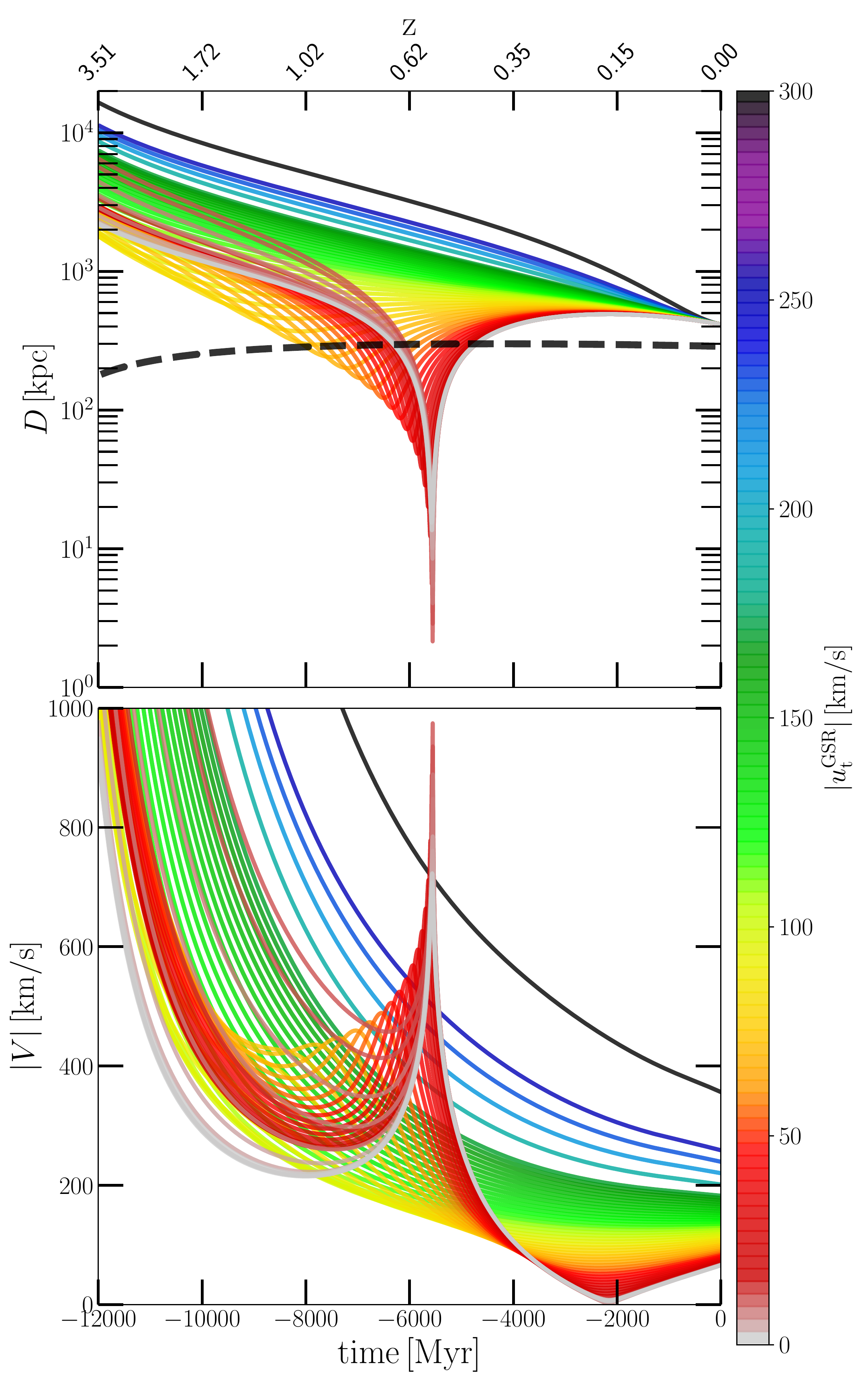}
\includegraphics[width=8.8cm]{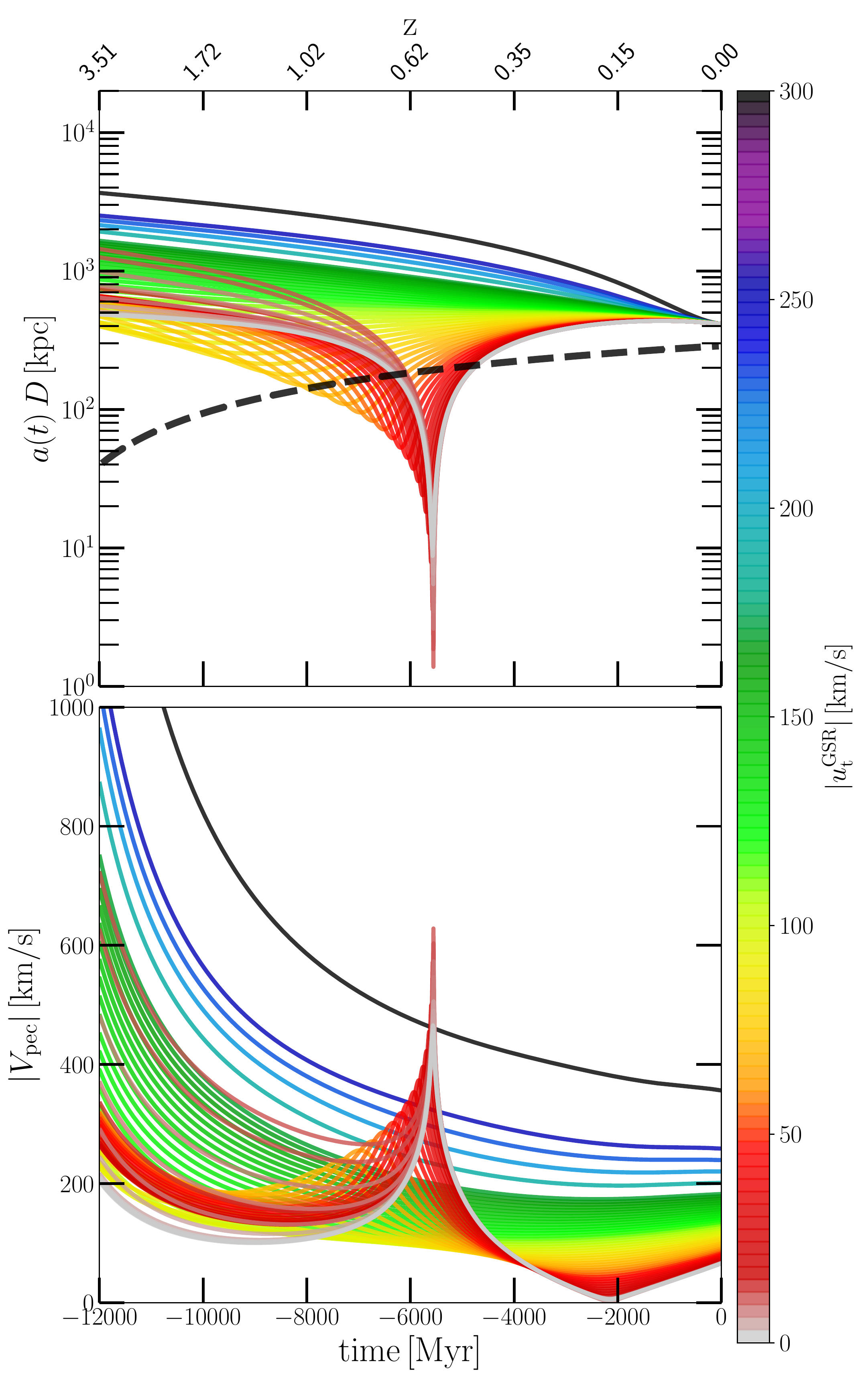}
\vspace{-0.2cm}
\caption{We show for the case cos4 (case 4 in an expanding space) 
the comoving Galactocentric distance and velocity (left panels) and the physical coordinate distance and peculiar velocity (right panels) as function of the lookback time (see Section \ref{sec:app:oint} and Eq. \ref{eq:cosmo2}).
Colours indicate the value of \Nutgsr, showing here only a selection of orbits coming from the direction of $\PA=3\degree$ 
to avoid overcrowding.
In the top panels is shown the virial radius as function of time in comoving coordinates ($R_{\rm vir}$) and physical coordinates ($r_{\rm vir}$) (curved dashed line).}
\label{fig:orb:case4cosmo:distvel}
\end{center}
\end{figure*}

In this section we show our orbits calculated backwards in time for Leo T, solving the equations of motion of an expanding Universe, as is explained in Section \ref{sec:met:orbit}.

\citet{Karachentsev2009} show how the kinematics of the galaxies in the Local Group (LG) transitions into the Hubble flow at a distance of $\si\!1\Mpc$ from the LG center, denoted the zero velocity radius (see their Fig. 1). 
They show how at those distances the LG gravitational potential perturbs the kinematics of the galaxies moving in the Hubble flow and, 
more relevant for this work, it is shown how the Hubble flow perturbs the dynamics of the LG and the distribution of its satellites in the outer regions of the LG.
\citet{Penarrubia2014} analysis also show with test particles that the cosmic expansion indeed affects the spatial and kinematical distribution of the satellites located at distances of about the separation between M31 and MW (currently ~0.78\Mpc) and larger, dominating the Keplerian potential. 
At smaller distances the quadrupole modes of the potentials contribute in addition to the monopole terms.
Hence, the total mass of the Local Group determines the extension of the influence on the satellites that transition into the Hubble flow.
Current Local Group mass estimates come from the timing argument, where most of the mass is from M31 and the MW.
The third most massive galaxy in the LG is the Triangulum galaxy (M33), with an estimated dynamical mass within its \HI distribution of $8\times10^{10}\sm$ \citep{Kam2017}.
The LG mass ranges in the literature mostly depend on the relative velocity between the M31 and MW systems \citepalias{Bland-Hawthorn2016}.
\citet{VanderMarel2012} determine a (virial) timing LG mass by selecting galaxy pairs in the Millennium simulations \citep{Li2008} finding $4.9\pm1.6\times10^{12}\sm$. After considering the orbit of M33 about M31, they estimate a timing mass of $3.2\pm0.6\times10^{12}\sm$.
\citet{Penarrubia2014} estimate a timing LG mass of $2.3\pm0.7\times10^{12}\sm$, similar to the values used in our models that include the MW and M31 (cases 3, 4, cos3 and cos4) with a total mass of $M_{\rm tot}=\left(1.1+1.3\right)\times10^{12}\sm=2.4\times10^{12}\sm$ or in subcases (b) $M_{\rm tot}=2.7\times10^{12}\sm$, which would then be within that range.

Therefore, dwarfs orbiting at large distances from the Local Group, such as Leo T or Cetus, located between $\si400$ and $\si700\kpc$, can experience deviations on their orbits by the cosmic expansion.
We observe this deviation in our orbital calculations, which is better revealed when we compare the distance and velocity of the orbits of our fiducial case 4 in Fig.\ref{fig:orb:case4:distvel} with 
the cosmological orbits of case cos4 in Fig. \ref{fig:orb:case4cosmo:distvel}, which is case 4 (MW accreting with M31 potentials), but in an expanding space time. 
In Fig. \ref{fig:orb:case4cosmo:distvel} we show the distance and velocity both in comoving coordinates ($\vec{X},\vec{V}$),
and in the physical position coordinate $\vec{r}$ and the peculiar velocity is $\vec{V}_{\rm pec}$, which are related by Eq. \ref{eq:cosmo2}, and the parameter $a\left(t\right)$ shown in Fig.\ref{fig:cosmopar}.

Similarly to other cases, we find that, qualitatively, the properties of the orbits are similar to that of the non-expanding cases, 
with the most noticeable differences listed here: 
\begin{enumerate}[leftmargin=0.5cm,itemsep=2pt,parsep=2pt]
\item We find in Fig. \ref{fig:orb:case4cosmo:distvel} for case cos4 that the distances reached by the backsplash and first in-fall orbits at -12\Gyr ($\z\e3.5$) are larger than in the non-expanding case, reaching distances between 1 and $5\Mpc$ in comoving coordinates or 0.4 and 1.5\Mpc in physical coordinates.
In units of virial radius this is roughly $D\si10\Rvir$ at that redshift, which is similar to the most distant backsplash orbits found in the NIHAO cosmological Milky Way-type simulations \citep[see their Fig. 7]{Buck2019}.

\item Shift of first pericenter passage: the first pericenter passage in case cos4 for the most radial backsplash orbits coming from large distances, is reached at $t\e-5.7\Gyr$, resulting in a time delay with a shift of $\Delta t\approx 2\Gyr$ later than in the non-expanding fiducial case 4, where the first pericenter passage is at $t\e-8.1\Gyr$. 
We also determined the median and maximum and minimum times for the backsplash threshold orbits for case 4 which is 
$t\left(D_{\rm min}^{\rm BAS}\right)\e-11.9^{+0.3}_{-0.1}\Gyr$,  
and for the ram pressure threshold is 
$t\left(D_{\rm min}^{\rm RPS,2}\right)\e-8.4^{+0.2}_{-0.4}\Gyr$.
While including the cosmic expansion delays these to 
 $t\left(D_{\rm min}^{\rm BAS}\right)\e-7.5^{+0.2}_{-0.6}\Gyr$ and 
 $t\left(D_{\rm min}^{\rm RPS,2}\right)\e-5.7^{+0.1}_{-0.2}\Gyr$, \ie a similar delay shift ranging $\Delta t \approx 3$ to 5\Gyr.

This results from the cosmic expansion at $\z>1$, where the backsplash orbital solutions for this dwarf predict larger distances from the Milky Way at early times, which then take longer to fall to the center, and also due to the Hubble flow that decelerate the in-falling satellites orbits when $H(\z)$ is large.
Then at closer distances ($D<2\Rvir$) and lower redshift ($\z<1$) the mass accretion of the Milky Way contributes in bringing the first in-falling satellites closer to the Milky Way center after their first in-fall.
\ch{We test the effects of the cosmic expansion without the effects of the Milky Way mass accretion or the dark halo extension. 
For this we calculate the orbits of Leo T backwards in time considering a Keplerian potential with constant total mass $M_{\rm vir}\e1.3\times10^{12}\sm$ with and without the cosmic expansion, also finding for the latter a time delay of $\si2\Gyr$.}

Also, in Section \ref{sec:res:dynfr} we show our analysis of the effects of dynamical friction on our backsplash orbits with the analytical Chandrasekhar approximation and full N-body simulations, which reveal that this mechanism is ineffective in bringing the satellite on backsplash orbits closer to the MW center when the pericenter distance and the maximum velocity are large.

\item The maximum velocity for the most radial orbits is $\si970\kms$ in comoving velocity $V$, or $\si600\kms$ in peculiar velocity, larger than in the non-expanding space case 4, that reaches $\si400\kms$. 
This results from a combined effects of the dwarf falling from a much larger distance,  as well as the first in-fall time delay that sets the pericenter passage 2\Gyr later, when the Milky Way has accreted more mass, assembling already 90 per cent if its present mass. 
The latter effect can be seen by comparing with case 1 in Fig.\ref{fig:orb:cases}, where the MW mass is constant, reaching the satellite a  velocity up to $500\kms$.

\end{enumerate}

\subsection{Dynamical friction and tidal effects}
\label{sec:res:dynfr}

We test the effects of dynamical friction including the Chandrasekhar approximation on the orbit calculation.
We find that cases 5 and 5c behave similarly to the cases without the friction term, with the largest effects for radial orbits that reach the center where the MW has the highest density.
To better estimate the effects of the dynamical friction on our orbits, we setup particle models for Leo T and the Milky Way to run full N-body simulations to explore orbits with different tangential velocities.

\subsubsection{Initial conditions and setup}
\label{sec:res:dynfr:setup}

\begin{figure}
\begin{center}
\includegraphics[width=8.cm]{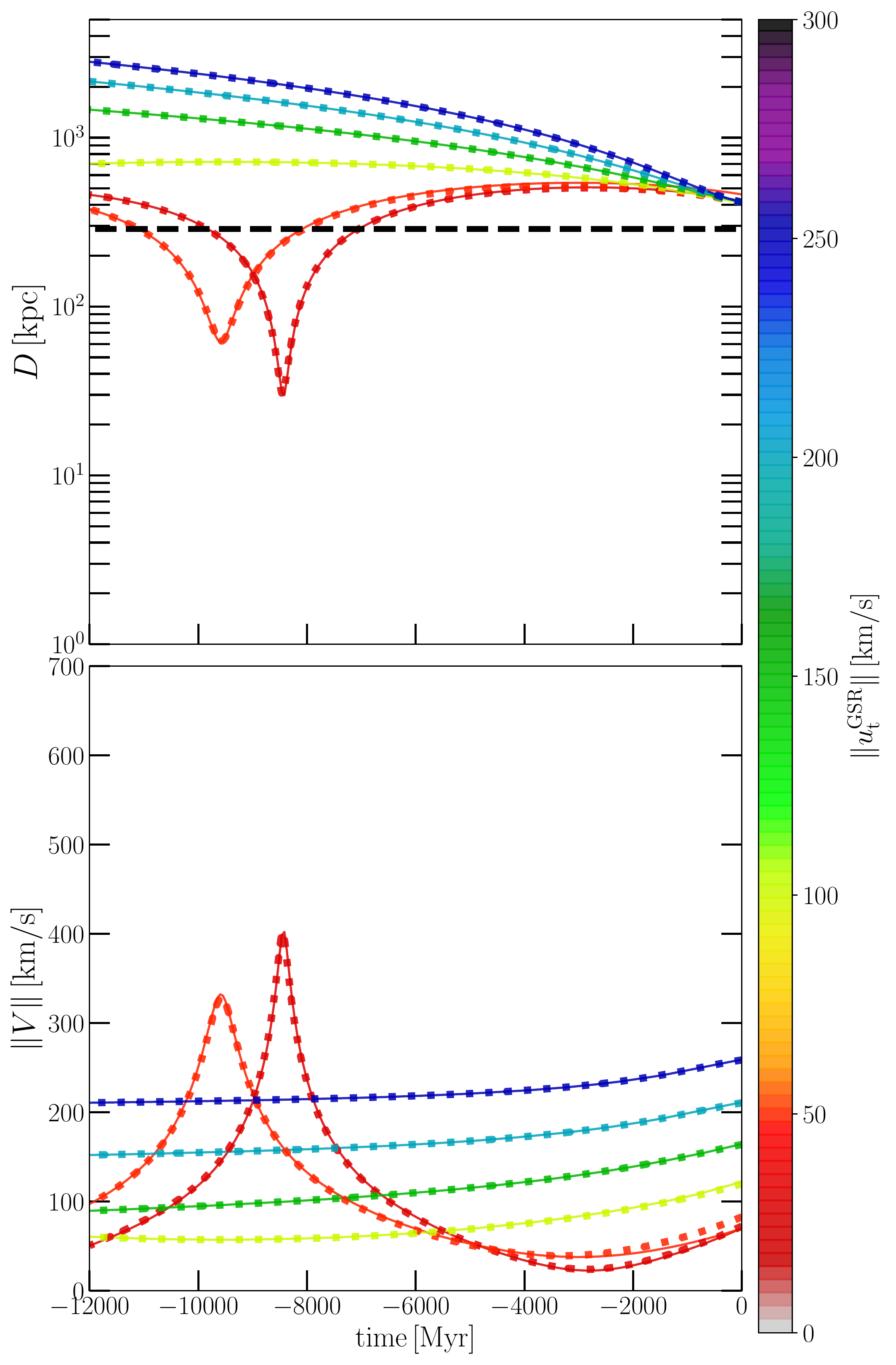}
\vspace{-0.4cm}
\caption{Relative distance and velocity of Leo T relative to the MW as function of lookback time 
for the full live N-body MW and Leo T simulation shown with solid curves, 
and for the frozen N-body MW potential, shown with dotted curves, almost always overlaying the solid curves. 
We show results for $\Nutgsr=(30,\,50,\,100,\,150,\,200,\,250\kms)$ (colours).}
\label{fig:orbit:nbody}
\end{center}
\end{figure}

We generate initial conditions for our N-body models with the open source program \textsc{dice} \citep{Perret2014,Perret2016}. 
For the Milky Way we set up three components with masses and scale parameters according to case 1 in Section \ref{sec:met:pot}. 
For the halo, disk and spheroid we use $3\times10^6$, $1.3\times10^5$ and $1.2\times10^4$ particles, respectively. 
Most orbits studied here have pericenters outside or in the outskirts of the disk and the inner spheroid, making a high number density unnecessary, while the potentials of these components are included.

For Leo T we model the stellar component with a Plummer profile with mass of $2\times10^{5}\sm$ and de-projected scale of $250\pc$. 
To consider LeoT's gas potential we also include a particle component for the gas that is modelled only as a collisionless component. 
We use the parameters of our fitted Plummer to the gas \ie a Plummer mass of $5.2\times10^5\sm$ and de-projected scale of $260\pc$.
For the dark matter halo we choose the cored \citet{Burkert1995, Burkert2015} profile, which is motivated by dwarf galaxy formation simulations that show that a bursty star formation can generate a cored dark matter profile \citep{Ogiya2014a,Read2016} 
with a core the size of the de-projected stellar half mass radius; here we choose $260\pc$.
For the dark matter mass we use the Leo T lowest estimate of \citep{Patra2018} of $M_{\rm DM}^{0.3\kpc}\e2.7\times10^6\sm$. $M_{\rm o}\e3.4\times10^{6}\sm$ is the dynamical mass within a $300\pc$ radius that includes the gaseous, stellar and a dark matter mass.
We use $5\times10^5$ particles for the dark matter and $2\times10^5$ for the gaseous and stellar components, which is enough to prevent an artificial tidal disruption \citep{Bosch2018b}. Our N-body models are not an exact match to Leo T, but sufficient as a metric to estimate the effects of dynamical friction and tidal disruption on our analytical orbits for a LeoT-type galaxy.

We also setup an NFW dark matter halo for Leo T to explore the effects of tidal disruption in a more concentrated profile. Following the main initial condition parameters of \citet{Read2016} for a Leo T-type dwarf, we setup a dwarf galaxy with a virial mass, concentration and particle number of $5\times10^8\sm$, $24.93$ and $10^6$ respectively, where the baryonic component parameters are the same as in the cored Leo T model. This model has a dynamical mass within $300\pc$ of $\si10^7\sm$  (Fig. \ref{fig:orbit:prof}), as high as estimates from the \HI kinematics \citet{Faerman2013,Adams2018}.

To virialise and relax the initial N-body models in isolation we use the one core tree-code \textsc{GyrFalcON} \citep{Dehnen2000} from \textsc{nemo} \citep{Teuben1995}.
We relax the N-body models for 12\Gyr, where the edges of the dark haloes spread, and in the case of the MW model, the disk develops a bar. For these calculations the bar orientation and pattern speed are left free, given that the studied orbits are far enough that the quadrupole potential of the bar is weak.

\subsubsection{Orbits in a frozen and a live MW potential}
\label{sec:res:dynfr:orb}

Our experiments consist of two steps. In the first step we coupled \textsc{delorean} with \textsc{GyrfalcON} to use the relaxed MW particle model to calculate several orbits for Leo T 12\Gyr backwards in time in a frozen potential as test particles, where the particles of the MW are not allowed to move in time.
We explored the values: 
$\Nutgsr=(10,\,20,\,30,\,40,\,50,\,60,\,70,\,80,\,90,\,100,\,150,\,200,\,250\kms)$, with a fixed direction of $\PA=-5\degree$ (see Section \ref{sec:res:main:sky}).
In the second step we use the coordinates and velocities found at -12\Gyr as initial positions and velocities to locate our relaxed N-body models of Leo T. 
Then we proceed to run it forward in time for 12\Gyr until the present ($t=0\Gyr$), but this time we evolve in time the MW and Leo T particles and potential.

In Fig.\ref{fig:orbit:nbody} are shown the distance and velocity of the orbits for the frozen and the live potentials. 
To calculate the orbits and the relative separation and velocity of Leo T and the MW in the live potentials, we keep track of the center of mass of the stellar component of Leo T and the MW. 
In Table \ref{tab:orbit:nbody} we present the difference in position and velocity between the orbits in a frozen potential and in a living potential for a selection of tangential velocities.
We find that at $t=0\Gyr$, even though the deviation between the frozen potential and the living potential orbits can be large for the backsplash solutions, \eg 52\kpc for a tangential velocity of 50\kms,
the distances between the satellite and the MW deviate only by 10 per cent or less, and the velocity deviates only by 15\kms or lower.
This means that, for the range of explored backsplash orbits, the dynamical friction can change the direction of the velocity and the direction of the orbit, but the magnitude of the velocity, reaching then similar large apocenter distances.
We also tested how well approximated would be the analytical orbits between the moment after the satellite passed through the MW center and its current position. 
For this we repeat the previous simulations, but now we position the N-body model of Leo T at $-7\Gyr$ instead of $-12\Gyr$, finding a deviation in distance for backsplash orbits of 2 per cent (Table \ref{tab:orbit:nbody}).\\

\begingroup
\setlength{\tabcolsep}{4pt} 
\renewcommand{\arraystretch}{1.} 
\begin{table}
\begin{threeparttable}
\caption{Orbits in a frozen and living N-body potential}
\begin{center}
\begin{tabular}{ccccccccc}\\\hline\hline
$\Nutgsr[\kms]$                                               &    30       & 50       & 100       &  150 & 200 & 30$^{(*)}$  & 50$^{(*)}$ \\\hline
$D_{\rm L}/D_{\rm F}$                                    &    1.01     & 1.11    &  0.98    &  1.0       &  1.0      &   0.98   & 0.99 \\
$\|\vec{D}_{\rm F}-\vec{D}_{\rm L}\|\kpc$       &    18        &  52      & 8          &   1      &    0.4       &   6     & 4  \\
$\|\vec{V}_{\rm F}-\vec{V}_{\rm L}\| \kms $     &    2.3      &  15      &   2.9      &   0.4    &   0.6      &    2.1     & 1.4 \\\hline
\end{tabular}
\begin{tablenotes}
\small
\item \textbf{Notes:} $\vec{V}_{\rm F}$ and $D_{\rm F}\e414\kpc$ are the relative distance and velocity between the satellite and the MW at $t\e0\Gyr$ for the frozen (F) N-body potential.
 $\vec{V}_{\rm F}$ and $D_{\rm L}$ are the same quantities calculated for the living (L) potential.
(*) Values calculated from orbits integrated from an initial condition at $-7\Gyr$ instead of $-12\Gyr$.
\end{tablenotes}
\end{center}
\label{tab:orbit:nbody}
\end{threeparttable}
\end{table}
\endgroup

How is the satellite affected by the tidal forces of the Milky Way for the different orbits?
As expected, the first in-fall orbits, with $\Nutgsr>90\kms$, do not experience strong tidal effects.
The backsplash solutions with wide orbits and velocities between $\Nutgsr>50\kms$ and 80\kms decrease the central stellar and dark matter densities, as shown in Fig. \ref{fig:orbit:prof}.
Backsplash orbits with $\Nutgsr<50\kms$ produce a decrease of the central density of one to two orders of magnitude below the initial profiles.
We note that even though the stellar density profile becomes shallower, the stars are still bound to the satellite, as the cumulative mass profile reveals in Fig. \ref{fig:orbit:prof}. 
However, an important fraction of dark matter (up to 50 per cent) can be stripped for extremely radial orbits.

Nonetheless, if we take a more concentrated and more massive dark matter halo model for Leo T, we find that it could survive to radial orbits with tangential velocities as low as 20\kms. 
In Fig. \ref{fig:orbit:prof} we show the density profiles of a dwarf model with the NFW dark matter profile.
As this profile has a cuspy density and a more massive dark matter halo, the tidal disruption is much weaker and the density in the central few hundred parsecs is only perturbed for more radial orbits with $\Nutgsr<20\kms$.
This is not surprising as known from several studies \citep{Penarrubia2008,Bosch2017}.
Furthermore, we note here that, while a progenitor for this dwarf galaxy could be found, using methods developed in \citet{Fellhauer2008a, Blana2015, Matus2019} \citep[see also][]{Dominguez2016}, this is beyond the scope of the present study.
The purpose of our numerical experiment here is to show that a satellite on such backsplash orbits could survive a tidal disruption by the Milky Way, and even provide a mechanism to generate a core in the central dark matter distribution.

\subsection{Satellites in the outer rim of the Milky Way halo}
\label{sec:disc:sat}

We also applied our method to find first in-fall and backsplash orbits to a selection of distant dwarfs located beyond the virial radius of the MW (288\kpc): Cetus, Eridanus II and Phoenix I.
We explored eight different directions of the tangential velocity \utgsr (every $\Delta \PA\e45\degree$), and 70 values of \Nutgsr, between 0 and 350\kms (with a higher sampling within 30\kms), exploring in total a velocity grid with 560 values.

Cetus and Eridanus II have no cold gas (only an upper limit \HI detectable threshold).
In the scenario that these satellites lost their gas due to ram pressure stripping,
how could we estimate the satellite's pre-existing cold gas, its density and its thermal pressure before it lost its gas?
We can use the \HI size-mass relation \citep{Begum2008,WangJ2016}, which is observed for galaxies covering a wide range of \HI masses, from Leo T, to the Magellanic Clouds, the MW, M31 and more massive galaxies \citep[see their Fig. 1]{WangJ2016}, with their parameters for the relation:
\begin{align}
\log\left(\frac{2\,R_{\rm HI}}{[\kpc]}\right)&= \left(0.506\!\pm\!0.003\right)\log\left(\frac{M_{\rm HI}}{[\sm]}\right)-\left(3.293\!\pm\!0.009\right)
\label{eq:HIrel}
\end{align}

The \HI-size relation implies a roughly constant surface gas density (or column density) $\Sigma_{\textsc{Hi}}\approx\Sigma_{\rm o}$. If we take a constant density core, it would relate the central gas density and its scale length in the form: $\Sigma_{\rm o}\propto M_{\textsc{Hi}} r_{\textsc{Hi}}^{-2}\propto \rho_{\textsc{Hi}}\times r_{\textsc{Hi}}$.  
Furthermore, \citet{Stevens2019} find that the \HI-size relation would be robust to ram-pressure-stripping a well.
As we find that in Leo T the \HI radius ($R_{\rm HI}$) is similar to the stellar half-light radius $R_{\rm Pl}^{\rm V}$ to within 4 per cent difference, we use the stellar half-light radius of these satellites as a proxy for their $R_{\rm HI}$ values to estimate the \HI mass that they would have contained. Then, assuming a Plummer profile, we estimate a central density and a \HI-mass-to-stellar light ratios for Cetus and Eridanus II of 1.5 and $17\sm\slu^{-1}$ respectively.
These assumptions, while arbitrary, would locate these pair on the gas rich side of what is observed for other distant gas rich dwarfs of the Local Group \citep{McConnachie2012,Spekkens2014}.
We test these results using the values of the central gas density and \HI mass of Leo T for these dwarfs, also finding ram pressure stripped orbits with similar values for the tangential velocities.\\

We present our findings below, and the main properties of the explored orbits in Tables \ref{tab:res:par:cet} and \ref{tab:res:par:eri} 
and in Fig. \ref{fig:orbs:par:dwarfs}:

\begingroup
\setlength{\tabcolsep}{6pt} 
\renewcommand{\arraystretch}{1.5} 
\begin{table}
\begin{threeparttable}
\caption{Orbital constraints for Cetus.}
\label{tab:res:par:cet}
\begin{center}
\begin{tabular}{ccccc}\hline \hline
Threshold: &  $\langle\Nutgsrbas\rangle$ & $\langle D_{\rm min}^{\rm BAS}\rangle$ & $\langle\Nutgsrrps\rangle$ & $\langle D_{\rm min}^{\rm RPS}\rangle$ \\
Case 	&  $\kms$                          & $\kpc$ & $\kms$ & $\kpc$  \\\hline
3 & $54^{+2}_{-5}$ & $287$ & $22^{+1}_{-8}$ & $55^{+1}_{-12}$\\
3a & $39^{+2}_{-3}$ & $263$ & $19^{+1}_{-7}$ & $54^{+1}_{-12}$\\
3b & $66^{+2}_{-7}$ & $308$ & $24^{+1}_{-8}$ & $56^{+1}_{-12}$\\
4 & $31^{+1}_{-3}$ & $184^{+5}_{-3}$ & $13^{+1}_{-5}$ & $50^{+0}_{-7}$\\
4a & $11^{+1}_{-1}$ & $164^{+2}_{-2}$ & (*) & (*) \\
4b & $43^{+2}_{-4}$ & $229^{+9}_{-4}$ & $16^{+1}_{-6}$ & $53^{+1}_{-9}$\\
cos4 & $42^{+2}_{-5}$ & $290^{+2}_{-2}$ & $16^{+2}_{-7}$ & $64^{+2}_{-15}$\\\hline
B.V($\pm\Delta$) & $31^{+37}_{-21}$ & $184^{+124}_{-19}$ & $13^{+13}_{-13}$ & $50^{+71}_{-9}$\\\hline\hline
\end{tabular}
\caption{Orbital constraints for Eridanus II.}
\label{tab:res:par:eri}
\begin{tabular}{ccccc}\hline \hline
Threshold: &  $\langle\Nutgsrbas\rangle$ & $\langle D_{\rm min}^{\rm BAS}\rangle$ & $\langle\Nutgsrrps\rangle$ & $\langle D_{\rm min}^{\rm RPS}\rangle$ \\
Case 	&  $\kms$                          & $\kpc$ & $\kms$ & $\kpc$  \\\hline
3 & $94^{+6}_{-11}$ & $287$ & $31^{+27}_{-17}$ & $38^{+20}_{-10}$\\
3a & $64^{+9}_{-10}$ & $263$ & $23^{+32}_{-15}$ & $35^{+21}_{-7}$\\
3b & $117^{+5}_{-10}$ & $308$ & $37^{+24}_{-18}$ & $40^{+19}_{-12}$\\
4 & $70^{+9}_{-10}$ & $211^{+3}_{-12}$ & $24^{+23}_{-14}$ & $40^{+17}_{-9}$\\
4a & $39^{+11}_{-10}$ & $166^{+1}_{-3}$ & $9^{+27}_{-9}$ & $36^{+14}_{-6}$\\
4b & $93^{+6}_{-11}$ & $257^{+6}_{-16}$ & $31^{+21}_{-16}$ & $41^{+16}_{-11}$\\
cos4 & $76^{+7}_{-10}$ & $284^{+1}_{-2}$ & $23^{+20}_{-13}$ & $42^{+18}_{-9}$\\\hline
B.V($\pm\Delta$) & $70^{+51}_{-41}$ & $211^{+97}_{-48}$ & $24^{+37}_{-24}$ & $40^{+20}_{-12}$\\\hline\hline
\end{tabular}
\begin{tablenotes}
\small
\item Notes: 
B.V. are the best values selected from case 4 with the errors estimated as in Table \ref{tab:res:par:leot}.
(*) There are no values for case 4a for Cetus which can produce ram pressure stripped orbits, as can be seen in Fig. \ref{fig:orbs:par:dwarfs}.
\end{tablenotes}
\end{center}
\end{threeparttable}
\end{table}
\endgroup

%

\begin{figure*}
\begin{center}
\includegraphics[width=7.8cm]{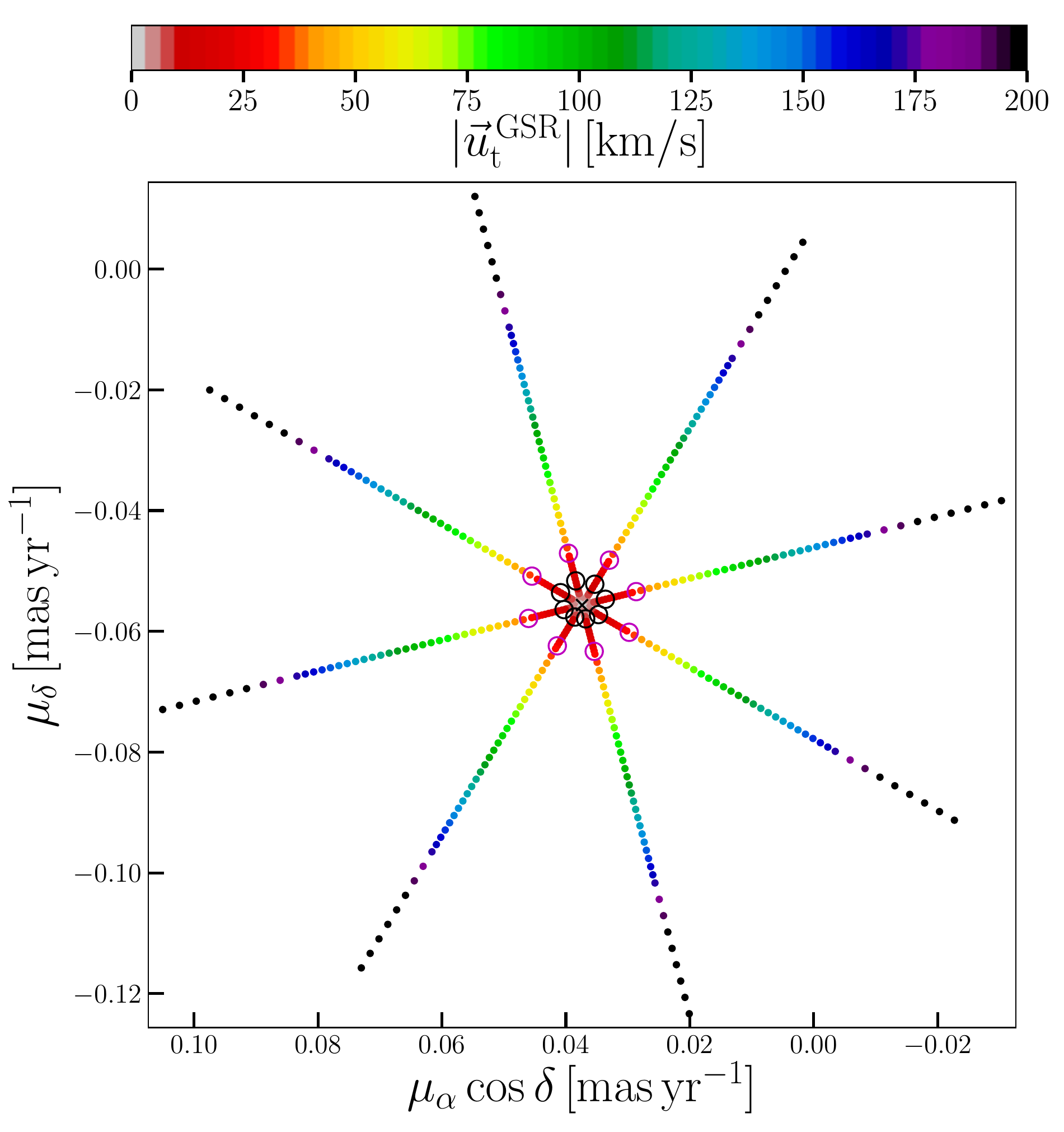}
\includegraphics[width=7.8cm]{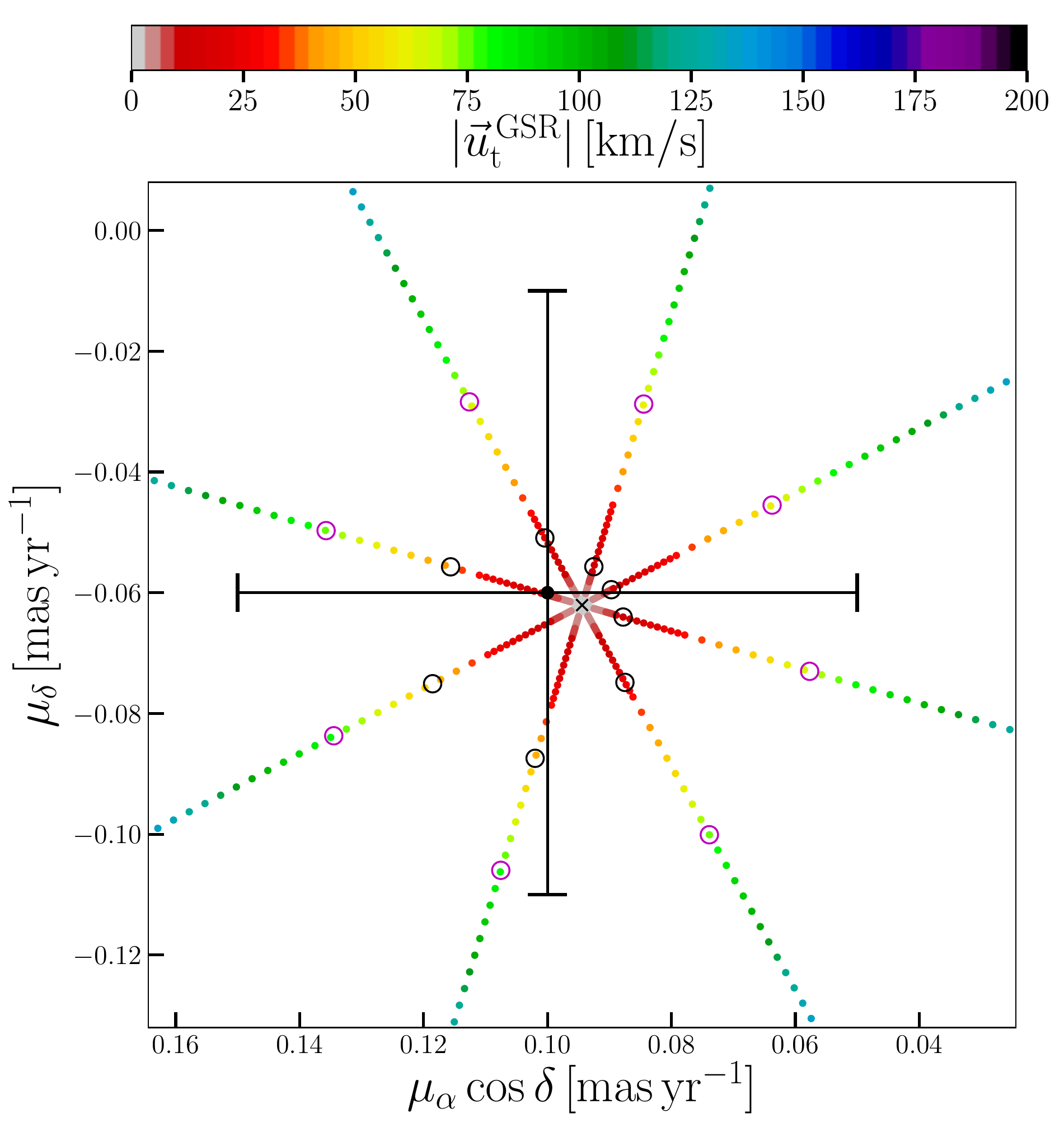}
\vspace{-0.3cm}
\caption{
Proper motions explored for Cetus (left panel) and Eridanus II (right panel) at their current positions for case 4 as function of \Nutgsr (colour bar). 
Each radial set of points corresponds to different directions of \utgsr. 
The black small circles surrounding the center mark the ram pressure stripped region due to $P_{\rm R,2}$,
and the magenta circles mark the backsplash solution region.
The point with error bars correspond to the observational estimate of \citet{McConnachie2020}.}
\label{fig:OA:propmotsat}
\end{center}
\end{figure*}

\begin{itemize}[leftmargin=0.3cm]
\item ~\textit{Cetus}:
this dwarf is located at $D_{\odot}\e755\!\pm\!24\kpc$ or $\si2.6\Rvir$ from the Milky Way \citep{McConnachie2006},
with a LOS GSR stellar velocity of $v^{\rm GSR}_{\rm los, \star} \e-14.85\pm1.7\kms$ \citep{Taibi2018}.
Its absolute magnitude is $M_{\rm V}\e-11.3\!\pm\!0.3\mags$ \citep{McConnachie2006}, 
and its stellar mass is $5.9\times10^6\sm$, that is $\si30$ times larger than in Leo T.
Its stellar velocity dispersion implies a dynamical mass of $M^{\rm dyn}_{\rm r_{3Dh}}\e67^{+19}_{-16}\times10^6\sm$ \citep{Taibi2018} within its 3D-half-light radius ($r_{\rm 3Dh}\!\approx\!4/3R_{\rm Pl}\e791\pm29\pc$) \citep{McConnachie2006}.

 \citet{Taibi2018} hypothesise that, given that Cetus has no detectable \HI \citep{Spekkens2014}, it could be a backsplash system that lost all its gas during a passage through the Milky Way. 
In Fig.\ref{fig:orbs:par:dwarfs} we show that surprisingly there is indeed a range of tangential velocities that result in backsplash solutions for Cetus, 
when $\Nutgsrbas\leq 31^{+37}_{-21}\kms$ (or see as proper motions in Fig. \ref{fig:OA:propmotsat}). 
which for case 4 happened at a range of
$t\left(D_{\rm min}^{\rm BAS}\right)\e-11.9\pm0.1\Gyr$  
while with the cosmic expansion for case cos4 was at
 $t\left(D_{\rm min}^{\rm BAS}\right)\e-7.4\pm0.1\Gyr$.

While several mechanisms to remove the cold gas in field dwarfs are possible,
such as UV background evaporation as shown in the NIHAO simulations \citep{Buck2019},
it is very interesting that we indeed find a range of ram pressure stripped backsplash orbits for $\Nutgsrrps\leq 13^{+12}_{-13}\kms$ (Fig. \ref{fig:OA:propmotsat}), 
which for case 4 happened at a range of
$t\left(D_{\rm min}^{\rm RPS,2}\right)\e-11.1^{+0.2}_{-0.1}\Gyr$
while for case cos4 was at
 $t\left(D_{\rm min}^{\rm RPS,2}\right)\e-6.6^{+0.1}_{-0.1}\Gyr$,
and for case 3 we find  $t\left(D_{\rm min}^{\rm RPS,2}\right)\e-10.2^{+0.3}_{-0.1}\Gyr$.
The exception where we find no ram pressure solutions is the extreme case 4a, 
where Cetus is too far and the time depending MW virial mass is too small at high redshift to have brought the dwarf close enough to strip its gas (see Table  \ref{tab:res:par:cet}). 
Of course, given that Cetus is far, we would expect the whole MW region attracting the satellite, a scenario better represented by case 1 or 3.
The maximum and minimum limits of the velocity thresholds are given by largest and smallest variations between cases, which are driven mostly by virial mass of the Milky Way, of case 3b and case 4a, respectively. 

We note that no GSR tangential velocity ($\Nutgsr\e0\kms$) would result in the proper motions of $\muasz\e0.037\masyr$ $\mudz\e-0.056\masyr$.

Given that Cetus currently is closer to M31 than to the Milky Way, and that it is presents no evidence for a stellar truncation radius, \citet{McConnachie2006} argue that this dwarf might have always been in isolation in the past. 
Here we report that we find no orbits passing near M31 in the past for this dwarf within the considered range of parameters.\\

\item ~\textit{Eridanus II}: 
it is located at $D_{\odot}\e366\!\pm  17\kpc$ from the MW $\si1.27\Rvir$ \citep{Crnojevic2016, Li2016a}, 
with a LOS GSR stellar velocity of $v^{\rm GSR}_{\rm los, \star} \e-76.8\pm2.0\kms$ \citep{Li2016a} \footnote{recomputed for the Galactic frame used here}.
It has a luminosity of $L_{\rm V}\e5.9^{+1.9}_{-1.4}\times10^4\slu$ with a 3D half-light radius of $r_{\rm 3D-half}\e369\pm18\pc$ and a dynamical mass within this of $M^{\rm dyn}_{\rm r_{3Dh}}\e12^{+4}_{-3}\times10^6\sm$ \citet{Li2016a}.
We also find a range of backsplash solutions for $\Nutgsr\leq 68^{+53}_{-39}\kms$ (see Fig.\ref{fig:orbs:par:dwarfs} and Table \ref{tab:res:par:eri})
which for case 4 happened at a range of
$t\left(D_{\rm min}^{\rm BAS}\right)\e-11.7\pm0.3\Gyr$,  
while including the cosmic expansion was at 
 $t\left(D_{\rm min}^{\rm BAS}\right)\e-6.8^{+1.2}_{-2.3}\Gyr$.
We also find a range of ram pressure stripped backsplash solutions for $\Nutgsr\leq 16^{+38}_{-16}\kms$.
\chn{Recent estimates from \citet{McConnachie2020} find new proper motion values of 
$\muas\e0.10\pm0.05\masyr$ and $\mud\e-0.06\pm0.05\masyr$ (prior A). 
We find that these values locate this satellite in the ram pressure stripped backsplash region, as shown in Fig. \ref{fig:OA:propmotsat}.
Therefore, it is possible that this dwarf could have passed close enough to the center of the Milky Way in the past to remove its gas.
We note however, that due to the large observational error some first in-fall solutions are also considered. 
Particularly, for case 4 the threshold for ram pressure stripped orbits happened at a range of
$t\left(D_{\rm min}^{\rm RPS,2}\right)\e-7.7^{+0.2}_{-0.6}\Gyr$
while for case cos4 was at
 $t\left(D_{\rm min}^{\rm RPS,2}\right)\e-5.4^{+0.1}_{-0.3}\Gyr$,
and for case 3 we find  $t\left(D_{\rm min}^{\rm RPS,2}\right)\e-7.2^{+0.3}_{-0.8}\Gyr$.
}
We note that a $\Nutgsr=0\kms$ translates into $\muasz=0.094\masyr$ $\mudz=-0.062\masyr$ for this satellite.\\

%

\item ~\textit{Phoenix I}:  
this dwarf is similar to Leo T in terms of mass and distance. 
It is located at $D_{\odot}\e409\pm 23\kpc$ ($\si1.4\Rvir$) from the MW \citep{Battaglia2012} (previous estimates are $D_{\odot}=415 \pm  19\kpc$ \citep{McConnachie2005}, $420 \pm  10\kpc$ \citep{Young2007}). 
Its projected half-light radius is $R^{\rm V}_{\rm h}\e2.3 \pm  0.07\arcmin\,(274 \pm  8\pc)$ \citep{Battaglia2012} and a luminosity of $L_{\rm V}\e7.6\times10^5\slu$ (re-scaled to a distance of 409\kpc). 
\citet{Kacharov2017} reports a stellar velocity dispersion of $\sigma_{\rm los,\star}\e9.2 \pm  0.7\kms$.
If we assume a constant dispersion profile, and we use the virial relation \ref{eq:mdyn} to find a dynamical mass of $M^{\rm dyn}_{\rm half}\e2.1\pm 0.3\times10^7\sm$ within the half-light radius and a mass-to-light ratio of $M/L_{\rm V}\e56\sm\slu^{-1}$.

Intriguingly, Phoenix also has an \HI cloud that is offset from the stellar center, with larger separation however, of about $300\arcsec$ ($595\pc$), with the \HI distribution not showing only a different systemic velocity from the stellar one, but it also shows a velocity gradient \citep{StGermain1999}.
\citet{Young2007} determine an \HI mass of $1.1\times10^5\sm$ and a central surface density of  $0.512\times10^{20} {\rm cm^{-2}}$ (or $0.55\sm\kpc^{-2}$, including He), where we have re-scaled it to a distance of $409\kpc$. 
Assuming that the central density follows a Plummer profile we find a gas core density and scale of $\rho_{\rm Pl}^{\rm gas}=1.4\times10^6\sm\kpc^{-3}$ and $R_{\rm Pl}^{\rm gas}=150\as\,(297\pc)$.

To calculate the orbits we use the stellar line-of-sight systemic velocity determined in \citet{Kacharov2017} with $v^{\odot}_{\rm los,\star}\e-21.2\pm1.0\kms$ that transforms to $v^{\rm GSR}_{\rm los,\star}\e-116.8\kms$, almost doubling the stellar LOS velocity of Leo T.
Exploring different magnitudes and directions of tangential velocities (\utgsr), we find almost exclusively first in-fall solutions.
Only in cases 3b and in case cos4 we find some backsplash solutions when the tangential velocity is lower than $30\pm 10\kms$. 
This occurs because these cases are extreme scenarios with the most massive Milky Way haloes.
Even for those cases the minimum distance that these orbits reach is half or 30 per cent the virial radius of the Milky Way. 
Furthermore, only in case cos4 we find some ram pressure stripped orbits for tangential velocities of $15\pm 5\kms$.
Therefore, we exclude backsplash orbits for Phoenix I, making this dwarf a first in-fall satellite candidate.

\citet{Fritz2018} also estimate the proper motion for Phoenix I finding 
$\muas\e0.079\pm 0.099\pm 0.04\masyr$ and $\mud\e-0.049\pm 0.12\pm 0.04\masyr$, which then converted to the GSR frame is $\Nutgsr\e68^{+406}_{-66}\kms$.
We note here that $\Nutgsr\e0\kms$ is equivalent to $\muasz\e0.083\masyr$ and $\mudz\e-0.084\masyr$.
We calculate the maximum ram pressure that the measured proper motion would imply.
The thermal pressure is  $7.0\times10^8\sm\kpc^{-3}\km^2\s^{-2}$, while the ram pressure is 
$1.6^{+19.1}_{-0.4}\times10^6\sm\kpc^{-3}\km^2\s^{-2}$ where the upper and lower error come from the velocity error range.
As the ratio is always lower than one, the ram pressure would be too weak to remove its gas at its current position.\newline
\chn{\citet{McConnachie2020} new estimates for the proper motion of Phoenix obtain similar values, 
finding $\muas\e0.08\pm0.05\masyr$ and $\mud\e-0.08\pm0.05\masyr$ (prior A).
Therefore, we reach the same conclusion, finding that Phoenix is a field dwarf on its first in-fall.}


\end{itemize}




 
\section{Summary and Conclusion}
\label{sec:con}

Leo T's interesting properties, such as a recent star formation episode, its location at a large distance from the Milky Way, its gas and dark matter content, makes this an excellent laboratory to study the formation and evolution of galaxies in the most idealised scenario.
In this paper we present a method to explore a wide range of orbits for the dwarf Leo T to determine if this dwarf could be a
backsplash satellite, \ie a dwarf that already passed through the Milky Way in the past,
or if it is approaching to the Milky Way for the first time.

For this purpose we developed the code \textsc{delorean} to explore a large number of proper motions and 
to calculate semi-analytical orbits backwards in time considering 
a number of scenarios for the gravitational potentials that include: 
an accreting MW halo, the potential of M31, dynamical friction and the cosmic expansion of the Universe.

Our main results are the following:
\begin{enumerate}[label*=\arabic*., leftmargin=0.5cm ,itemsep=3pt,parsep=3pt]
\item{We find backsplash orbital solutions for Leo T when the tangential velocity in the Galactic Standard of Rest is lower than $\Nutgsr \leq 63^{+47}_{-39}\kms$.
We find that backsplash solutions with $\Nutgsr>21^{+33}_{-21}\kms$ would allow the satellite to survive the ram pressure stripping of the Milky Way. 
This suggests that Leo T could be on a backsplash orbit, but in a wide orbit that could have reached a minimum distance of $D_{\rm min} \geq 38^{+26}_{-16}\kpc$ in the past that allowed it to keep its gas.
Therefore, Leo T could be first in-fall satellite or also a wide-orbit backsplash dwarf with a moderate interaction with the Milky Way in the past.
The velocity uncertainty range is dominated at first order by the uncertainty in the virial mass of the Milky Way, where we explored a 20 per cent virial mass variations, 
followed by the uncertainties in the Milky Way mass accretion history.}

\item{When comparing with the \HI morphology of Leo T \citep{Adams2018} we can select the orbits that better align with the observed \HI-stellar offset, the \HI inner flattening or the \HI tail, 
obtaining a backsplash velocity threshold of $\Nutgsr \leq 69\pm39\kms$.
This corresponds to a range of proper motions for different orbital solutions:
\vspace{-0.2cm}
\begin{enumerate}[leftmargin=0.3cm]
\item{first in-fall: $\mud<-0.1507[\masyr]$.}

\item{unstripped gas backsplash: $-0.0100\leq\muas/[\masyr]\leq-0.0127$ and $-0.1507\leq\mud/[\masyr]\leq-0.1318$.}

\item{gas stripped backsplash: $-0.0127\leq\muas/[\masyr]\leq-0.0150$ and $-0.1318\leq\mud/[\masyr]\leq-0.1153$.}
\end{enumerate}
In Fig. \ref{fig:OA:propmot} we provide proper motion constraints for all directions, and in Table \ref{tab:res:par:leot} we include error ranges for different cases.
A comparison with full hydrodynamical simulations that reproduce the \HI features will better constrain the proper motion and evolution of Leo T (future publication).
\chn{We note that new proper motions estimates for Leo T were recently published by \citet{McConnachie2020}, which locate Leo T in our region of gas surviving backsplash orbital solutions, but due to the large observational errors, some first in-fall solutions can also be considered.}}
%
%

\item{We also calculate orbits backwards in time with a cosmological scheme where space changes according to a cosmology with parameters from the \citet{PlanckCollaboration2015}. 
We find that the backsplash orbital solutions explored for Leo T experience a delay on their first in-fall of $\si 2\Gyr$ later than in the non-expanding fiducial case. 
This results from the cosmic expansion at $\z>1$, where backsplash orbital solutions are found at larger distances from the Milky Way at early times, which then take longer to fall, contributing to the first in-fall delay the Hubble flow that decelerate the in-falling satellites when $H(\z)$ was large.}

\item{We applied our method to the distant dwarfs Cetus, {$\rm Eridanus\,II$} and Phoenix I.
We find a range of backsplash solutions for Cetus where the gas ram pressure of the Milky Way could have been large enough to strip its gas, 
providing an explanation for its current absence of cold gas.
\chn{Considering the new proper motion estimates for the gas-less dwarf Eridanus II \citep{McConnachie2020}, 
we also find ram pressure stripped backsplash orbits that could explain its lack of cold gas.}
For Phoenix I we find neither ram pressure stripped nor backsplash solutions. This is due to the large line-of-sight velocity, suggesting that this dwarf is a pristine dwarf on its first in-fall.
Finally, we also use our cosmological orbital calculation for these dwarfs, where we also find a time delay of the first pericenter passage in the backsplash solutions of $\Delta t\approx2\Gyr$ for Eridanus II and of $\Delta t \approx5\Gyr$ for Cetus.}
\end{enumerate}

We conclude that, \chn{while the \HI rich dwarf Leo T could be a backsplash system with some mild interaction with the Milky Way in the past, 
Phoenix I is likely a pristine system, and therefore, both systems are interesting laboratories to test formation and evolution models of dwarf galaxies in isolation or interaction regimes}.
Certainly, cosmological galaxy simulations predict that 30 to 50 per cent of the satellites 
between one and twice the virial radius of host galaxies like the Milky Way or Andromeda, 
are a backsplash population, while the remaining are first in-fall dwarfs.
This raises up great expectations that, with the upcoming telescopes, such as the Vera C. Rubin Observatory \footnote{https://www.lsst.org}, 
we will discover many more distant dwarf satellites, including backsplash population members.



\section*{Acknowledgements}
\label{sec:ack}
We thank the referee for the thoughtful review and comments which have greatly improved the manuscript.
We also thank E. Adams, T. Oosterloo and their team for making the \HI data of Leo T publicly available.
MB would like to deeply thank the enlightening comments shared by Aaron Dutton, Ariel S\'anchez, Carolina Agurto, Cesar Mu\~noz, Christopher Wegg, Claudio Leopoldo Bla\~na de la Cruz, Eva Grebel, Go Ogyia, Klaus Dolag, Manuel Behrendt, Marcel Lotz, Rhea Remus, Tobias Buck, among others.
MB also acknowledges ANID (CONICYT) for the awarded postdoctoral fellowship BECA DE POSTDOCTORADO EN EL EXTRANJERO CONVOCATORIA 2018 folio 74190011 resoluci\'on exenta N$^{\rm o}$ 8772/2018.
This research was supported by the Excellence Cluster ORIGINS which is founded by the Deutsche Forschungsgemeinschaft (DFG, German Research Foundation) under Germany's Excellence Strategy – EXC-2094 – 390783311.
MF acknowledges funding through FONDECYT regular N$^{\rm o}$ 1180291, BASAL N$^{\rm o}$ AFB-170002 (CATA), CONICYT PII20150171 and QUIMAL 17001.

\subsection*{ORCID iDs}
\label{sec:orcid}
Mat\'ias Bla\~{n}a \orcidicon{0000-0003-2139-0944}\,\footnotesize{\url{https://orcid.org/0000-0003-2139-0944}}\\
Andreas Burkert\,\orcidicon{0000-0001-6879-9822}\,\footnotesize{\url{https://orcid.org/0000-0001-6879-9822}}\\
Michael Fellhauer\,\orcidicon{0000-0002-3989-4115}\,\footnotesize{{\url{https://orcid.org/0000-0002-3989-4115}}\\
Marc Schartmann\,\orcidicon{0000-0003-1318-8631}\,\footnotesize{\url{https://orcid.org/0000-0003-1318-8631}}

\subsection*{Data availability}
\label{sec:data}
Data available on request \footnote{\href{http://matiasblana.github.io}{http://matiasblana.github.io}}.


\bibliographystyle{mnras}
\bibliography{paper_LeoT_v10} 

\begin{thebibliography}{}
\makeatletter
\relax
\def\mn@urlcharsother{\let\do\@makeother \do\$\do\&\do\#\do\^\do\_\do\%\do\~}
\def\mn@doi{\begingroup\mn@urlcharsother \@ifnextchar [ {\mn@doi@}
  {\mn@doi@[]}}
\def\mn@doi@[#1]#2{\def\@tempa{#1}\ifx\@tempa\@empty \href
  {http://dx.doi.org/#2} {doi:#2}\else \href {http://dx.doi.org/#2} {#1}\fi
  \endgroup}
\def\mn@eprint#1#2{\mn@eprint@#1:#2::\@nil}
\def\mn@eprint@arXiv#1{\href {http://arxiv.org/abs/#1} {{\tt arXiv:#1}}}
\def\mn@eprint@dblp#1{\href {http://dblp.uni-trier.de/rec/bibtex/#1.xml}
  {dblp:#1}}
\def\mn@eprint@#1:#2:#3:#4\@nil{\def\@tempa {#1}\def\@tempb {#2}\def\@tempc
  {#3}\ifx \@tempc \@empty \let \@tempc \@tempb \let \@tempb \@tempa \fi \ifx
  \@tempb \@empty \def\@tempb {arXiv}\fi \@ifundefined
  {mn@eprint@\@tempb}{\@tempb:\@tempc}{\expandafter \expandafter \csname
  mn@eprint@\@tempb\endcsname \expandafter{\@tempc}}}

\bibitem[\protect\citeauthoryear{Adams \& Oosterloo}{Adams \&
  Oosterloo}{2018}]{Adams2018}
Adams E. A.~K.,  Oosterloo T.~A.,  2018, \mn@doi [A{\&}A]
  {10.1051/0004-6361/201732017}, 612, 12

\bibitem[\protect\citeauthoryear{Battaglia et~al.,}{Battaglia
  et~al.}{2012}]{Battaglia2012}
Battaglia G.,  et~al., 2012, \mn@doi [MNRAS]
  {10.1111/j.1365-2966.2012.21286.x}, 424, 1113

\bibitem[\protect\citeauthoryear{Begum, Chengalur, Karachentsev, Sharina  \&
  Kaisin}{Begum et~al.}{2008}]{Begum2008}
Begum A.,  Chengalur J.~N.,  Karachentsev I.~D.,  Sharina M.~E.,   Kaisin
  S.~S.,  2008, \mn@doi [MNRAS] {10.1111/j.1365-2966.2008.13150.x}, 386, 1667

\bibitem[\protect\citeauthoryear{Belokurov, Erkal, Deason, Koposov, {De
  Angeli}, Evans, Fraternali  \& Mackey}{Belokurov
  et~al.}{2017}]{Belokurov2017}
Belokurov V.,  Erkal D.,  Deason A.~J.,  Koposov S.~E.,  {De Angeli} F.,  Evans
  D.~W.,  Fraternali F.,   Mackey D.,  2017, \mn@doi [MNRAS]
  {10.1093/mnras/stw3357}, 466, 4711

\bibitem[\protect\citeauthoryear{Belokurov et~al.,}{Belokurov
  et~al.}{2018}]{Belokurov2018}
Belokurov V.,  et~al., 2018, \mn@doi [MNRAS] {10.1093/mnras/sty982}, 478, 611

\bibitem[\protect\citeauthoryear{Binney \& Tremaine}{Binney \&
  Tremaine}{2008}]{Binney2008}
Binney J.,  Tremaine S.,  2008, \mn@doi [Phys. Today] {10.1063/1.3141945}, 62,
  56

\bibitem[\protect\citeauthoryear{Blana, Fellhauer, Smith, Candlish, Cohen  \&
  Farias}{Blana et~al.}{2015}]{Blana2015}
Blana M.,  Fellhauer M.,  Smith R.,  Candlish G.~N.,  Cohen R.,   Farias J.~P.,
   2015, \mn@doi [MNRAS] {10.1093/mnras/stu1797}, 446, 144

\bibitem[\protect\citeauthoryear{Bland-Hawthorn \& Gerhard}{Bland-Hawthorn \&
  Gerhard}{2016}]{Bland-Hawthorn2016}
Bland-Hawthorn J.,  Gerhard O.,  2016, \mn@doi [Annu. Rev. Astron. Astrophys.]
  {10.1146/annurev-astro-081915-023441}, 54, 529

\bibitem[\protect\citeauthoryear{Blitz \& Robishaw}{Blitz \&
  Robishaw}{2000}]{Blitz2000}
Blitz L.,  Robishaw T.,  2000, \mn@doi [ApJ] {10.1086/309457}, 541, 675

\bibitem[\protect\citeauthoryear{Bond, Cole, Efstathiou  \& Kaiser}{Bond
  et~al.}{1991}]{Bond1991}
Bond J.~R.,  Cole S.,  Efstathiou G.,   Kaiser N.,  1991, \mn@doi [ApJ]
  {10.1086/170520}, 379, 440

\bibitem[\protect\citeauthoryear{Bovy}{Bovy}{2015}]{Bovy2015}
Bovy J.,  2015, \mn@doi [ApJSS.] {10.1088/0067-0049/216/2/29}, 216, 29

\bibitem[\protect\citeauthoryear{Bovy \& Rix}{Bovy \& Rix}{2013}]{Bovy2013}
Bovy J.,  Rix H.-W.,  2013, \mn@doi [ApJ] {10.1088/0004-637X/779/2/115}, 779,
  30

\bibitem[\protect\citeauthoryear{Broeils \& Rhee}{Broeils \&
  Rhee}{1997}]{Broeils1997}
Broeils A.~H.,  Rhee M.~H.,  1997, A{\&}A, 324, 877

\bibitem[\protect\citeauthoryear{Buck, Macci{\`{o}}, Dutton, Obreja  \&
  Frings}{Buck et~al.}{2019}]{Buck2019}
Buck T.,  Macci{\`{o}} A.~V.,  Dutton A.~A.,  Obreja A.,   Frings J.,  2019,
  \mn@doi [MNRAS] {10.1093/mnras/sty2913}, 483, 1314

\bibitem[\protect\citeauthoryear{Burkert}{Burkert}{1995}]{Burkert1995}
Burkert A.,  1995, \mn@doi [ApJL] {10.1086/309560}, 447, L25

\bibitem[\protect\citeauthoryear{Burkert}{Burkert}{2015}]{Burkert2015}
Burkert A.,  2015, \mn@doi [ApJ] {10.1088/0004-637x/808/2/158}, 808, 10

\bibitem[\protect\citeauthoryear{Clementini, Cignoni, Ramos, Federici, Ripepi,
  Marconi, Tosi  \& Musella}{Clementini et~al.}{2012}]{Clementini2012}
Clementini G.,  Cignoni M.,  Ramos R.~C.,  Federici L.,  Ripepi V.,  Marconi
  M.,  Tosi M.,   Musella I.,  2012, \mn@doi [ApJ]
  {10.1088/0004-637X/756/2/108}, 756, 108

\bibitem[\protect\citeauthoryear{Correa, Wyithe, Schaye  \& Duffy}{Correa
  et~al.}{2015a}]{Correa2015a}
Correa C.~A.,  Wyithe J. S.~B.,  Schaye J.,   Duffy A.~R.,  2015a, \mn@doi
  [MNRAS] {10.1093/mnras/stv689}, 450, 1514

\bibitem[\protect\citeauthoryear{Correa, Wyithe, Schaye  \& Duffy}{Correa
  et~al.}{2015b}]{Correa2015b}
Correa C.~A.,  Wyithe J. S.~B.,  Schaye J.,   Duffy A.~R.,  2015b, \mn@doi
  [MNRAS] {10.1093/mnras/stv697}, 450, 1521

\bibitem[\protect\citeauthoryear{Correa, Stuart, Wyithe, Schaye  \&
  Duffy}{Correa et~al.}{2015c}]{Correa2015c}
Correa C.~A.,  Stuart J.,  Wyithe B.,  Schaye J.,   Duffy A.~R.,  2015c,
  \mn@doi [MNRAS] {10.1093/mnras/stv1363}, 452, 1217

\bibitem[\protect\citeauthoryear{Courteau \& van~den Bergh}{Courteau \& van~den
  Bergh}{1999}]{Courteau1999}
Courteau S.,  van~den Bergh S.,  1999, \mn@doi [Astron. Journal, Vol. 118,
  Issue 1, pp. 337-345.] {10.1086/300942}, 118, 337

\bibitem[\protect\citeauthoryear{Crnojevi{\'{c}}, Sand, Zaritsky, Spekkens,
  Willman  \& Hargis}{Crnojevi{\'{c}} et~al.}{2016}]{Crnojevic2016}
Crnojevi{\'{c}} D.,  Sand D.~J.,  Zaritsky D.,  Spekkens K.,  Willman B.,
  Hargis J.~R.,  2016, \mn@doi [ApJL] {10.3847/2041-8205/824/1/L14}, 824, 6

\bibitem[\protect\citeauthoryear{Deason, Belokurov, Koposov  \&
  Lancaster}{Deason et~al.}{2018}]{Deason2018}
Deason A.~J.,  Belokurov V.,  Koposov S.~E.,   Lancaster L.,  2018, \mn@doi
  [ApJL] {10.3847/2041-8213/aad0ee}, 862, L1

\bibitem[\protect\citeauthoryear{Deason, Fattahi, Frenk, Grand, Oman,
  Garrison-Kimmel, Simpson  \& Navarro}{Deason et~al.}{2020}]{Deason2020}
Deason A.~J.,  Fattahi A.,  Frenk C.~S.,  Grand R. J.~J.,  Oman K.~A.,
  Garrison-Kimmel S.,  Simpson C.~M.,   Navarro J.~F.,  2020

\bibitem[\protect\citeauthoryear{Dehnen}{Dehnen}{2000}]{Dehnen2000}
Dehnen W.,  2000, \mn@doi [ApJSS.] {10.1086/312724}, 536, L39

\bibitem[\protect\citeauthoryear{Dominguez, Fellhauer, Bla{\~{n}}a, Farias,
  Dabringhausen, Candlish, Smith  \& Choque}{Dominguez
  et~al.}{2016}]{Dominguez2016}
Dominguez R.,  Fellhauer M.,  Bla{\~{n}}a M.,  Farias J.-P.,  Dabringhausen J.,
   Candlish G.~N.,  Smith R.,   Choque N.,  2016, \mn@doi [MNRAS]
  {10.1093/mnras/stw1559}, 461, 3630

\bibitem[\protect\citeauthoryear{Faerman, Sternberg  \& McKee}{Faerman
  et~al.}{2013}]{Faerman2013}
Faerman Y.,  Sternberg A.,   McKee C.~F.,  2013, \mn@doi [ApJ]
  {10.1088/0004-637X/777/2/119}, 777, 14

\bibitem[\protect\citeauthoryear{Fellhauer et~al.,}{Fellhauer
  et~al.}{2006}]{Fellhauer2006}
Fellhauer M.,  et~al., 2006, \mn@doi [ApJ] {10.1086/507128}, 651, 167

\bibitem[\protect\citeauthoryear{Fellhauer, Wilkinson, Evans, Belokurov, Irwin,
  Gilmore, Zucker  \& Kleyna}{Fellhauer et~al.}{2008}]{Fellhauer2008a}
Fellhauer M.,  Wilkinson M.~I.,  Evans N.~W.,  Belokurov V.,  Irwin M.~J.,
  Gilmore G.,  Zucker D.~B.,   Kleyna J.~T.,  2008, \mn@doi [MNRAS]
  {10.1111/j.1365-2966.2008.12921.x}, 385, 1095

\bibitem[\protect\citeauthoryear{Fritz, Battaglia, Pawlowski, Kallivayalil,
  van~der Marel, Sohn, Brook  \& Besla}{Fritz et~al.}{2018}]{Fritz2018}
Fritz T.~K.,  Battaglia G.,  Pawlowski M.~S.,  Kallivayalil N.,  van~der Marel
  R.,  Sohn T.~S.,  Brook C.,   Besla G.,  2018, \mn@doi [A{\&}A]
  {10.1051/0004-6361/201833343}, 619, A103

\bibitem[\protect\citeauthoryear{Garrison-Kimmel et~al.,}{Garrison-Kimmel
  et~al.}{2017}]{Garrison-Kimmel2017}
Garrison-Kimmel S.,  et~al., 2017, \mn@doi [MNRAS] {10.1093/mnras/stx1710},
  471, 1709

\bibitem[\protect\citeauthoryear{Gatto, Fraternali, Read, Marinacci, Lux  \&
  Walch}{Gatto et~al.}{2013}]{Gatto2013}
Gatto A.,  Fraternali F.,  Read J.~I.,  Marinacci F.,  Lux H.,   Walch S.,
  2013, \mn@doi [MNRAS] {10.1093/mnras/stt896}, 433, 2749

\bibitem[\protect\citeauthoryear{Gill, Knebe, Gibson, Gill, Knebe  \&
  Gibson}{Gill et~al.}{2005}]{Gill2005}
Gill S. P.~D.,  Knebe A.,  Gibson B.~K.,  Gill S. P.~D.,  Knebe A.,   Gibson
  B.~K.,  2005, \mn@doi [MNRAS] {10.1111/j.1365-2966.2004.08562.x}, 356, 1327

\bibitem[\protect\citeauthoryear{Grcevich \& Putman}{Grcevich \&
  Putman}{2009}]{Grcevich2009}
Grcevich J.,  Putman M.~E.,  2009, \mn@doi [ApJ] {10.1088/0004-637X/696/1/385
  10.1088/0004-637X/721/1/922}, 696, 385

\bibitem[\protect\citeauthoryear{Haggar, Gray, Pearce, Knebe, Cui, Mostoghiu
  \& Yepes}{Haggar et~al.}{2020}]{Haggar2020}
Haggar R.,  Gray M.~E.,  Pearce F.~R.,  Knebe A.,  Cui W.,  Mostoghiu R.,
  Yepes G.,  2020, MNRAS

\bibitem[\protect\citeauthoryear{Hausammann, Revaz  \& Jablonka}{Hausammann
  et~al.}{2019}]{Hausammann2019}
Hausammann L.,  Revaz Y.,   Jablonka P.,  2019, A{\&}A

\bibitem[\protect\citeauthoryear{Haywood, {Di Matteo}, Lehnert, Snaith,
  Khoperskov  \& G{\'{o}}mez}{Haywood et~al.}{2018}]{Haywood2018}
Haywood M.,  {Di Matteo} P.,  Lehnert M.~D.,  Snaith O.,  Khoperskov S.,
  G{\'{o}}mez A.,  2018, \mn@doi [ApJ] {10.3847/1538-4357/aad235}, 863, 113

\bibitem[\protect\citeauthoryear{Helmi et~al.,}{Helmi et~al.}{2018}]{Helmi2018}
Helmi A.,  et~al., 2018, \mn@doi [Natur] {10.1038/s41586-018-0625-x}, 563, 85

\bibitem[\protect\citeauthoryear{Irwin et~al.,}{Irwin et~al.}{2007}]{Irwin2007}
Irwin M.~J.,  et~al., 2007, \mn@doi [ApJ] {10.1086/512183}, 656, L13

\bibitem[\protect\citeauthoryear{Jedrzejewski}{Jedrzejewski}{1987}]{Jedrzejewski1987}
Jedrzejewski R.~I.,  1987, \mn@doi [MNRAS] {10.1093/mnras/226.4.747}, 226, 747

\bibitem[\protect\citeauthoryear{Kacharov et~al.,}{Kacharov
  et~al.}{2017}]{Kacharov2017}
Kacharov N.,  et~al., 2017, \mn@doi [MNRAS] {10.1093/mnras/stw3188}, 466, 2006

\bibitem[\protect\citeauthoryear{Kalberla \& Kerp}{Kalberla \&
  Kerp}{2009}]{Kalberla2009}
Kalberla P.~M.,  Kerp J.,  2009, \mn@doi [Annu. Rev. Astron. Astrophys.]
  {10.1146/annurev-astro-082708-101823}, 47, 27

\bibitem[\protect\citeauthoryear{Kam, Carignan, Chemin, Foster, Elson  \&
  Jarrett}{Kam et~al.}{2017}]{Kam2017}
Kam S.~Z.,  Carignan C.,  Chemin L.,  Foster T.,  Elson E.,   Jarrett T.~H.,
  2017, \mn@doi [AJ] {10.3847/1538-3881/aa79f3}, 154, 41

\bibitem[\protect\citeauthoryear{Karachentsev, Kashibadze, Makarov  \&
  Tully}{Karachentsev et~al.}{2009}]{Karachentsev2009}
Karachentsev I.~D.,  Kashibadze O.~G.,  Makarov D.~I.,   Tully R.~B.,  2009,
  \mn@doi [MNRAS] {10.1111/j.1365-2966.2008.14300.x}, 393, 1265

\bibitem[\protect\citeauthoryear{Li \& White}{Li \& White}{2008}]{Li2008}
Li Y.-S.,  White S. D.~M.,  2008, \mn@doi [MNRAS]
  {10.1111/j.1365-2966.2007.12748.x}, 384, 1459

\bibitem[\protect\citeauthoryear{Li et~al.,}{Li et~al.}{2017}]{Li2016a}
Li T.~S.,  et~al., 2017, \mn@doi [ApJ] {10.3847/1538-4357/aa6113}, 838, 15

\bibitem[\protect\citeauthoryear{Lotz et~al.,}{Lotz et~al.}{2019}]{Lotz2019}
Lotz M.,  et~al., 2019, \mn@doi [MNRAS] {10.1093/mnras/stz2070}, 488, 5370

\bibitem[\protect\citeauthoryear{{Matus Carillo}, Fellhauer, {Alarcon Jara},
  Aravena  \& {Urrutia Zapata}}{{Matus Carillo} et~al.}{2019}]{Matus2019}
{Matus Carillo} D.~R.,  Fellhauer M.,  {Alarcon Jara} A.~G.,  Aravena C.~A.,
  {Urrutia Zapata} F.,  2019, A{\&}A

\bibitem[\protect\citeauthoryear{McConnachie}{McConnachie}{2012}]{McConnachie2012}
McConnachie A.~W.,  2012, \mn@doi [AJ] {10.1088/0004-6256/144/1/4}, 144, 36

\bibitem[\protect\citeauthoryear{McConnachie \& Irwin}{McConnachie \&
  Irwin}{2006}]{McConnachie2006}
McConnachie A.,  Irwin M.,  2006, \mn@doi [MNRAS]
  {10.1111/j.1365-2966.2005.09806.x}, 365, 1263

\bibitem[\protect\citeauthoryear{McConnachie \& Venn}{McConnachie \&
  Venn}{2020}]{McConnachie2020}
McConnachie A.~W.,  Venn K.~A.,  2020

\bibitem[\protect\citeauthoryear{McConnachie, Irwin, Ferguson, Ibata, Lewis  \&
  Tanvir}{McConnachie et~al.}{2005}]{McConnachie2005}
McConnachie A.~W.,  Irwin M.~J.,  Ferguson A. M.~N.,  Ibata R.~A.,  Lewis
  G.~F.,   Tanvir N.,  2005, \mn@doi [MNRAS]
  {10.1111/j.1365-2966.2004.08514.x}, 356, 979

\bibitem[\protect\citeauthoryear{Miller \& Bregman}{Miller \&
  Bregman}{2015}]{Miller2015}
Miller M.~J.,  Bregman J.~N.,  2015, \mn@doi [ApJ]
  {10.1088/0004-637X/800/1/14}, 800, 19

\bibitem[\protect\citeauthoryear{Mo, van~den Bosch  \& White}{Mo
  et~al.}{2010}]{MoBoWh2010}
Mo H.,  van~den Bosch F.~C.,   White S.,  2010, Galaxy Form. Evol. by Houjun Mo
  , Frank van den Bosch , Simon White, Cambridge, UK Cambridge Univ. Press.
  2010

\bibitem[\protect\citeauthoryear{Mori \& Burkert}{Mori \&
  Burkert}{2000}]{Mori2000}
Mori M.,  Burkert A.,  2000, \mn@doi [ApJ] {10.1086/309140}, 538, 559

\bibitem[\protect\citeauthoryear{Niederste-Ostholt, Belokurov, Evans,
  Penarrubia, Niederste-Ostholt, Belokurov, Evans  \&
  Pe{\~{n}}arrubia}{Niederste-Ostholt et~al.}{2010}]{Niederste-Ostholt2010}
Niederste-Ostholt M.,  Belokurov V.,  Evans N.~W.,  Penarrubia J.,
  Niederste-Ostholt M.,  Belokurov V.,  Evans N.~W.,   Pe{\~{n}}arrubia J.,
  2010, \mn@doi [ApJ] {10.1088/0004-637X/712/1/516}, 712, 516

\bibitem[\protect\citeauthoryear{Nulsen}{Nulsen}{1982}]{Nulsen1982}
Nulsen P. E.~J.,  1982, \mn@doi [MNRAS] {10.1093/mnras/198.4.1007}, 198, 1007

\bibitem[\protect\citeauthoryear{Ogiya, Mori, Ishiyama  \& Burkert}{Ogiya
  et~al.}{2014}]{Ogiya2014a}
Ogiya G.,  Mori M.,  Ishiyama T.,   Burkert A.,  2014, \mn@doi [MNRAS Lett.]
  {10.1093/mnrasl/slu023}, 440, L71

\bibitem[\protect\citeauthoryear{Patra}{Patra}{2018}]{Patra2018}
Patra N.~N.,  2018, MNRAS, 480, 4369

\bibitem[\protect\citeauthoryear{Penarrubia, Navarro  \&
  McConnachie}{Penarrubia et~al.}{2008}]{Penarrubia2008}
Penarrubia J.,  Navarro J.~F.,   McConnachie A.~W.,  2008, \mn@doi [ApJ]
  {10.1086/523686}, 673, 226

\bibitem[\protect\citeauthoryear{Pe{\~{n}}arrubia, Ma, Walker  \&
  McConnachie}{Pe{\~{n}}arrubia et~al.}{2014}]{Penarrubia2014}
Pe{\~{n}}arrubia J.,  Ma Y.-Z.,  Walker M.~G.,   McConnachie A.,  2014, \mn@doi
  [MNRAS] {10.1093/mnras/stu879}, 443, 2204

\bibitem[\protect\citeauthoryear{Perret}{Perret}{2016}]{Perret2016}
Perret V.,  2016, Astrophys. Source Code Libr. Rec. ascl1607.002

\bibitem[\protect\citeauthoryear{Perret, Renaud, Epinat, Amram, Bournaud,
  Contini, Teyssier  \& Lambert}{Perret et~al.}{2014}]{Perret2014}
Perret V.,  Renaud F.,  Epinat B.,  Amram P.,  Bournaud F.,  Contini T.,
  Teyssier R.,   Lambert J.~C.,  2014, \mn@doi [A{\&}A]
  {10.1051/0004-6361/201322395}, 562, 39

\bibitem[\protect\citeauthoryear{{Planck Collaboration} et~al.,}{{Planck
  Collaboration} et~al.}{2015}]{PlanckCollaboration2015}
{Planck Collaboration} P.,  et~al., 2015, \mn@doi [A{\&}A]
  {10.1051/0004-6361/201525830}, 594, 63

\bibitem[\protect\citeauthoryear{Press \& Schechter}{Press \&
  Schechter}{1974}]{Press1974}
Press W.~H.,  Schechter P.,  1974, \mn@doi [ApJ] {10.1086/152650}, 187, 425

\bibitem[\protect\citeauthoryear{Quinn, Katz, Stadel  \& Lake}{Quinn
  et~al.}{1997}]{Quinn1997}
Quinn T.,  Katz N.,  Stadel J.,   Lake G.,  1997

\bibitem[\protect\citeauthoryear{Read, Agertz  \& Collins}{Read
  et~al.}{2016}]{Read2016}
Read J.~I.,  Agertz O.,   Collins M. L.~M.,  2016, \mn@doi [MNRAS]
  {10.1093/mnras/stw713}, 459, 2573

\bibitem[\protect\citeauthoryear{{Rodriguez Wimberly}, Cooper, Fillingham,
  Boylan-Kolchin, Bullock  \& Garrison-Kimmel}{{Rodriguez Wimberly}
  et~al.}{2018}]{Rodriguez2018}
{Rodriguez Wimberly} M.~K.,  Cooper M.~C.,  Fillingham S.~P.,  Boylan-Kolchin
  M.,  Bullock J.~S.,   Garrison-Kimmel S.,  2018, eprint arXiv:1806.07891

\bibitem[\protect\citeauthoryear{Ruiz-Lara, Gallart, Bernard  \&
  Cassisi}{Ruiz-Lara et~al.}{2020}]{Ruiz-Lara2020}
Ruiz-Lara T.,  Gallart C.,  Bernard E.~J.,   Cassisi S.,  2020, \mn@doi [NatAs]
  {10.1038/S41550-020-1097-0}

\bibitem[\protect\citeauthoryear{Ryan-Weber, Begum, Oosterloo, Pal, Irwin,
  Belokurov, Evans  \& Zucker}{Ryan-Weber et~al.}{2008}]{RyanWeber2008}
Ryan-Weber E.~V.,  Begum A.,  Oosterloo T.,  Pal S.,  Irwin M.~J.,  Belokurov
  V.,  Evans N.~W.,   Zucker D.~B.,  2008, \mn@doi [MNRAS]
  {10.1111/j.1365-2966.2007.12734.x}, 384, 535

\bibitem[\protect\citeauthoryear{Salem, Besla, Bryan, Putman, van~der Marel  \&
  Tonnesen}{Salem et~al.}{2015}]{Salem2015}
Salem M.,  Besla G.,  Bryan G.,  Putman M.,  van~der Marel R.~P.,   Tonnesen
  S.,  2015, \mn@doi [ApJ] {10.1088/0004-637X/815/1/77}, 815, 77

\bibitem[\protect\citeauthoryear{Sawala, Scannapieco  \& White}{Sawala
  et~al.}{2012}]{Sawala2012}
Sawala T.,  Scannapieco C.,   White S.,  2012, \mn@doi [MNRAS]
  {10.1111/j.1365-2966.2011.20181.x}, 420, 1714

\bibitem[\protect\citeauthoryear{Simon \& Geha}{Simon \&
  Geha}{2007}]{Simon2007}
Simon J.~D.,  Geha M.,  2007, \mn@doi [ApJ] {10.1086/521816}, 670, 313

\bibitem[\protect\citeauthoryear{Simpson, Grand, G{\'{o}}mez, Marinacci,
  Pakmor, Springel, Campbell  \& Frenk}{Simpson et~al.}{2018}]{Simpson2018}
Simpson C.~M.,  Grand R. J.~J.,  G{\'{o}}mez F.~A.,  Marinacci F.,  Pakmor R.,
  Springel V.,  Campbell D. J.~R.,   Frenk C.~S.,  2018, \mn@doi [MNRAS]
  {10.1093/mnras/sty774}, 478, 548

\bibitem[\protect\citeauthoryear{Spekkens, Urbancic, Mason, Willman  \&
  Aguirre}{Spekkens et~al.}{2014}]{Spekkens2014}
Spekkens K.,  Urbancic N.,  Mason B.~S.,  Willman B.,   Aguirre J.~E.,  2014,
  \mn@doi [ApJ] {10.1088/2041-8205/795/1/L5}, 795, L5

\bibitem[\protect\citeauthoryear{Springel, {Di Matteo}  \& Hernquist}{Springel
  et~al.}{2005}]{Springel2005}
Springel V.,  {Di Matteo} T.,   Hernquist L.,  2005, \mn@doi [MNRAS]
  {10.1111/j.1365-2966.2005.09238.x}, 361, 776

\bibitem[\protect\citeauthoryear{St-Germain, Carignan, Coˆte  \&
  Oosterloo}{St-Germain et~al.}{1999}]{StGermain1999}
St-Germain J.,  Carignan C.,  Coˆte S.,   Oosterloo T.,  1999, \mn@doi [AJ]
  {10.1086/301021}, 118, 1235

\bibitem[\protect\citeauthoryear{Stevens, Diemer, Lagos, Nelson, Obreschkow,
  Wang  \& Marinacci}{Stevens et~al.}{2019}]{Stevens2019}
Stevens A. R.~H.,  Diemer B.,  Lagos C. d.~P.,  Nelson D.,  Obreschkow D.,
  Wang J.,   Marinacci F.,  2019, \mn@doi [MNRAS] {10.1093/mnras/stz2513}, 490,
  96

\bibitem[\protect\citeauthoryear{Taibi et~al.,}{Taibi et~al.}{2018}]{Taibi2018}
Taibi S.,  et~al., 2018, \mn@doi [A{\&}A] {10.1051/0004-6361/201833414}, 618,
  22

\bibitem[\protect\citeauthoryear{Tamm, Tempel, Tenjes, Tihhonova  \&
  Tuvikene}{Tamm et~al.}{2012}]{Tamm2012}
Tamm a.,  Tempel E.,  Tenjes P.,  Tihhonova O.,   Tuvikene T.,  2012, \mn@doi
  [A{\&}A] {10.1051/0004-6361/201220065}, 546, A4

\bibitem[\protect\citeauthoryear{Teuben}{Teuben}{1995}]{Teuben1995}
Teuben P.,  1995, Astron. Data Anal. Softw. Syst. IV, 77

\bibitem[\protect\citeauthoryear{Teyssier, Johnston  \& Kuhlen}{Teyssier
  et~al.}{2012}]{TeyssierM2012}
Teyssier M.,  Johnston K.~V.,   Kuhlen M.,  2012, \mn@doi [MNRAS]
  {10.1111/j.1365-2966.2012.21793.x}, 426, 1808

\bibitem[\protect\citeauthoryear{{The Astropy Coll.} et~al.,}{{The Astropy
  Coll.} et~al.}{2013}]{TheAstropyCollaboration2013}
{The Astropy Coll.} et~al., 2013, \mn@doi [A{\&}A]
  {10.1051/0004-6361/201322068}, 558, 9

\bibitem[\protect\citeauthoryear{{The Astropy Coll.} et~al.,}{{The Astropy
  Coll.} et~al.}{2018}]{TheAstropyCollaboration2018}
{The Astropy Coll.} et~al., 2018, \mn@doi [AJ] {10.3847/1538-3881/aabc4f}, 156,
  19

\bibitem[\protect\citeauthoryear{Tonnesen}{Tonnesen}{2019}]{Tonnesen2019}
Tonnesen S.,  2019, \mn@doi [ApJ] {10.3847/1538-4357/ab0960}, 874, 161

\bibitem[\protect\citeauthoryear{Wang, Koribalski, Serra, van~der Hulst,
  Roychowdhury, Kamphuis  \& Chengalur}{Wang et~al.}{2016}]{WangJ2016}
Wang J.,  Koribalski B.~S.,  Serra P.,  van~der Hulst T.,  Roychowdhury S.,
  Kamphuis P.,   Chengalur J.~N.,  2016, \mn@doi [MNRAS]
  {10.1093/mnras/stw1099}, 460, 2143

\bibitem[\protect\citeauthoryear{Watkins, Evans  \& An}{Watkins
  et~al.}{2010}]{Watkins2010}
Watkins L.~L.,  Evans N.~W.,   An J.~H.,  2010, \mn@doi [MNRAS]
  {10.1111/j.1365-2966.2010.16708.x}, 406, 264

\bibitem[\protect\citeauthoryear{Weisz et~al.,}{Weisz et~al.}{2012}]{Weisz2012}
Weisz D.~R.,  et~al., 2012, \mn@doi [ApJ] {10.1088/0004-637X/748/2/88}, 748, 6

\bibitem[\protect\citeauthoryear{Wolf, Martinez, Bullock, Kaplinghat, Geha,
  Munoz, Simon  \& Avedo}{Wolf et~al.}{2009}]{Wolf2010}
Wolf J.,  Martinez G.~D.,  Bullock J.~S.,  Kaplinghat M.,  Geha M.,  Munoz
  R.~R.,  Simon J.~D.,   Avedo F.~F.,  2009, \mn@doi [MNRAS]
  {10.1111/j.1365-2966.2010.16753.x}, 406, 1220

\bibitem[\protect\citeauthoryear{Young, Skillman, Weisz, Dolphin, Young,
  Skillman, Weisz  \& Dolphin}{Young et~al.}{2007}]{Young2007}
Young L.~M.,  Skillman E.~D.,  Weisz D.~R.,  Dolphin A.~E.,  Young L.~M.,
  Skillman E.~D.,  Weisz D.~R.,   Dolphin A.~E.,  2007, \mn@doi [ApJ]
  {10.1086/512153}, 659, 331

\bibitem[\protect\citeauthoryear{de Jong et~al.,}{de~Jong
  et~al.}{2008}]{DeJong2008}
de Jong J. T.~A.,  et~al., 2008, \mn@doi [ApJ] {10.1086/587835}, 680, 1112

\bibitem[\protect\citeauthoryear{van~den Bosch \& Ogiya}{van~den Bosch \&
  Ogiya}{2018}]{Bosch2018b}
van~den Bosch F.~C.,  Ogiya G.,  2018, \mn@doi [MNRAS] {10.1093/mnras/sty084},
  475, 4066

\bibitem[\protect\citeauthoryear{van~den Bosch, Ogiya, Hahn  \&
  Burkert}{van~den Bosch et~al.}{2017}]{Bosch2017}
van~den Bosch F.~C.,  Ogiya G.,  Hahn O.,   Burkert A.,  2017, \mn@doi [MNRAS]
  {10.1093/mnras/stx2956}, 474, 3043

\bibitem[\protect\citeauthoryear{van~den Bosch, Ogiya, Hahn  \&
  Burkert}{van~den Bosch et~al.}{2018}]{VandenBosch2018}
van~den Bosch F.~C.,  Ogiya G.,  Hahn O.,   Burkert A.,  2018, \mn@doi [MNRAS]
  {10.1093/mnras/stx2956}, 474, 3043

\bibitem[\protect\citeauthoryear{van~der Marel, Besla, Cox, Sohn  \&
  Anderson}{van~der Marel et~al.}{2012}]{VanderMarel2012}
van~der Marel R.~P.,  Besla G.,  Cox T.~J.,  Sohn S.~T.,   Anderson J.,  2012,
  \mn@doi [ApJ] {10.1088/0004-637X/753/1/9}, 753, 9

\bibitem[\protect\citeauthoryear{van~der Marel, Fardal, Sohn, Patel, Besla, del
  Pino-Molina, Sahlmann  \& Watkins}{van~der Marel
  et~al.}{2018}]{VanderMarel2019}
van~der Marel R.~P.,  Fardal M.~A.,  Sohn S.~T.,  Patel E.,  Besla G.,  del
  Pino-Molina A.,  Sahlmann J.,   Watkins L.~L.,  2018, \mn@doi [ApJ]
  {10.3847/1538-4357/ab001b}, 872, 24

\makeatother
\end{thebibliography}



%


\appendix
\section{Tools of \sc{delorean}}
\label{sec:ap:int}

\begin{figure}
\begin{center}
\includegraphics[width=8.3cm]{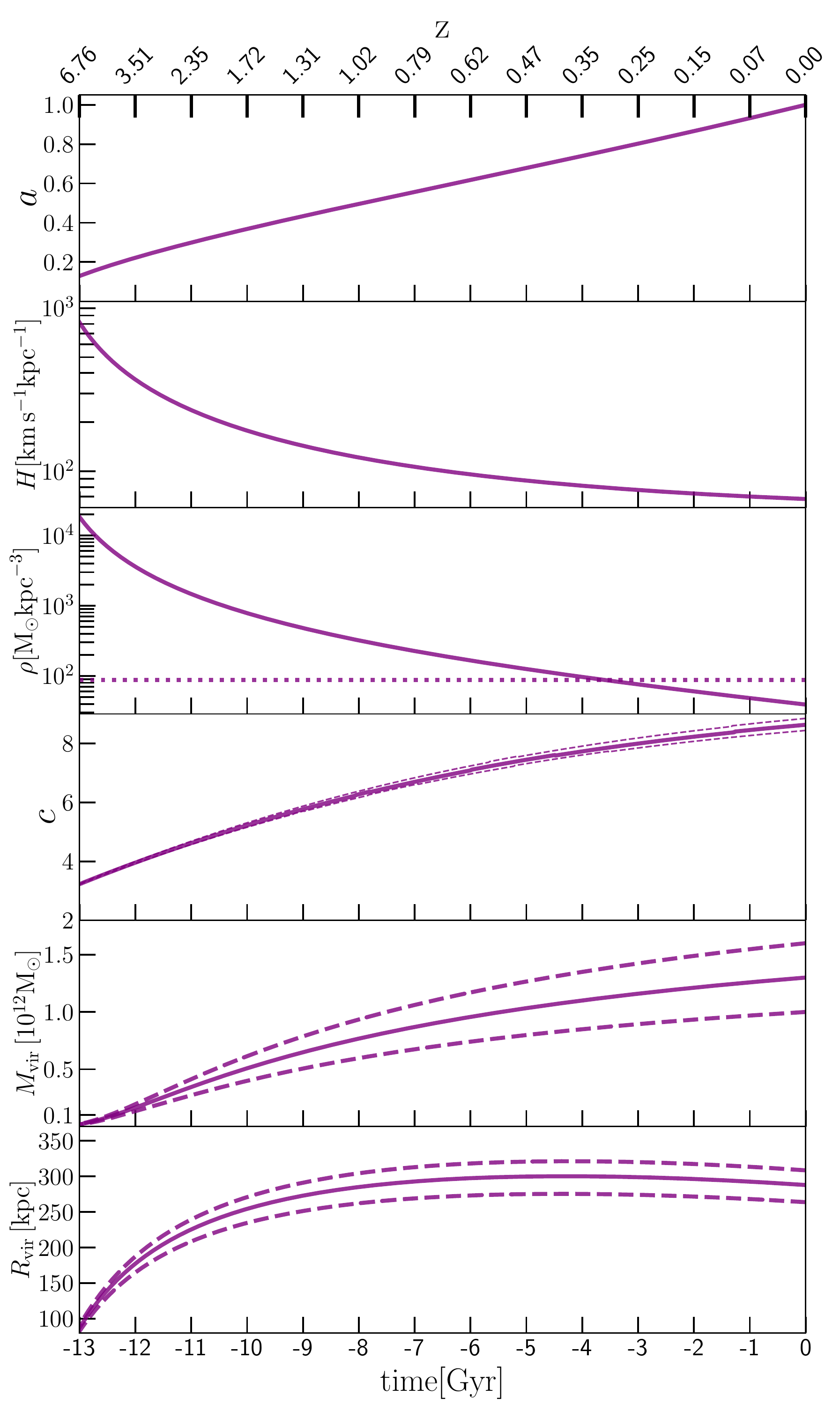}
\vspace{-0.5cm}
\caption{Top to third panel, are shown as function of lookback time (redshift): scale factor, expansion factor, matter density ($\rho_{\rm m}=\rho_{\rm crit}\Omega_{\rm m}$) (solid curve) and dark energy density (dotted line). 
The values are calculated from cosmological parameters according to the \citet{PlanckCollaboration2015}{ Paper XIII} taken from \textsc{astropy} (Planck15).
Third to bottom panels: halo concentration, virial mass and comoving virial radius (Eq. \ref{eq:rvir}) as function of redshift according to the software \textsc{commah} \citep{Correa2015a,Correa2015b,Correa2015c}, 
which come from semi-analytical extended Press-Schechter and halo mass accretion models fitted to cosmological simulations.
The values used at redshift zero for the concentration, the virial mass and radius are $8.6$, $1.3\times10^{12}\sm$, $288\kpc$ (solid curve),
and the upper and lower dashed curves correspond to $8.8$, $1.6\times10^{12}\sm$, $308\kpc$ 
and $8.4$, $1.0\times10^{12}\sm$ and $264\kpc$, respectively.}
\label{fig:cosmopar}
\end{center}
\end{figure}

\subsection{Proper motion exploration}
\label{sec:app:pm}
Our code \textsc{delorean} has the option to calculate orbits providing directly heliocentric proper motion values pre-defined by the user.
The code can also construct the vector $\utgsr$ which is tangential to the LOS velocity in the GSR frame at the current position of the object.
It starts by converting $\vlosgsr$ to GC coordinates $\vlosgc$.
Then, it builds two orthonormal vectors bases of $\utgsr$ in the GC frame, which are perpendicular to the line-of-sight velocities $\vlosgc$.
For this it uses the angular momentum from the known LOS velocity in the GC frame:
\begin{align}
\utgc&=\lambda_{1}\hat{l}_{\rm los}+\lambda_{2}\hat{l}_{\rm los}\times\unvlosgc,
\end{align}
where $\hat{l}_{\rm los}$ is the unitarian vector of $\vec{l}_{\rm los}=\vec{D}\times\vlosgc$, which is a component of the total specific orbital angular momentum:
$\vec{l}=\vec{D}\times(\vlosgc+\utgc)$, with $\vec{D}$ being the object's Galactocentric coordinate.
The lambda parameters define the magnitude of \utgsr: $\Nutgsr=\sqrt{\lambda_{1}^2+\lambda_{2}^2}$, and its direction is given by the angle $\theta\e\arctan(\lambda_{2}/\lambda_{1})$, which is measured on the plane of the sky, and can be converted to the position angle after conversion to equatorial coordinates ($\PA\e\theta-\theta_{\rm o}$).
Once a value of $\utgsr$ is defined, it can be converted into proper motions.

\subsection{Orbital integration with cosmic expansion}
\label{sec:app:oint}
\textsc{delorean} can also integrate orbits in a cosmological framework.
For this we pre-compute the factors in the drift $(D)$ and kick $(K)$ operators that depend on the cosmological parameters, using here the values from the \citet{PlanckCollaboration2015}{(Paper XIII)} from \textsc{astropy} (Planck15) to obtain the cosmological variables \z, $a(\z)$, $H(\z)$, $\rho_{\rm crit}(\z)$, $\rho_{\Lambda}(\z)$, $\rho_{\rm m}(\z)$ and $\Omega_{\rm m}(\z)$, corresponding to the redshift, scale factor, expansion factor, critical density, matter density, dark energy density, and matter to critical density ratio respectively, as functions of the lookback time $(t)$ \ie $\z=\z\left(t\right)$. We tabulate them to interpolate the values when needed (Fig.\ref{fig:cosmopar}). These tables can be recomputed for any cosmology.
To calculate the orbits we use the symplectic form (ergo time reversible) of the equations of motion in a Hamiltonian formalism given by \citet{Quinn1997} \citep[as applied in \textsc{gadget-2}][]{Springel2005}, which uses the leapfrog method, which is time-reversible, to calculate the orbits in comoving coordinates ($\vec{X},\vec{V}$) as:
\begin{align}
D\left(\Delta t\right)&=\vec{X}_{t+\Delta t}=\vec{X}_{t} + \vec{P}\int^{t+\Delta t}_{t}\frac{dt}{a^2}\\
K\left(\Delta t\right)&=\vec{P}_{t+\Delta t}=\vec{P}_{t} - \int^{t+\Delta t}_{t} dt \frac{\vec{\nabla} \psi}{a\left(t\right)} 
\label{eq:cosmo1}
\end{align}
where $\vec{P}=a^2\vec{V}$ is the specific canonical momentum and $\vec{\nabla}\psi$ the gradient of the peculiar potential in comoving coordinates for the periodic boundary solution given as:
\begin{align}
\vec{\nabla} \psi/a &=\vec{\nabla} \Phi/a - \frac{\ddot{a}}{a}a^2 \vec{X}
\label{eq:cosmo1b}
\end{align}
with $\vec{\nabla} \Phi$ being the Newtonian gravitational potential gradient in comoving coordinates, where we used analytical potentials (\ie disk, bulge, dark halo), and where we use the second Friedman equation in the matter dominated epoch to calculate the term with the second time derivative of the scale factor:
\begin{align}
\frac{\ddot{a}}{a} &= -\frac{4\pi G}{3} \left(\rho_{\rm m} - 2\rho_{\Lambda}\right)
\label{eq:cosmo1c}
\end{align}

The relation between comoving ($\vec{X},\vec{V}$) and physical coordinates ($\vec{r},\vec{v}$) is given by
\begin{align}
\vec{r}&=a\left(t\right)\,\vec{X}\\
\vec{v}&= H\left(t\right)\,\vec{r} + a\left(t\right)\,\vec{V}
\label{eq:cosmo2}
\end{align}
The peculiar velocity is then $\vec{V}_{\rm pec}\e a\,\vec{V}$.
In cases where the potential of M31 is included, the orbits of MW and M31 are pre-computed with the cosmological scheme.

\begin{figure}
\begin{center}
\includegraphics[width=8.5cm]{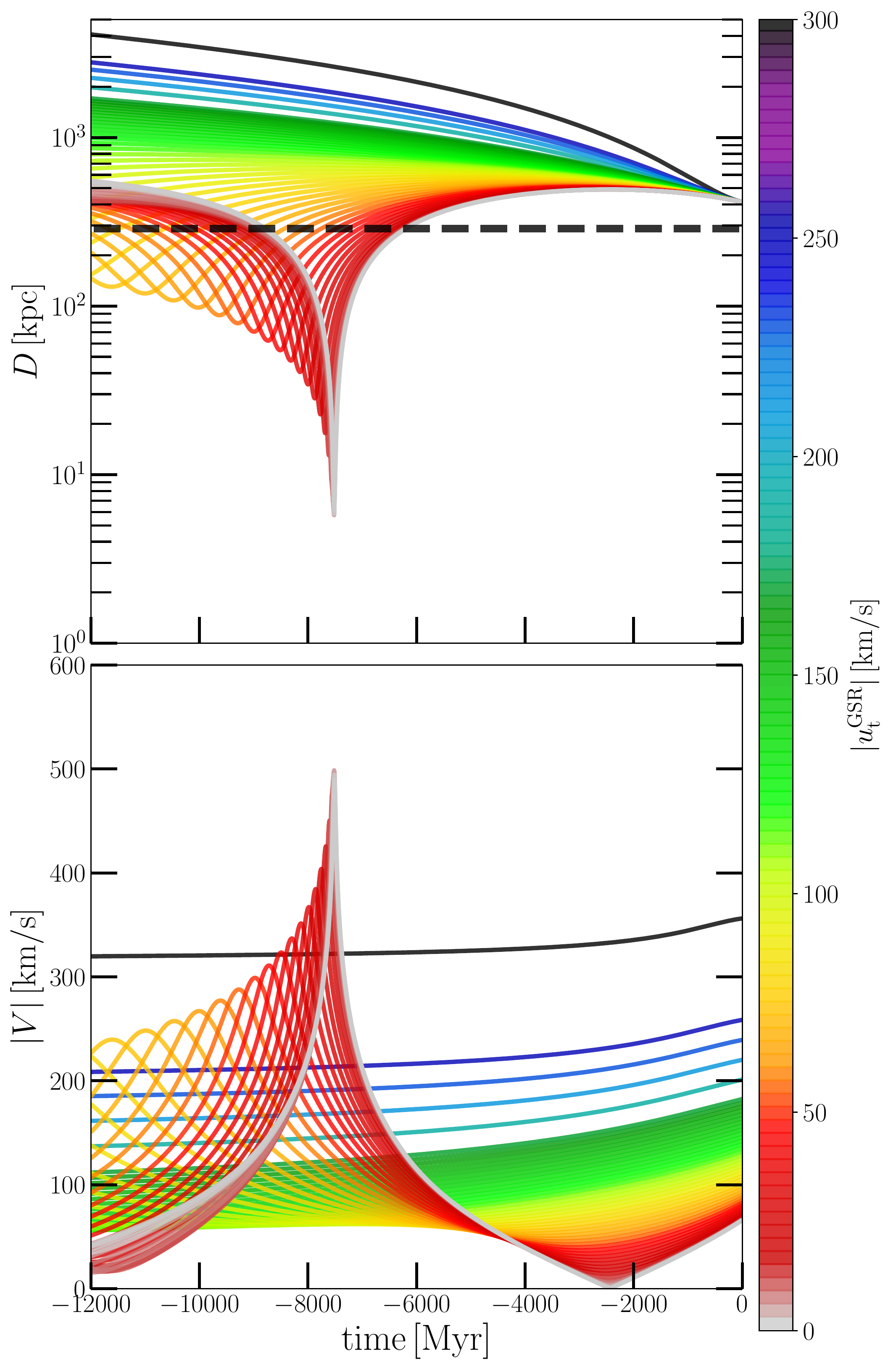}\\
\vspace{-0.4cm}
\caption{Orbits integrated $12\Gyr$ backwards in time for case 1 (constant MW $M_{\rm vir}$),
showing a range of values of \Nutgsr with selection of orbits in the direction of $\PAOV$ to avoid an overcrowding. 
Top panel: distance as function of time showing the constant virial radius (288\kpc)  (black dashed line).
Bottom panel: velocity as function of time.}
\label{fig:orb:cases}
\end{center}
\end{figure}

\begin{figure*}
\begin{center}
\includegraphics[width=15.cm]{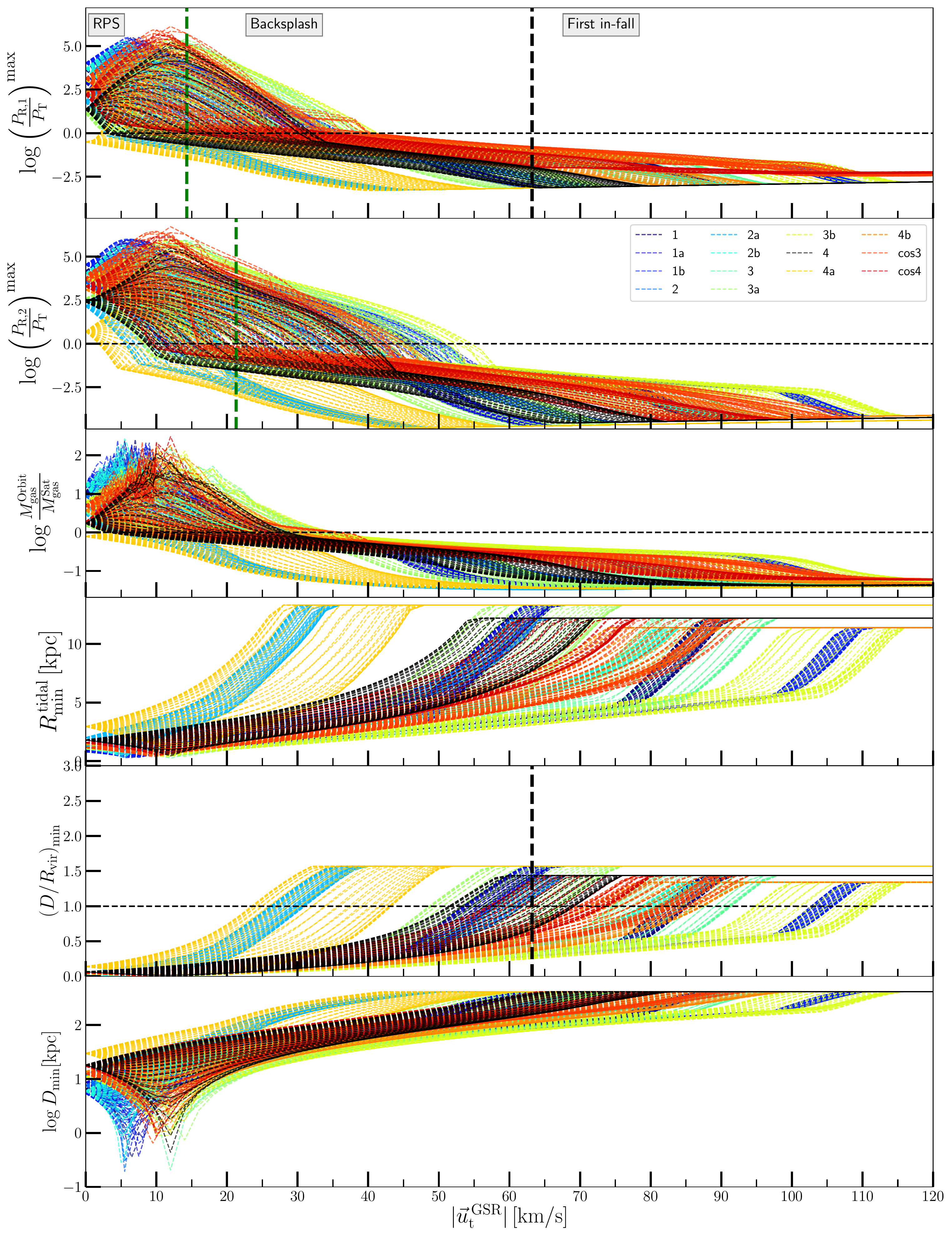}
\vspace{-0.4cm}
\caption{The main parameters for different directions of $\utgsr$ and different cases case (Table \ref{tab:cases}) as function of \Nutgsr. 
See the main text in Section \ref{sec:res:main} for the parameters's definitions.
Each coloured line corresponds to a case labeled in the second top panel, where we note
that our fiducial case 4 (MW accreting and M31 potential) is shown in black. 
From this figure we identify three main regions in \utgsr;
The region RPS where the MW ram pressure is larger than the thermal pressure of the satellite.
Each direction and case has a slightly different value of \Nutgsrrps.
We mark the median of $P_{\rm R}$ of our fiducial case 4 taking all directions of the tangential velocities finding $\Nutgsr=14^{+26}_{-14}\kms$ for $P_{\rm R,1}$ (top panel)
and $21^{+33}_{-21}\kms$ for $P_{\rm R,2}$ (second panel) (dashed green vertical lines).
The error range considers the maximum and minimum threshold velocity values found among different directions of \utgsr and different cases.
The backsplash region with orbits that passed through the MW dark matter halo,
which we mark for case 4 at the median value and range of $\Nutgsr=63^{+47}_{-39}\kms$ (dashed black vertical line).
And the first in-fall region where orbits never entered the halo.}
\label{fig:orbs:par:leot:ang}
\end{center}
\end{figure*}


\begin{figure*}
\begin{center}
\includegraphics[width=5.6cm]{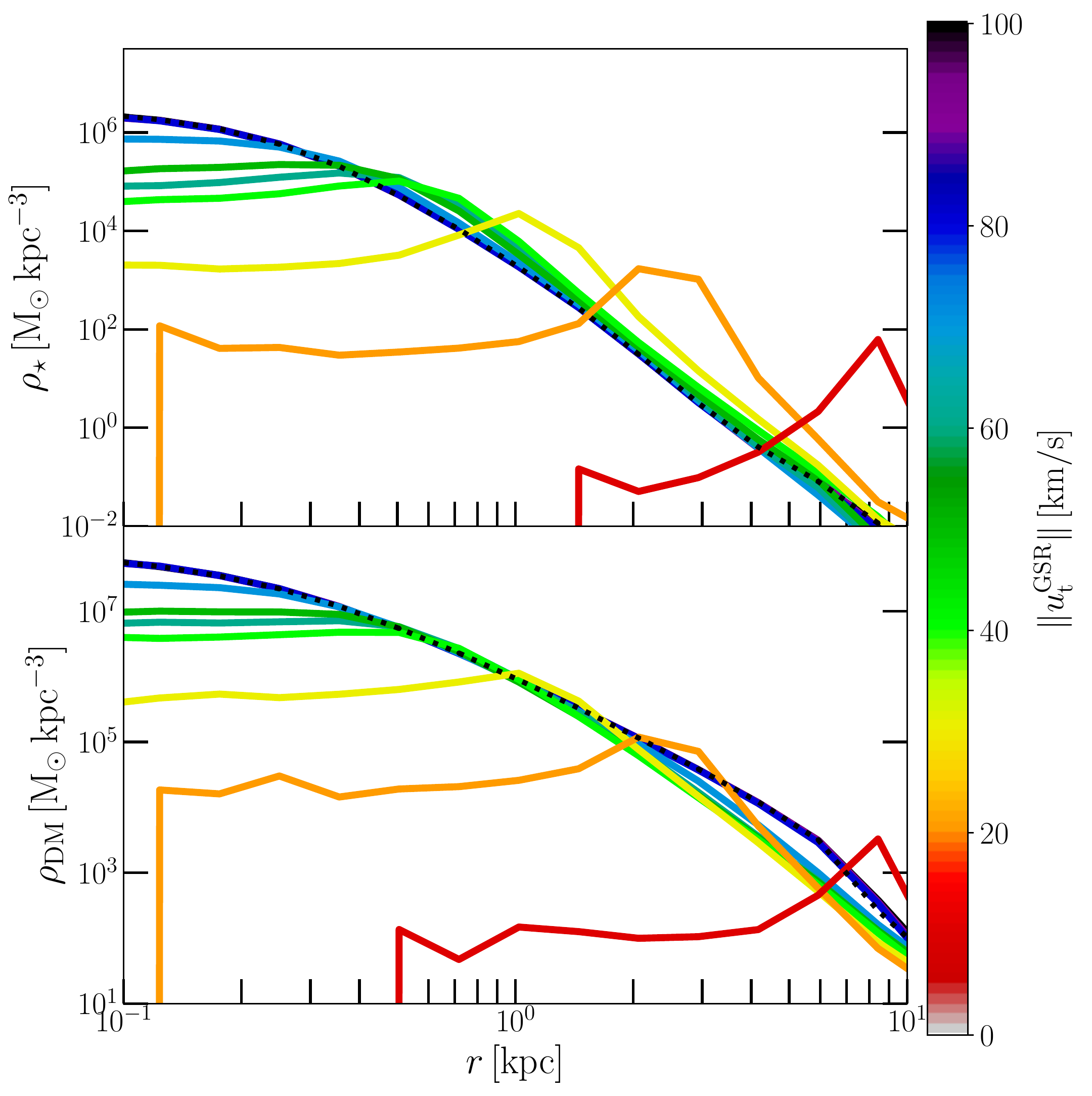}
\includegraphics[width=5.6cm]{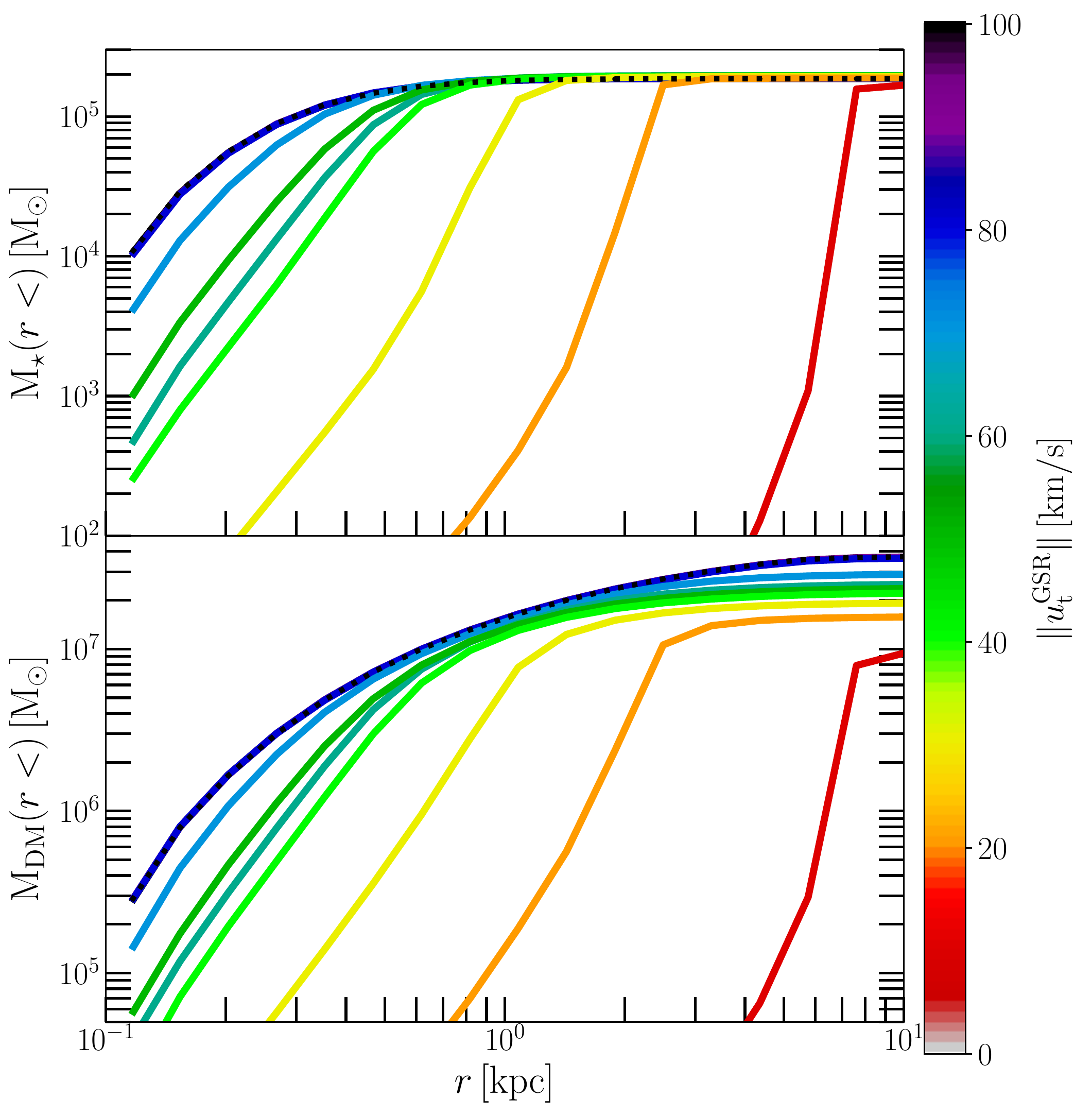}
\includegraphics[width=5.6cm]{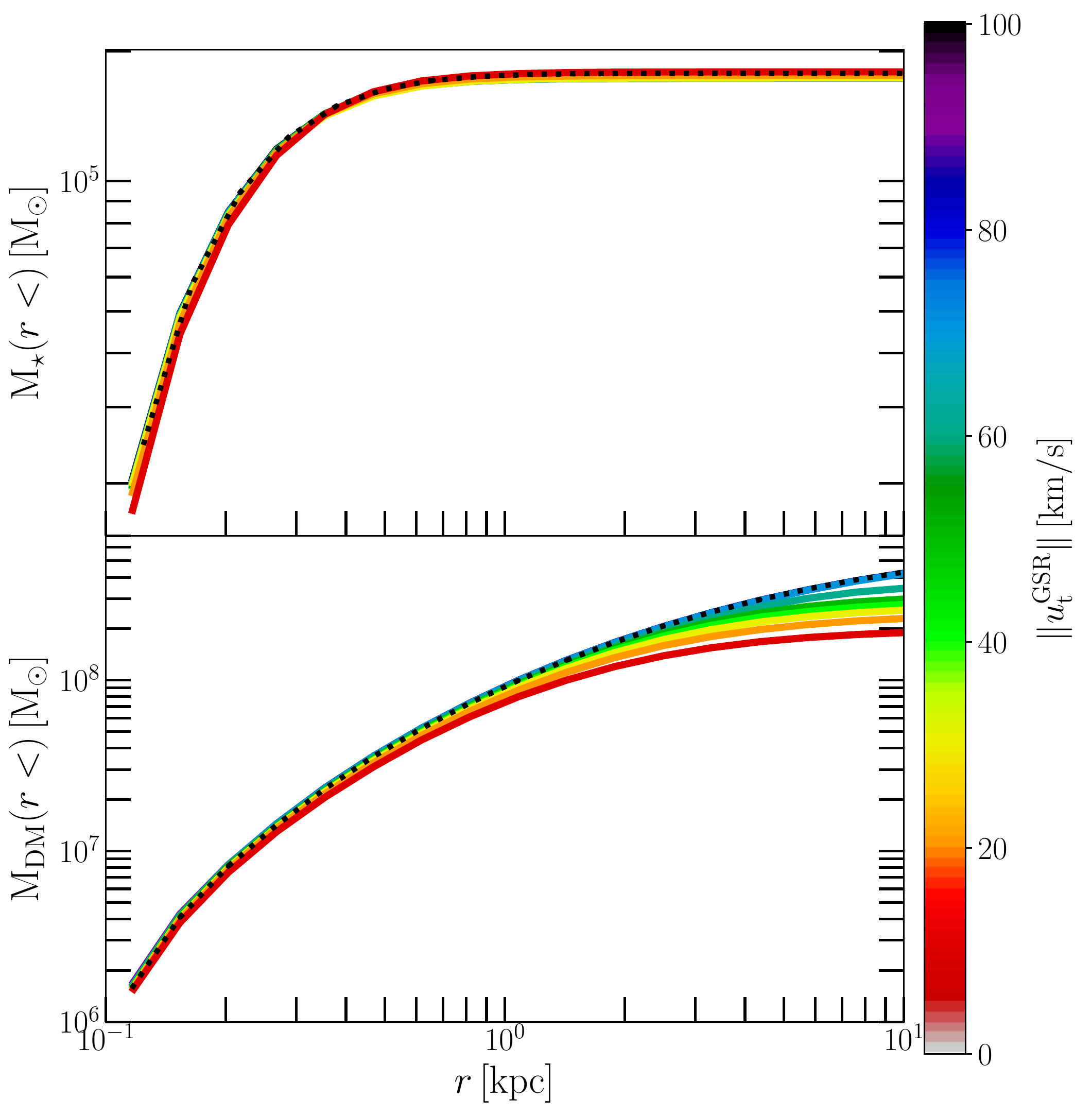}
\caption{
Left panels: Density profiles for model 1 showing the stellar mass profile (top panel) and dark matter profile (bottom panel) for N-body simulations of Leo T for different orbits.
Middle panels: Cumulative mass profiles of model 1 showing the stellar (top) and dark matter components (bottom) for the N-body models of Leo T at different snapshots. 
Right panels: same as middle panels but for model 2 (NFW dark halo).
The initial profiles after the relaxation are shown with dotted curves, and the profiles of 8 simulations after a 12\Gyr orbit, showing the orbits with $\Nutgsr=30\kms$ up to 100\kms.}
\label{fig:orbit:prof}
\end{center}
\end{figure*}

\begin{figure*}
\begin{center}
\includegraphics[width=8.cm]{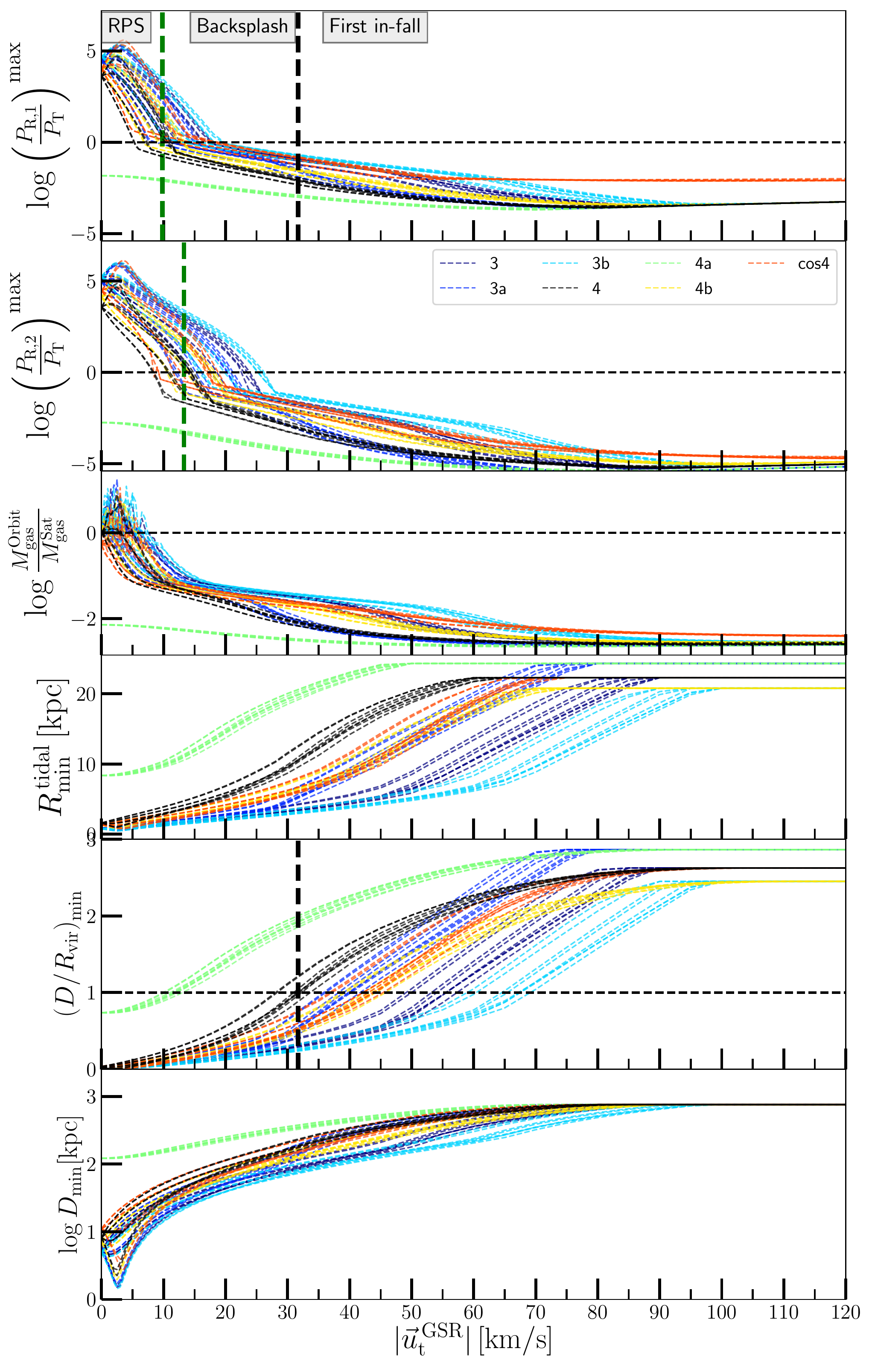}
\includegraphics[width=8.cm]{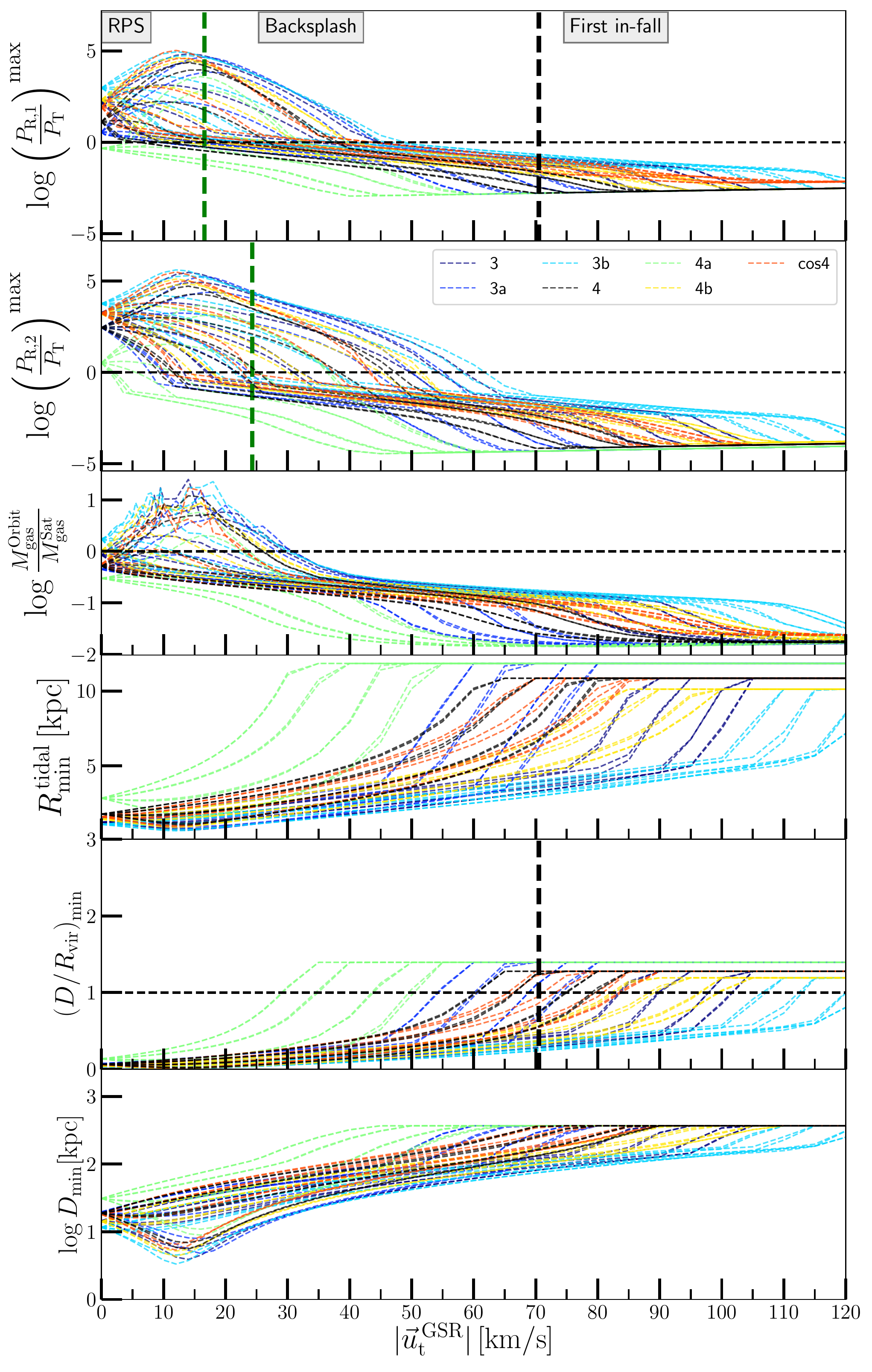}
\vspace{-0.3cm}
\caption{Main parameters explored for Cetus (left panel) and Eridanus II (right panel) for different directions of \utgsr orbit and different cases as function of \Nutgsr. 
See the main text in Section \ref{sec:res:main} for the parameters's definitions.}
\label{fig:orbs:par:dwarfs}
\end{center}
\end{figure*}

\label{lastpage}
\end{document}